\documentclass[]{emulateapj}
\usepackage{color}
\usepackage{natbib}

\newcommand{\ms}{m\ s\ensuremath{^{-1}}}
\newcommand{\mass}{\mathcal{M}}
\newcommand{\radius}{\mathcal{R}}
\newcommand{\vsini}{\ensuremath{V\sin i}}
\newcommand{\msun}{\ensuremath{\mass_\sun}}
\newcommand{\mearth}{\ensuremath{\mass_\earth}}
\newcommand{\mjup}{\ensuremath{\mass_{J}}}
\newcommand{\rsun}{\ensuremath{\radius_\sun}}
\newcommand{\rhk}{\ensuremath{\log R'_{hk}}}
\newcommand{\scale}{0.47}
\begin{document}
\shorttitle{Four New Exoplanet Candidates}

\shortauthors{Meschiari et al.}

\title{The Lick-Carnegie Survey: Four New Exoplanet Candidates}

\author{Stefano Meschiari\altaffilmark{1}, Gregory Laughlin\altaffilmark{1}, Steven S. Vogt\altaffilmark{1}, R. Paul Butler\altaffilmark{2}, Eugenio J. Rivera\altaffilmark{1},  Nader Haghighipour\altaffilmark{3} and Peter Jalowiczor}
\altaffiltext{1}{UCO/Lick Observatory, 
Department of Astronomy and Astrophysics, 
University of California at Santa Cruz,
Santa Cruz, CA 95064}
\altaffiltext{2}{Department of Terrestrial Magnetism, Carnegie Institute of Washington, Washington, DC 20015}
\altaffiltext{3}{Institute for Astronomy and NASA Astrobiology Institute,
University of Hawaii-Manoa, Honolulu, HI 96822
}

\begin{abstract}
We present  new precise HIRES radial velocity (RV) data sets of five nearby stars obtained at Keck Observatory.
HD 31253, HD 218566, HD 177830, HD 99492 and HD 74156 are host stars of spectral classes F through K and show radial velocity 
variations consistent with new or additional planetary companions in  Keplerian motion. The orbital parameters of the candidate planets 
in the five planetary systems span minimum masses of $\mass\sin i = 27.43 \mearth$ to $8.28 \mjup$, 
periods of 17.05 to 4696.95 days and eccentricities ranging from circular to extremely eccentric ($e \approx 0.63$).

{The 5th star, HD 74156, was known to have both a 52-day and a 2500-day planet, and was claimed to also harbor a 3rd planet at 336d, in apparent support of the ``Packed Planetary System'' hypothesis. Our greatly expanded data set for HD 74156 provides strong confirmation of both the 52-day and 2500-d planets, but strongly contradicts the existence of a 336-day planet, and offers no significant evidence for any other planets in the system.
}
\end{abstract}


\section{Introduction}\label{sec:intro}
The planetary census has reached an impressive 496 extrasolar planets. 
Planetary companions have been successfully detected using a variety of techniques, primarily 
radial velocities \citep[463; see e.g.][]{MayorQueloz95, Butler06, Udry07} and transit
photometry \citep[106; see e.g.][]{Henry00, Charbonneau00, Charbonneau07}. Other 
techniques employed include microlensing \citep{Bennett09}, astrometry \citep{Benedict02, Bean09}, 
stellar pulsations \citep{Silvotti07} and even direct imaging \citep{Chauvin05, Kalas08, Marois08}
\footnote{Source: \url{http://www.exoplanet.eu}, retrieved on 06/14/2010}.

The radial velocity method has been used to either detect or characterize more than 90\% of all currently known planets,
and continues to be a very important technique. Both its continued
productivity \citep[e.g.][]{FischerValenti05} and its ability to accurately probe
planetary architectures into the vicinity of the terrestrial mass region
\citep[e.g][]{Rivera05, Mayor08, Vogt10} are a testament to the
rapid technological advances.

We have been monitoring a large set of nearby stars under precise radial velocity survey with the
High Resolution Echelle Spectrometer (HIRES) at Keck for the past 17 years. In this paper, 
we present new radial velocity (RV) observations for five of our target stars: HD 31253, HD 218566, HD 177830, HD 99492 and HD 74156. 

The plan of this paper is as follows. In Section \ref{sec:rv}, we discuss the procedure followed to obtain 
and reduce the RV dataset. In Sections \ref{sec:1} through \ref{sec:5} we describe the main
stellar properties, derive model Keplerian fits with associated parameter uncertainties and discuss
the planetary systems they imply. Finally, we discuss the new planetary companions in Section \ref{sec:conc}.

\section{Radial Velocity observations and target stars}\label{sec:rv}
\begin{deluxetable*}{lccccc}
\tabletypesize{\footnotesize}
\tablecaption{Stellar parameters\label{tab:allstars}}
\tablecolumns{6}
\tablehead{{Parameter}&{HD 31253}&{HD 218566}&{HD 177830}&{HD 99492}&{HD 74156}}
\startdata
Spec. Type	& F8		&K3V		&K0IV		&K2V		&G1V	\\
$M_v$	& 3.48	 \tablenotemark{a}	&6.21	 \tablenotemark{b}	&3.3		&6.3		&3.56	 \tablenotemark{c}\\
$B-V$	& 0.58		&1.013		&1.091		&1.053		&0.581	\\
$V$	& 7.133		&8.628		&7.177		&7.383		&7.614	\\
Mass (\msun)	& 1.23 [0.05] 		&0.85 [0.03] 	 \tablenotemark{d}	&1.47		&0.83	 \tablenotemark{d}	&1.24	\\
Radius (\rsun)	& 1.71 [0.17] 		&0.86 [0.08] 		&2.62 [0.06] 		&0.96 [0.11] 		&1.64 [0.19] 	\\
Luminosity ($L_\sun$)	& 3.286 [0.446] 		&0.353 [0.032] 		&4.842 [1.003] 		&0.418 [0.057] 		&3.037 [0.485] 	\\
Distance (pc)	& 53.82 [3.45] 		&29.94 [1.11] 		&59.03 [2.77] 		&17.99 [1.14] 		&64.56 [4.93] 	\\
$V \sin i$ (km s$^{-1}$)	& 3.8		&0.0		&2.5 	&1.4		&4.3	\\
$S_{hk}$	& 0.141 [0.018] 		&0.297		&0.125 [0.016] 		&0.254 [0.033] 		& 0.144	\\
$\log R_{hk}$	& -5.11		&-4.88		&-5.37		&-4.84		&-5.08	\\
Age (Gyr)	& 3		&8.5		&2.2 - 6.6		&4		&3.7 [0.4] 	 \tablenotemark{d}\\
$[Fe/H]$	& 0.16		&0.38		&0.545 [0.03]		&0.36		&0.13	\\
$T_{eff}$ ($K$)	& 5960.0		&4820.0		&4949		&4740.0		&5960.0 [100.0] 	\\
$\log g$	& 4.1		&4.81		&3.65		&4.77		&4.4 [0.15] 	 \tablenotemark{c}\\
$P_{rot}$	& 23		&-		&65 \tablenotemark{f}		&45	 \tablenotemark{e}	&-	\\
\enddata
\tablenotetext{a}{\citet{Nordstrom04}}\tablenotetext{b}{\citet{Wright04}}\tablenotetext{c}{\citet{Naef04}}\tablenotetext{d}{\citet{Takeda07}}\tablenotetext{e}{\citet{Marcy05}}\tablenotetext{f}{\citet{Barnes01}}
\end{deluxetable*}

\begin{deluxetable*}{lccccc}
\tablecaption{Keplerian orbital solutions\label{tab:allfits}}
\tablecolumns{6}
\tablewidth{0pt}
\tabletypesize{\footnotesize}
\tablehead{{Parameter}&{HD 31253}&{HD 218566}&{HD 177830}&{HD 99492}&{HD 74156}}
\startdata
$P$ (d) & 	466 [3]  $\dagger$ & 	225.7 [0.4]  $\dagger$ & 	406.6 [0.4] & 	17.054 [0.003] & 	2520 [15] \\ 
 & 	 & 	 & 	110.9 [0.1]  $\dagger$& 	4970 [744]  $\dagger$& 	51.638 [0.004] \\ 
$e$ & 	0.3 [0.2] & 	0.3 [0.1] & 	0.009 [0.004] & 	0.13 [0.07] & 	0.38 [0.02] \\ 
 & 	 & 	 & 	0.3 [0.1] & 	0.1 [0.2] & 	0.63 [0.01] \\ 
$K$ (m s$^{-1}$) & 	12 [2] & 	8.3 [0.7] & 	31.6 [0.6] & 	7.9 [0.6] & 	115 [3] \\ 
 & 	 & 	 & 	5.1 [0.8] & 	4.9 [0.7] & 	108 [4] \\ 
$T_{peri}$ - 2,440,000 (JD) & 	10660 [19] & 	10360 [154] & 	10154 [35] & 	10449 [2] & 	8416 [33] \\ 
 & 	 & 	 & 	10179 [7] & 	9636 [2210] & 	10793.3 [0.2] \\ 
$\varpi$ (deg) & 	244 [23] & 	36 [24] & 	85 [31] & 	196 [32] & 	268 [4] \\ 
 & 	 & 	 & 	110 [29] & 	38 [64] & 	174 [2] \\ 
$M\sin i$ ($\mjup$) & 	0.50 [0.07] & 	0.21 [0.02] & 	1.49 [0.03] & 	0.087 [0.006] & 	8.2 [0.2] \\ 
 & 	 & 	 & 	0.15 [0.02] & 	0.36 [0.06] & 	1.78 [0.04] \\ 
$a$ (AU) & 	1.260 [0.006] & 	0.6873 [0.0008] & 	1.2218 [0.0008] & 	0.12186 [0.00002] & 	3.90 [0.02] \\ 
 & 	 & 	 & 	0.5137 [0.0003] & 	5.4 [0.5] & 	0.29169 [0.00001] \\ 
$P_{tr}$ & 	0.004 & 	0.007 & 	0.004 & 	0.04 & 	0.001 \\ 
 & 	 & 	 & 	0.010 & 	0.0008 & 	0.03 \\ 
$N_{obs}$ [Median uncertainty] & 	KECK: 39 [1.59] & 	KECK: 56 [1.27] & 	KECK: 88 [1.06] & 	KECK: 93 [1.36] & 	CORALIE04: 44 [7] \\ 
 & 	 & 	 & 	 & 	 & 	ELODIE04: 51 [12] \\ 
 & 	 & 	 & 	 & 	 & 	HET09: 82 [8] \\ 
 & 	 & 	 & 	 & 	 & 	KECK: 29 [1.98] \\ 
Epoch (JD) & 	2450838.7519 & 	2450366.8550 & 	2450276.0239 & 	2450462.1140 & 	2450823.5570 \\ 
$\chi^2$ & 	8.87 & 	8.41 & 	15.31 & 	7.17 & 	3.09 \\ 
RMS (m s$^{-1}$) & 	4.23 & 	3.48 & 	3.85 & 	3.22 & 	12.80 \\ 
Jitter (m s$^{-1}$) & 	3.92 & 	3.23 & 	3.71 & 	2.94 & 	8.59 \\ 
\enddata
\tablecomments{{$\dagger$} {New planet candidate}}
\end{deluxetable*}

The HIRES spectrometer \citep{Vogt94} of the Keck-I telescope was used for all the new RVs presented in this paper. Doppler shifts were measured in the usual manner \citep{Butler96} by placing an Iodine absorption cell just ahead of the spectrometer slit in the converging beam from the telescope. This gaseous Iodine absorption cell superimposes a rich forest of Iodine lines on the stellar spectrum, providing a wavelength calibration and proxy for the point spread function (PSF) of the spectrometer. The Iodine cell is sealed and temperature-controlled to 50 $\pm$ 0.1 C such that the column density of Iodine remains constant.  For the Keck planet search program, we operate the HIRES spectrometer at a spectral resolving power R $\approx$ 70,000 and wavelength range of 3700-8000\,\AA, though only the region 5000-6200\,\AA\ (with Iodine lines) is used in the present Doppler analysis. A block of the spectrum containing the Iodine region is divided into $\sim$700 chunks of 2\,\AA\ each.  Each chunk produces an independent measure of the wavelength, PSF, and Doppler shift. The final measured velocity is the weighted mean of the velocities of the individual chunks. All radial velocities have been corrected to the solar system barycenter, but are not tied to any absolute radial velocity system. As such, they are ``relative'' radial velocities, with a zero point that is usually set simply to the mean of each set.

The internal uncertainties quoted for all the RV's in this paper reflect only one term in the overall error budget, and results from a host of systematic errors from characterizing and determining the PSF, detector imperfections, optical aberrations, effects of under-sampling the Iodine lines, etc. Two additional major sources of error are photon statistics and stellar jitter. The latter varies widely from star to star, and can be mitigated to some degree by selecting magnetically-inactive older stars and by time-averaging over the star's unresolved low-degree surface p-modes. All observations have been further binned on 2-hour timescales.

We present in Table \ref{tab:allstars} a few basic  parameters (and uncertainties, where available) for all the host stars considered in this paper. 
Unless otherwise noted, the data are mostly as listed in the SPOCS database \citep{FischerValenti05}
and the NASA NStED database\footnote{http://nsted.ipac.caltech.edu/}. 

Table \ref{tab:allfits} summarizes all the Keplerian fits for the target stars in this paper; they will be discussed in more detail in the following sections. The orbital fits were derived using the Systemic Console \citep{Meschiari09}\footnote{Downloadable at \url{http://www.oklo.org}}.
The errors on each parameter are estimated using the bootstrap technique with 5000 scrambled realizations of the RV datasets. For each planet, we list best-fit period ($P$), eccentricity ($e$), semi-amplitude ($K$), time of periastron passage ($T_{peri}$), longitude of pericenter ($\varpi$), minimum mass ($\mass\ \sin i$) and semi-major axis ($a$). Additionally, we report approximate estimates of the transit probability calculated as part of the Monte-Carlo modeling, assuming a putative radius $\radius = \radius_{JUP}$.

\section{HD 31253 (HIP 22826)}\label{sec:1}
\subsection{Stellar properties}\label{sec:HD31253_star}
HD 31253 is a V = 7.133 magnitude star of spectral class F8. Relative to the Sun, HD 31253 is modestly metal-rich ([Fe/H] = 0.16).  

{This 1.23$\mass_\sun$ star has a reported \vsini{} of 3.8 \ms{} which, in conjunction with its derived radius, implies a maximum rotation period of about 23 days. Our measurement of \rhk = -5.13 agrees well with that listed on the NASA NStED site, and leads to an estimate of $\sim$2.3 for the expected radial velocity jitter due to stellar surface activity \citep{Wright05}. 
}

\subsection{Keplerian solution}

Table \ref{tab:rvdata_HD31253} shows the complete set of 39 relative radial 
velocity observations for HD 31253.  The radial velocity coverage spans 
approximately 13 years of RV monitoring.
The median internal uncertainty for our observations is 1.59 \ms, and the peak-to-peak velocity 
variation is 36.37 \ms. The velocity scatter around the average RV in our 
measurements is 8.92 \ms. 

\begin{figure}
\plotone{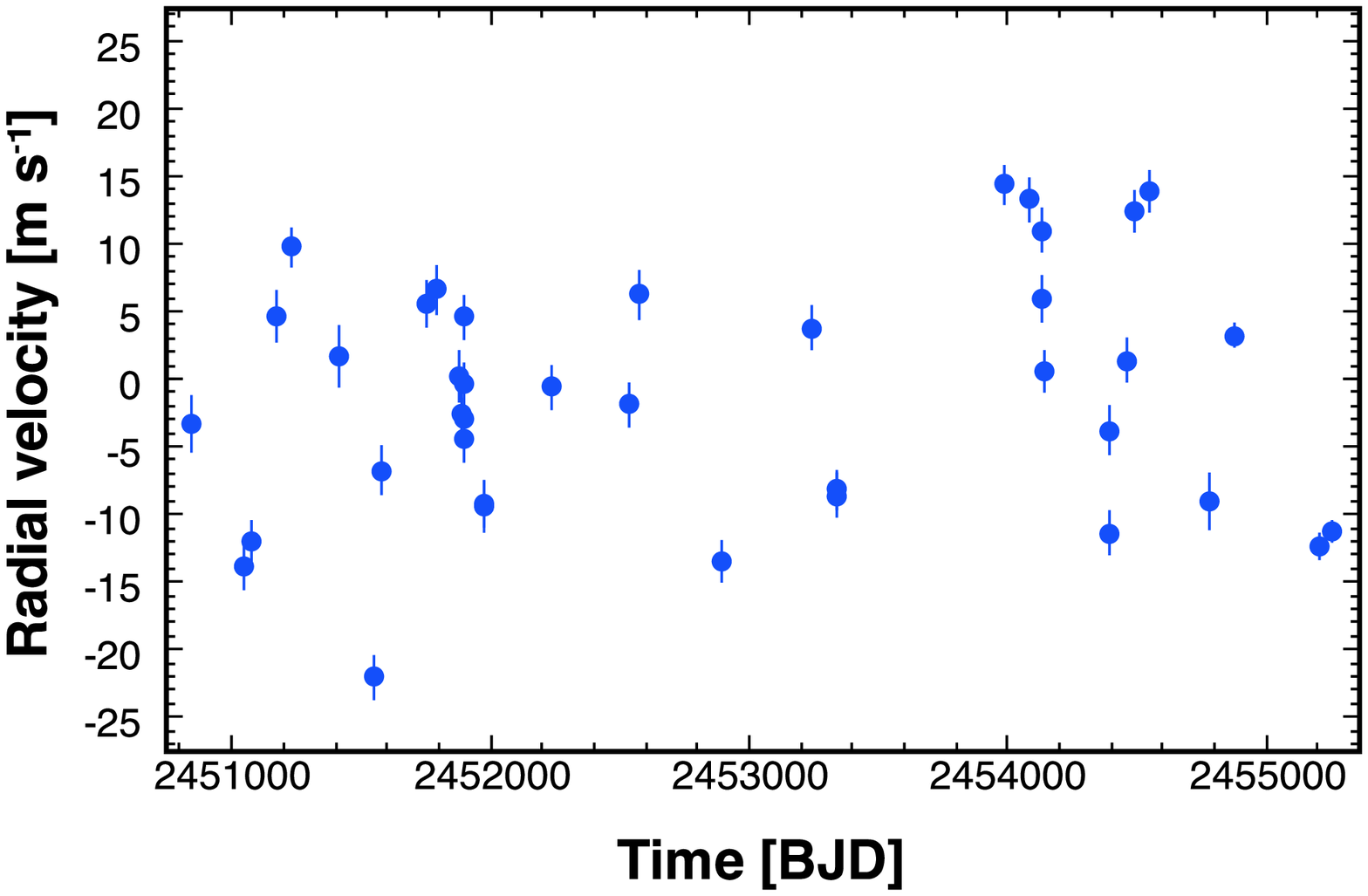}\\
\plotone{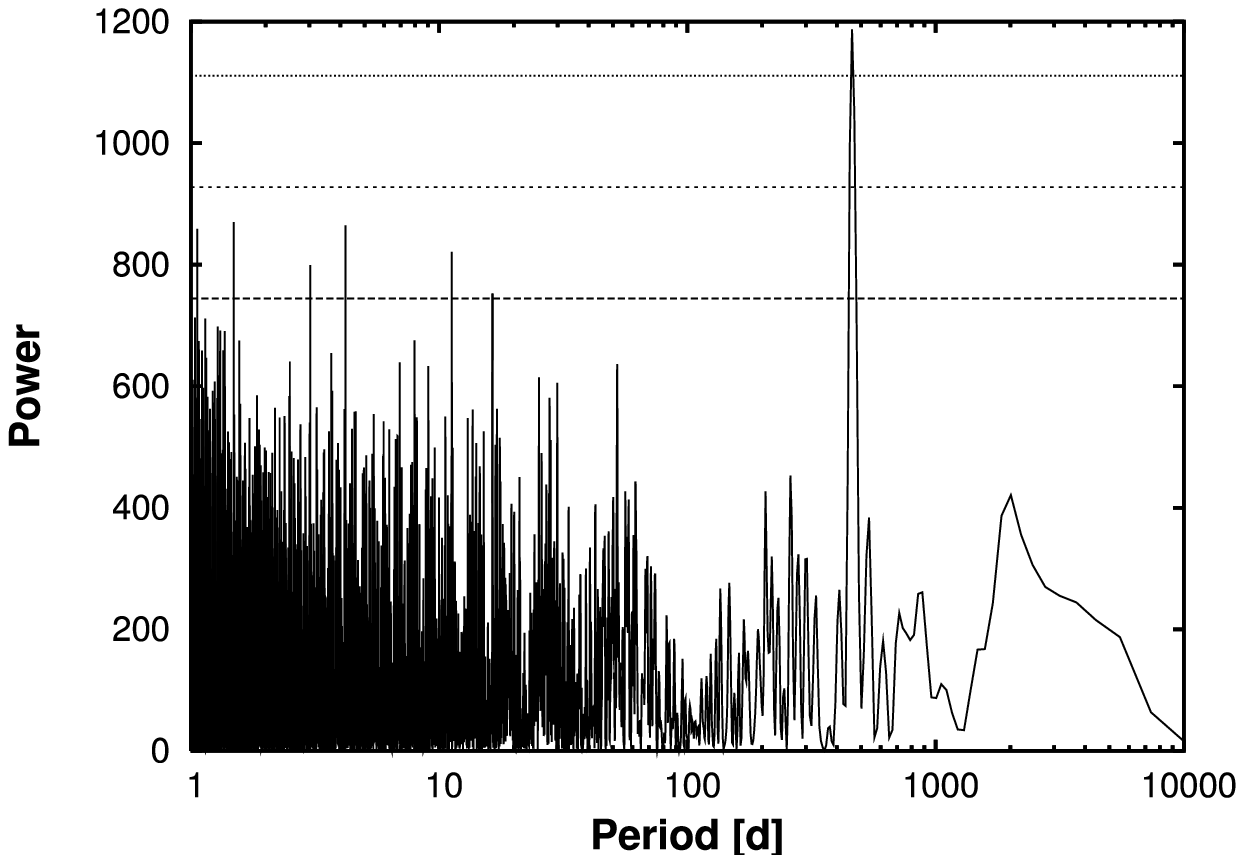}\\
\plotone{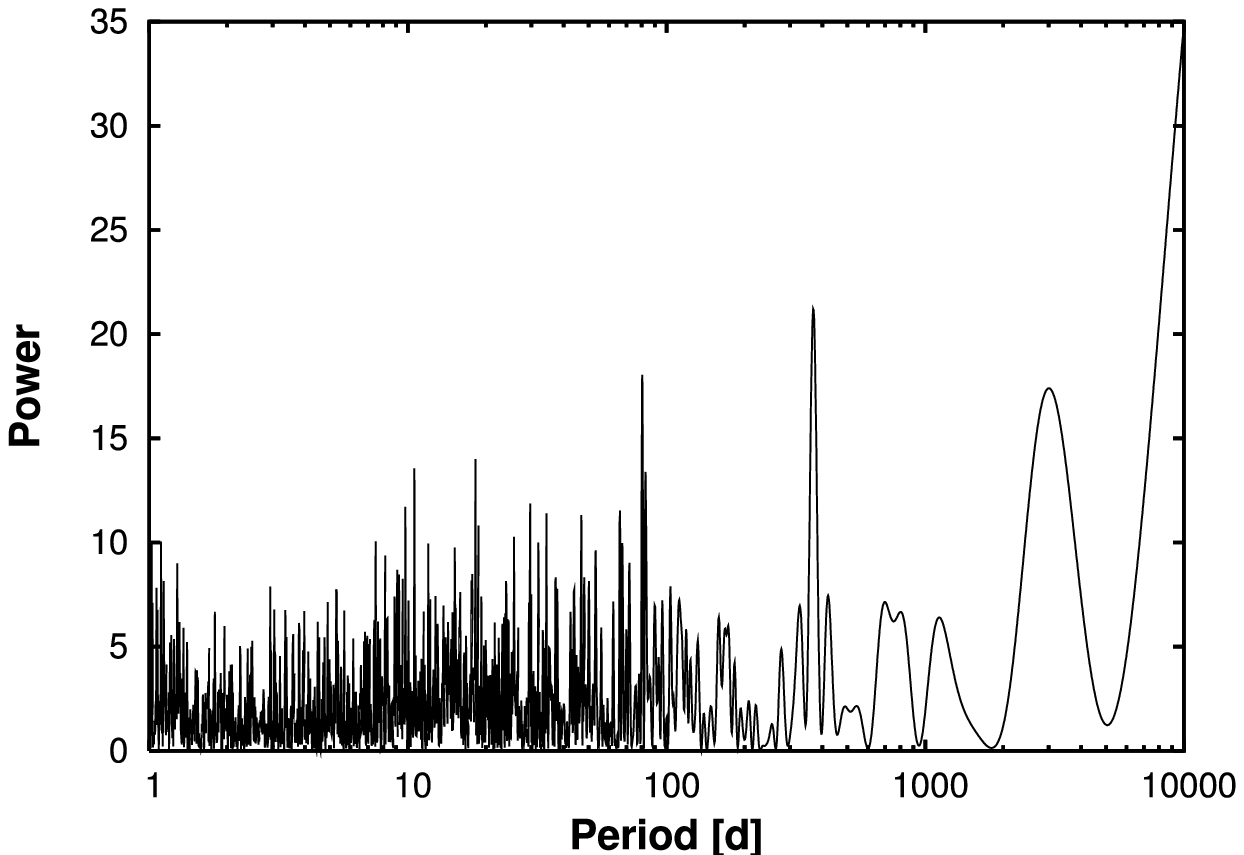}\\
\caption{Radial velocity data and periodograms for HD 31253. \textit{Top panel:} Relative radial velocity data obtained by KECK. \textit{Middle panel: } Error-weighted Lomb-Scargle periodogram of the radial velocity data. \textit{Bottom panel: } Power spectral window.}\label{fig:data_HD31253}
\end{figure}

The top panel of Figure \ref{fig:data_HD31253} shows the individual RV observations for HD 31253 
. The middle panel shows the error-weighted Lomb-Scargle (LS) periodogram of the full RV dataset \citep{Gilliland87}.
The three horizontal lines in this figure and other comparable plots represent, from top to bottom, 
the 0.1\%, 1.0\%, and 10.0\% analytic False Alarm Probability (FAP) levels, respectively. The analytic FAPs
are computed using a straightforward approach, where we estimate the number of independent frequencies
by analyzing a set of 1,000 gaussian deviates with the same timestamps as the original dataset \citep{Press}.

For the highest peak, the quoted FAP is estimated using a more robust Monte Carlo approach, which consists of
generating sets of scrambled realizations of the dataset and determining the maximum periodogram power
for each \citep[e.g.][]{Marcy05}. For all the datasets presented in this paper, we analyze $3\times 10^5$ scrambled
datasets.

The computed FAP for the strong Keplerian signal at $P  = $ 
460.32 days in the  RV dataset indicates an estimated FAP $\approx 4 \times 10^{-5}$. 
Finally, the lower panel of Figure \ref{fig:data_HD31253} shows the spectral window.
A peak at frequency, $f_s$ in the spectral window function can be associated with aliases occurring at $\left| f_p \pm f_s\right|$, where $f_p$ is a true periodicity of the input signal. For more details, see \citet{DawsonFabrycky10}. Peaks in the spectral window function are often associated with relatively immutable periodicities in the observational cadence, such as those arising from the sidereal and solar day, the lunar synodic month and the solar year.
The strongest peak in the periodogram is well-fit by a Keplerian orbit of period 465.54
days and semi-amplitude $K = 12.22$ \ms. 
Together with the assumed stellar mass of 1.23 $\msun$,
this amplitude implies a minimum mass  
of $\mass \sin i = 0.50 \mjup$. The best-fit orbit for the planet is mildly eccentric ($e \approx 0.34$). 
This 1-planet fit achieves a reduced $\chi^2 = 8.87 $, with an RMS of 4.23 \ms. The expected jitter of HD 31253 (that is, the amount of jitter required to bring the reduced $\chi^2$ of the best-fit solution to 1.0) is 3.92 \ms. 

\begin{figure}
	\plotone{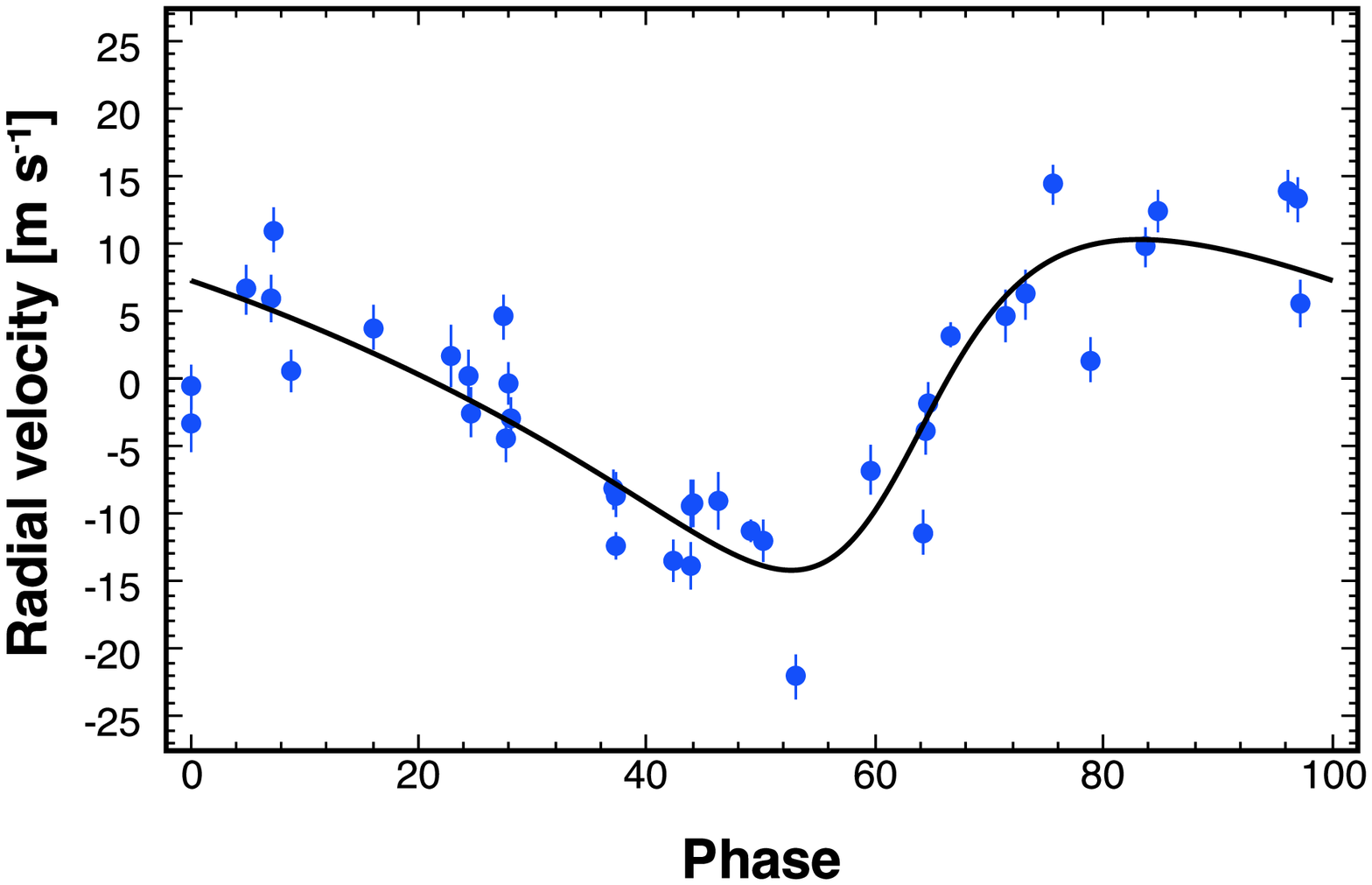}\\
\plotone{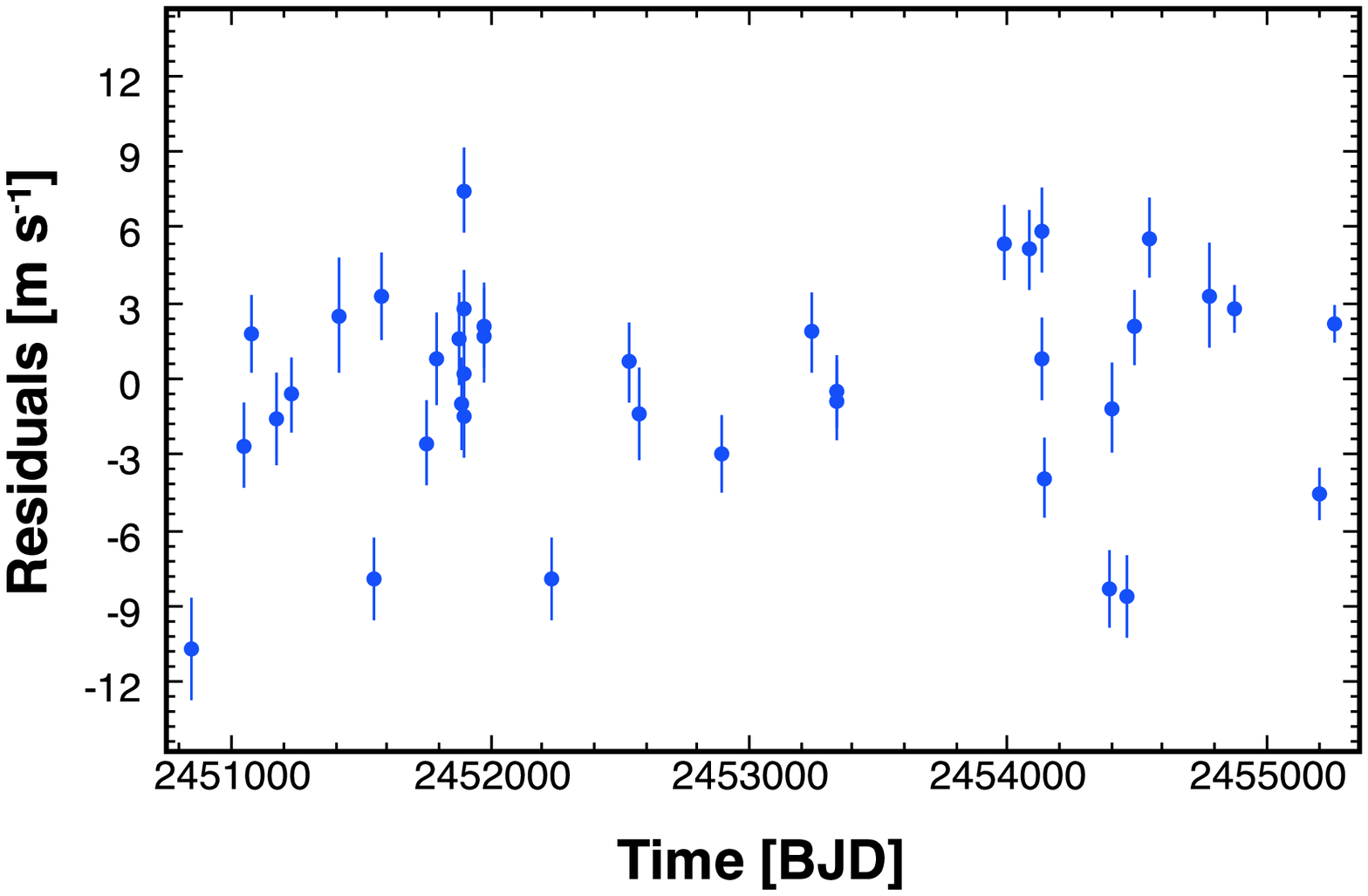}\\
\plotone{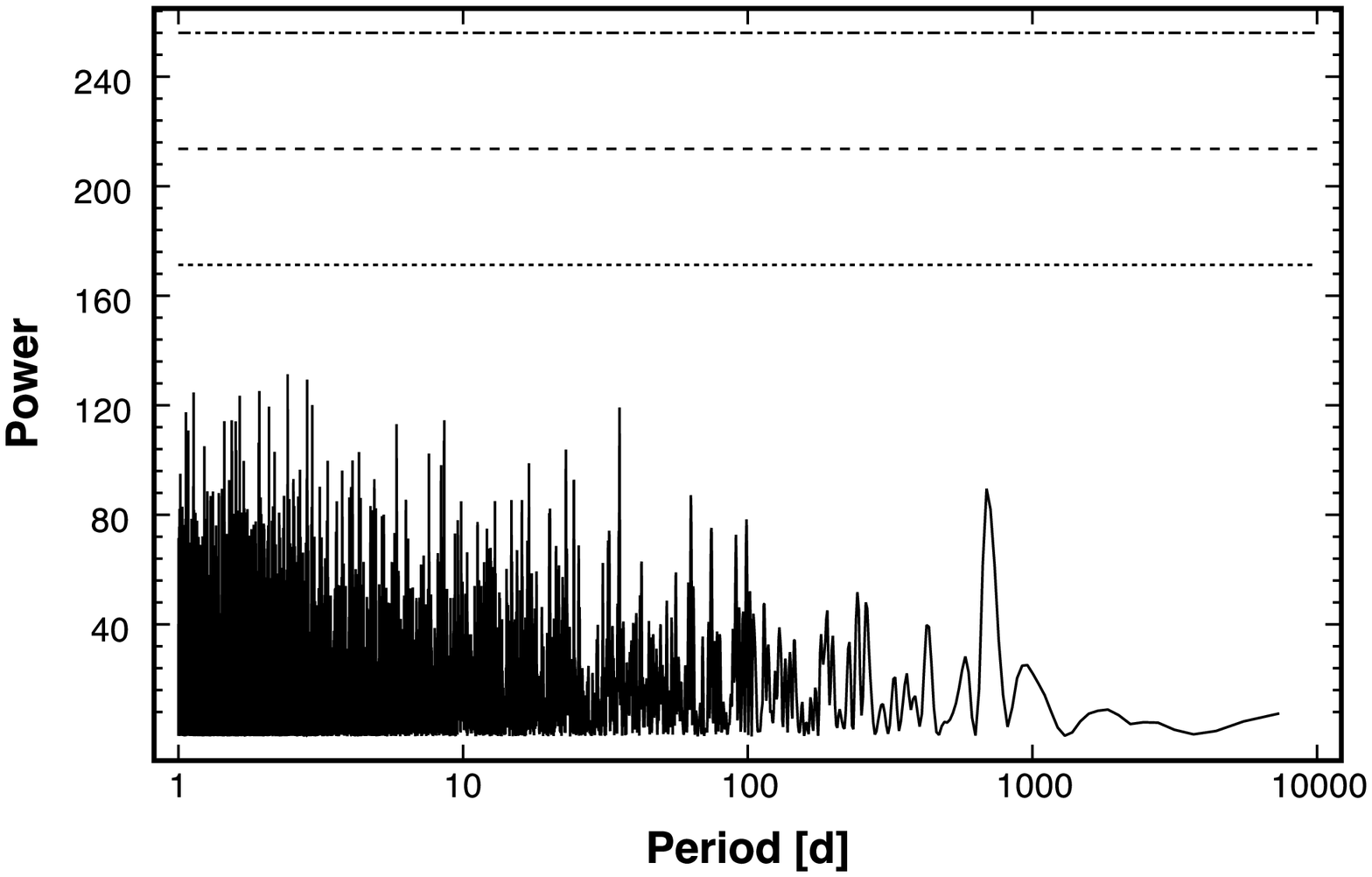}\\
\caption{Keplerian solution and residuals periodogram for HD 31253.
\textit{Top panel:} Phased Keplerian fit. \textit{Middle panel: } Residuals to the 1-planet Keplerian fit. \textit{Bottom panel: } Periodogram of the residuals to the 1-planet best-fit solution.}
\label{fig:bestfit_HD31253}
\end{figure}

The top panel of Figure \ref{fig:bestfit_HD31253} shows the phased stellar reflex velocity of HD 31253 compared to the 
RV dataset. The middle panel shows the residuals to the 1-planet solution. 
Finally, the bottom panel shows the periodogram of the residuals of the best-fit solution. No 
interesting peaks are evident, indicating that the present 
data set provides no strong support for additional planets in the system.

{We do not have photometry of HD 31253 that might conclusively rule out stellar rotation signatures as a cause of the RV variations. But 
this star does have a measured \vsini{} of 3.8 \ms{} which implies a maximum rotation period of about 23 days, much shorter than the 466-d Keplerian period. The semi-amplitude of the observed variations is 12 \ms, whereas a 466-d rotation period, combined with a stellar radius of 1.71 \rsun, would not produce radial velocity effects above a few tenths of a \ms. Therefore,  stellar rotation can sensibly be ruled out as being responsible for the observed RV variations.

}

\begin{deluxetable}{ccc}
		\tablewidth{0pt}
		\tablecaption{KECK radial velocities for HD 31253 (\textit{Sample: full table in electronic version})
		\label{tab:rvdata_HD31253}}
		\tablecolumns{3}
		\tablehead{{Barycentric JD}&{RV [\ms]}&{Uncertainty [\ms]}}
		\startdata
		2450838.75 & -1.44 & 1.97\\ 
2451043.12 & -11.98 & 1.64\\ 
2451073.03 & -10.12 & 1.43\\ 
2451170.91 & 6.52 & 1.78\\ 
2451228.79 & 11.64 & 1.38\\ 
2451411.13 & 3.54 & 2.19\\ 
2451550.87 & -20.13 & 1.58\\ 
2451581.86 & -4.86 & 1.69\\ 
2451757.13 & 7.45 & 1.58\\ 
2451793.12 & 8.50 & 1.78\\ 
2451883.00 & 2.05 & 1.80\\ 
2451884.08 & -0.63 & 1.73\\ 
2451898.01 & 6.44 & 1.61\\ 
2451899.00 & -2.59 & 1.54\\ 
2451899.99 & 1.53 & 1.48\\ 
2451901.01 & -1.13 & 1.47\\ 
2451973.75 & -7.57 & 1.83\\ 
2451974.76 & -7.26 & 1.63\\ 
2452235.85 & 1.28 & 1.62\\ 
2452536.09 & 0.00 & 1.55\\ 

		\enddata
		\end{deluxetable}

\section{HD 218566 (HIP 114322)}\label{sec:2}
\subsection{Stellar properties}\label{sec:HD218566_star}
HD 218566 is a V = 8.628 magnitude star of spectral class 
K3V. In comparison to the Sun, HD 218566 is quite metal-rich ([Fe/H] = 0.38).  Table \ref{tab:allstars} reports some of the salient stellar properties, as reported by NStEd, \citet{Wright04} and \citet{Takeda07}. 

\subsection{Keplerian solution}
		Table \ref{tab:rvdata_HD218566} shows the 56 relative radial 
velocity observations for HD 218566.  The radial velocity coverage spans 
approximately 14 years of RV monitoring.
The median internal uncertainty for our observations is 1.27 \ms, and the peak-to-peak velocity 
variation is 28.46 \ms. The velocity scatter around the mean RV in our 
measurements is 7.18 \ms. 

\begin{figure}
\plotone{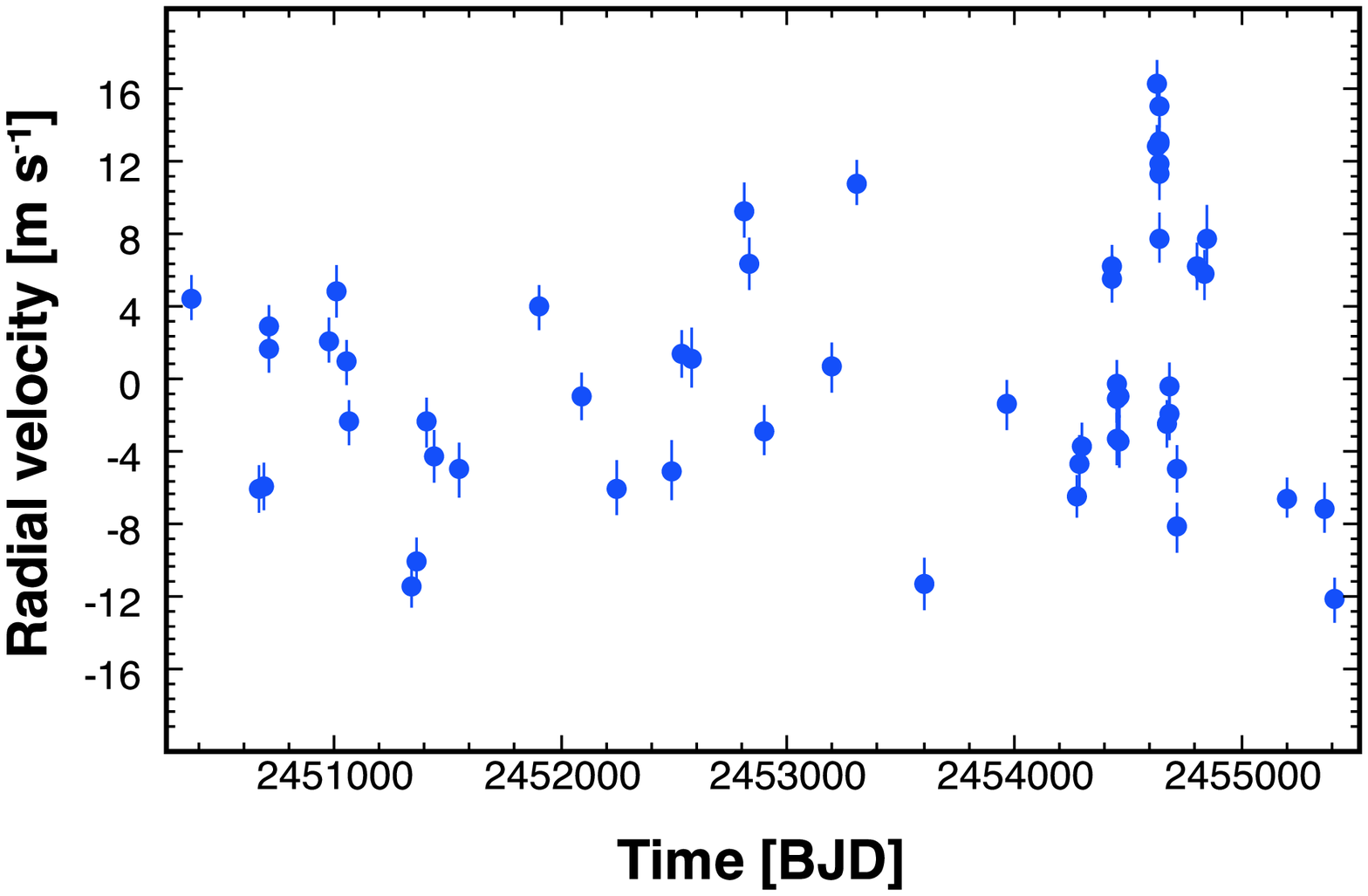}\\
\plotone{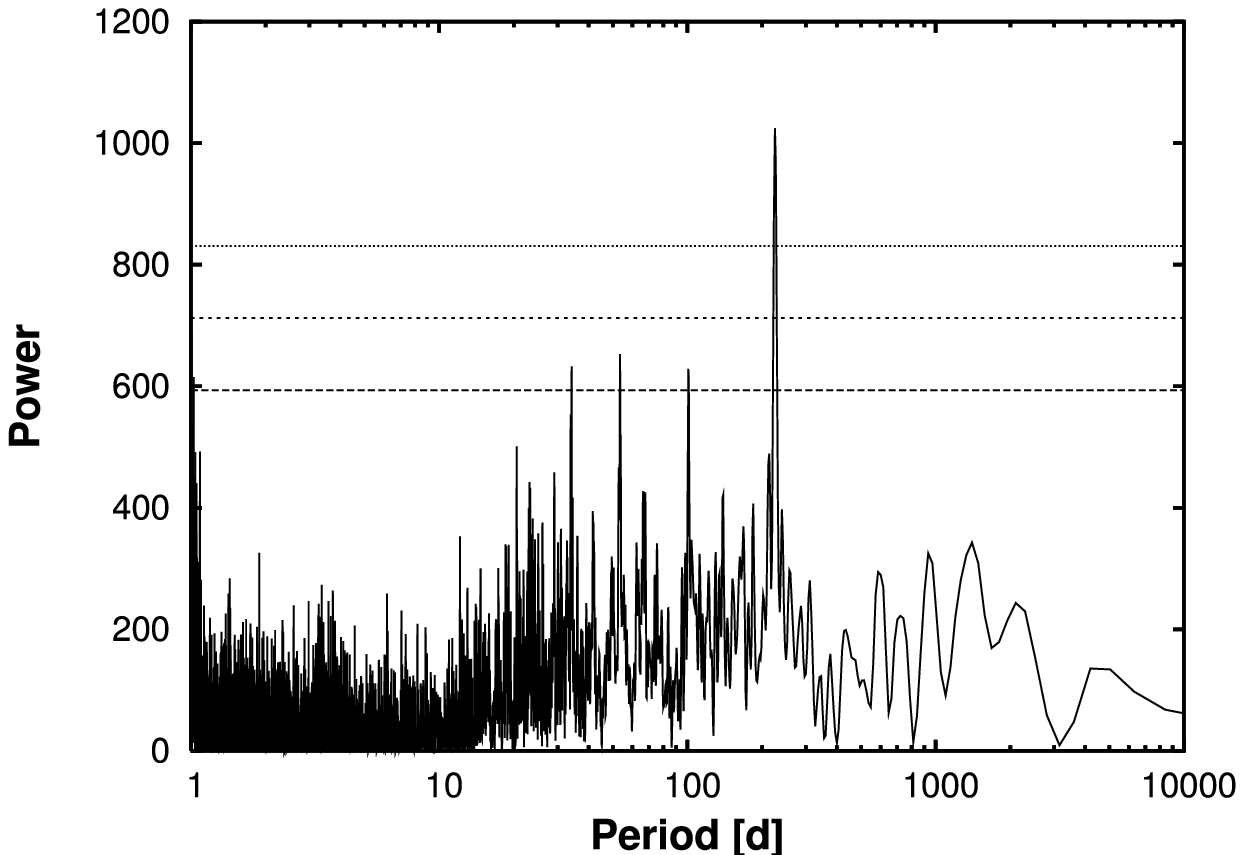}\\
\plotone{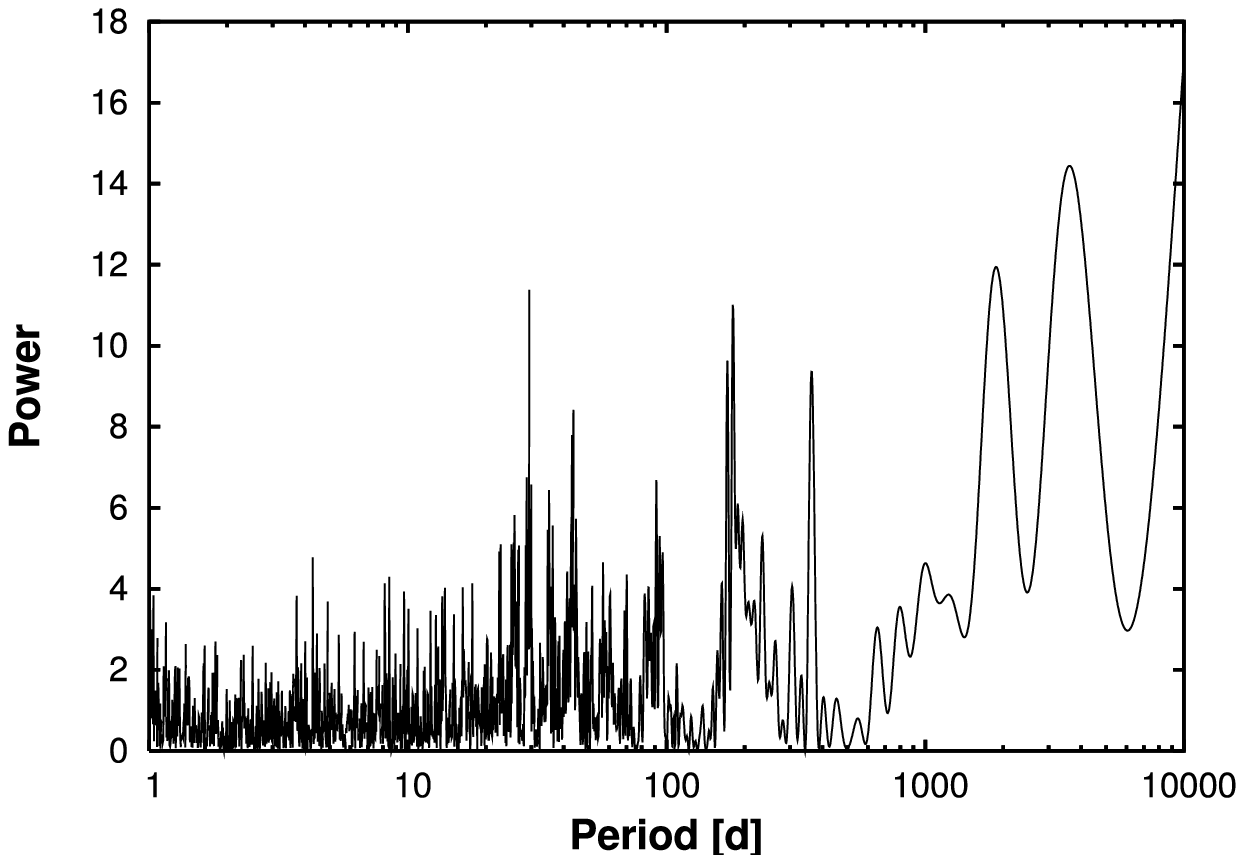}\\
\caption{Radial velocity data and periodograms for HD 218566. \textit{Top panel:} Relative radial velocity data obtained by KECK. \textit{Middle panel: } Error-weighted Lomb-Scargle periodogram of the radial velocity data. \textit{Bottom panel: } Power spectral window.}\label{fig:data_HD218566}
\end{figure}

The top panel of Figure \ref{fig:data_HD218566} shows the individual RV observations for HD 218566. The middle panel shows the error-weighted Lomb-Scargle (LS) periodogram of the full RV data set, while the bottom figure shows the spectral window.
The FAP calculation for the strong Keplerian signal at $P  = $ 
225.06 days in the  RV dataset indicates an estimated FAP $\approx < 4 \times 10^{-6}$. 
The dominant peak in the periodogram can be explained by a Keplerian orbit of period 225.73
days and semi-amplitude $K = 8.34 $ \ms. 
This amplitude suggests a minimum mass  
of $\mass \sin i = 0.21 \mjup$ (assuming a stellar mass of 0.88 $\msun$). The best-fit orbit for the planet is moderately eccentric ($e \approx 0.37$). 
This fit achieves a reduced $\chi^2 = 8.41$, with an RMS of 3.48 \ms. The expected jitter of HD 218566 (that is, the amount of jitter required to bring the reduced $\chi^2$ of the best-fit solution to 1.0) is 3.23 \ms. 

\begin{figure}
	\plotone{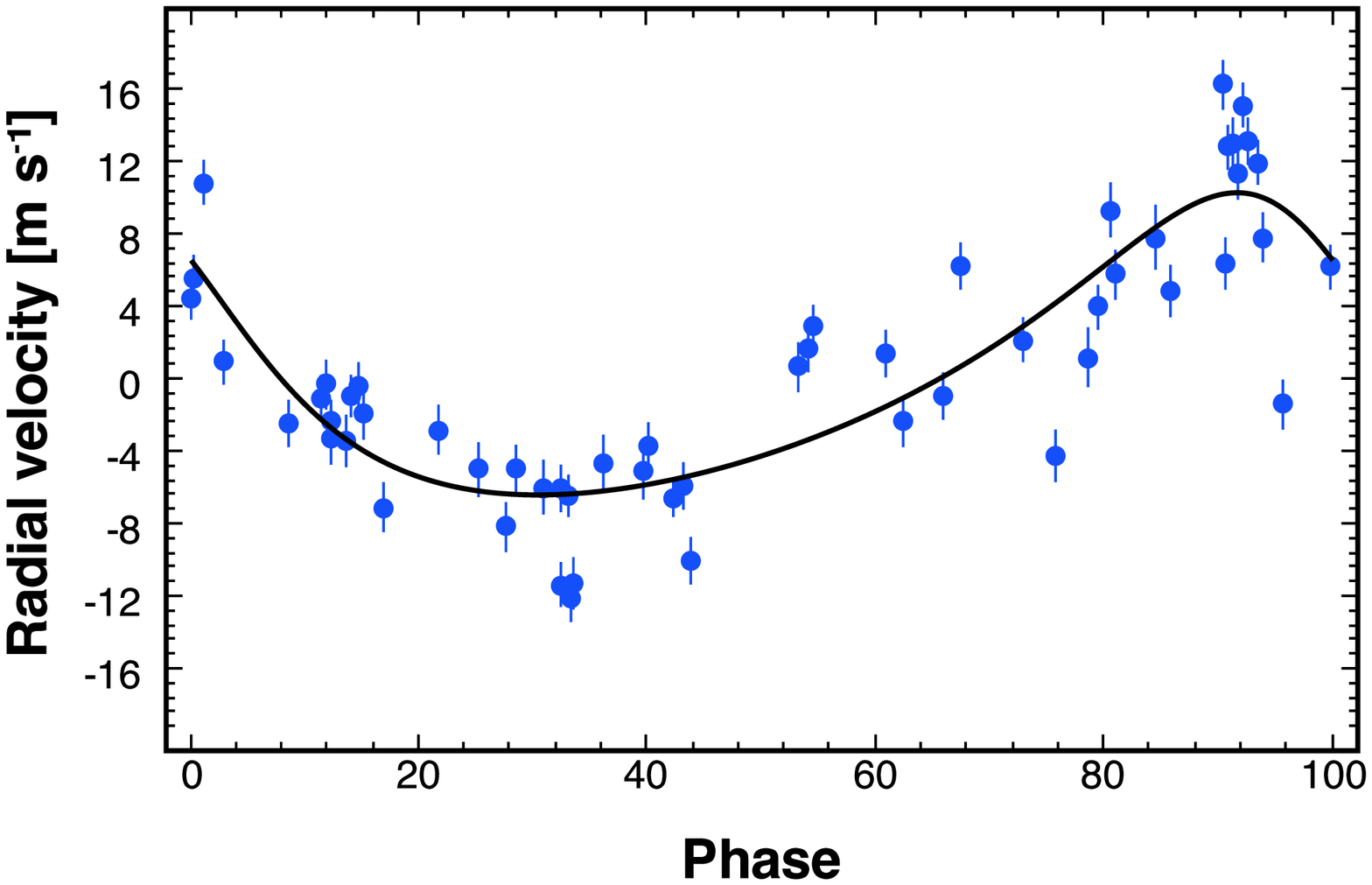}\\
\plotone{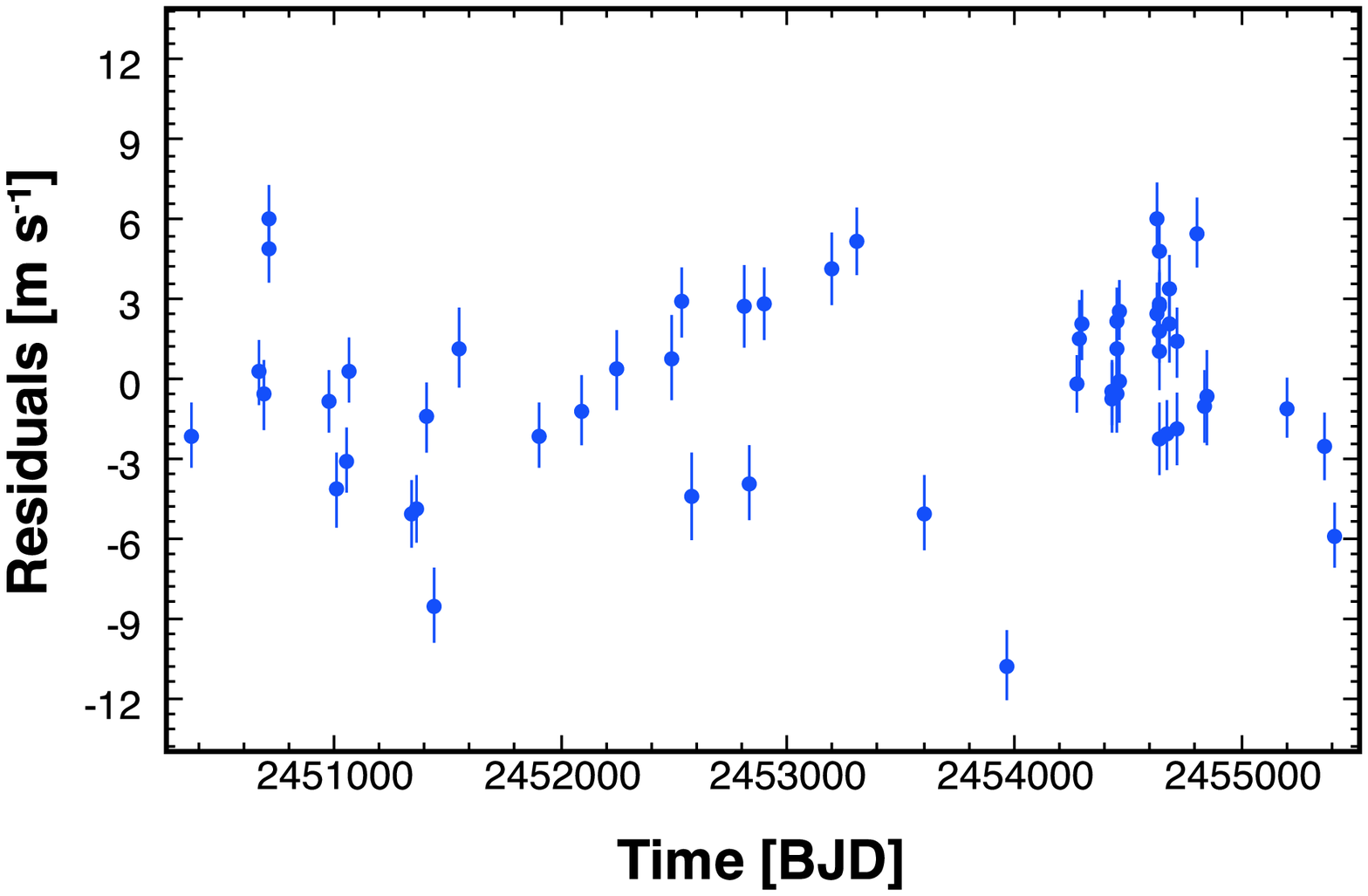}\\
\plotone{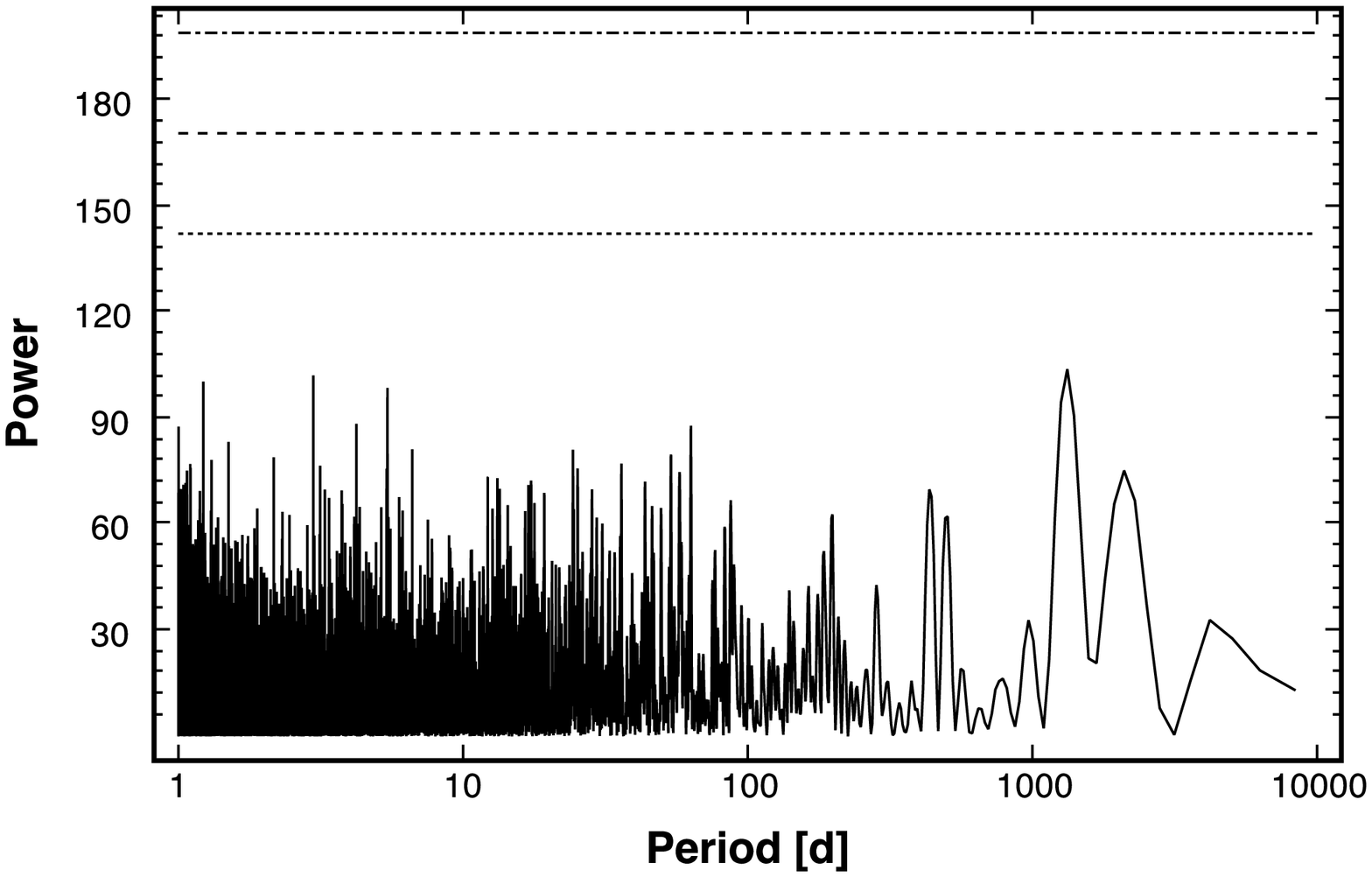}
\caption{Keplerian solution and residuals periodogram for HD 218566.
\textit{Top panel:} Phased Keplerian fit. \textit{Middle panel: } Residuals to the 1-planet Keplerian fit. \textit{Bottom panel: } Periodogram of the residuals to the 1-planet best-fit solution.}
\label{fig:bestfit_HD218566}
\end{figure}

The top panel of Figure \ref{fig:bestfit_HD218566} shows the phased stellar reflex velocity of HD 218566 compared to the 
RV dataset. The middle panel shows the residuals to the 1-planet solution. The periodogram of the residuals to the best-fit solution, shown in the bottom panel,  displays no strong peaks that would support the evidence for additional planets in the system.

{}

\begin{deluxetable}{ccc}
		\tablewidth{0pt}
		\tablecaption{KECK radial velocities for HD 218566 (\textit{Sample: full table in electronic version})
		\label{tab:rvdata_HD218566}}
		\tablecolumns{3}
		\tablehead{{Barycentric JD}&{RV [\ms]}&{Uncertainty [\ms]}}
		\startdata
		2450366.85 & 4.96 & 1.16\\ 
2450666.09 & -5.59 & 1.20\\ 
2450690.03 & -5.47 & 1.27\\ 
2450714.99 & 2.14 & 1.22\\ 
2450715.94 & 3.39 & 1.16\\ 
2450983.10 & 2.63 & 1.14\\ 
2451012.04 & 5.34 & 1.35\\ 
2451050.94 & 1.41 & 1.24\\ 
2451071.96 & -1.88 & 1.19\\ 
2451343.04 & -10.88 & 1.26\\ 
2451369.04 & -9.64 & 1.23\\ 
2451410.99 & -1.87 & 1.24\\ 
2451440.89 & -3.74 & 1.36\\ 
2451552.73 & -4.50 & 1.49\\ 
2451900.74 & 4.46 & 1.17\\ 
2452096.07 & -0.44 & 1.28\\ 
2452242.73 & -5.51 & 1.47\\ 
2452488.05 & -4.53 & 1.57\\ 
2452535.87 & 1.94 & 1.27\\ 
2452575.74 & 1.66 & 1.63\\ 

		\enddata
		\end{deluxetable}

\section{HD 177830 (HIP 93746)}\label{sec:3}
\subsection{Stellar properties}\label{sec:HD177830_star}
HD 177830 is a V = 7.177 magnitude star of spectral class K0IV. Relative to the Sun, HD 177830 is quite metal-rich ([Fe/H] = 0.55).  

{HD 177830 is a subgiant with $M_v$ = 3.32, close to giant status. The star has a known early M stellar companion with a projected separation of 97 AU \citep{Eggenberger07}. It was first reported by \citet{Vogt00} to host a 392-day Jovian-mass planet in an eccentric (e=0.42) orbit. Updates to the orbit were provided by \citet{Butler06}. \cite{Wright07} noted a possible 111-day or 46.8-day signal but suggested that it could be correlated noise. \citet{Tanner09} studied the star using \emph{Spitzer} to place limits on the amount of dust in the system, and concluded that no significant excess emission at 160 $\mu$m was detected \citep[see also][]{Trilling08, Bryden09}. 

The stellar parameters for this star listed in Table \ref{tab:allstars} are a compilation of various results, mostly from the SPOCS database \citet{FischerValenti05} with additions from the NStED database. The values for the stellar mass in Table \ref{tab:allstars} are the lower and upper limits of the isochrone mass listed in the SPOCS database. We find a current \rhk value of -5.37. HD 177830 has a derived rotation period of 65 days \citep{Barnes01}.
}

\subsection{Keplerian solution}

We show the 88 Keck radial velocity measurements in Table \ref{tab:rvdata_HD177830}, spanning 
approximately 15 years of RV monitoring.
The median internal uncertainty for our observations is 1.05 \ms, and the peak-to-peak velocity 
variation is 87.15 \ms. The velocity scatter around the average RV in our 
observations is 24.68 \ms. 

\begin{figure}
\plotone{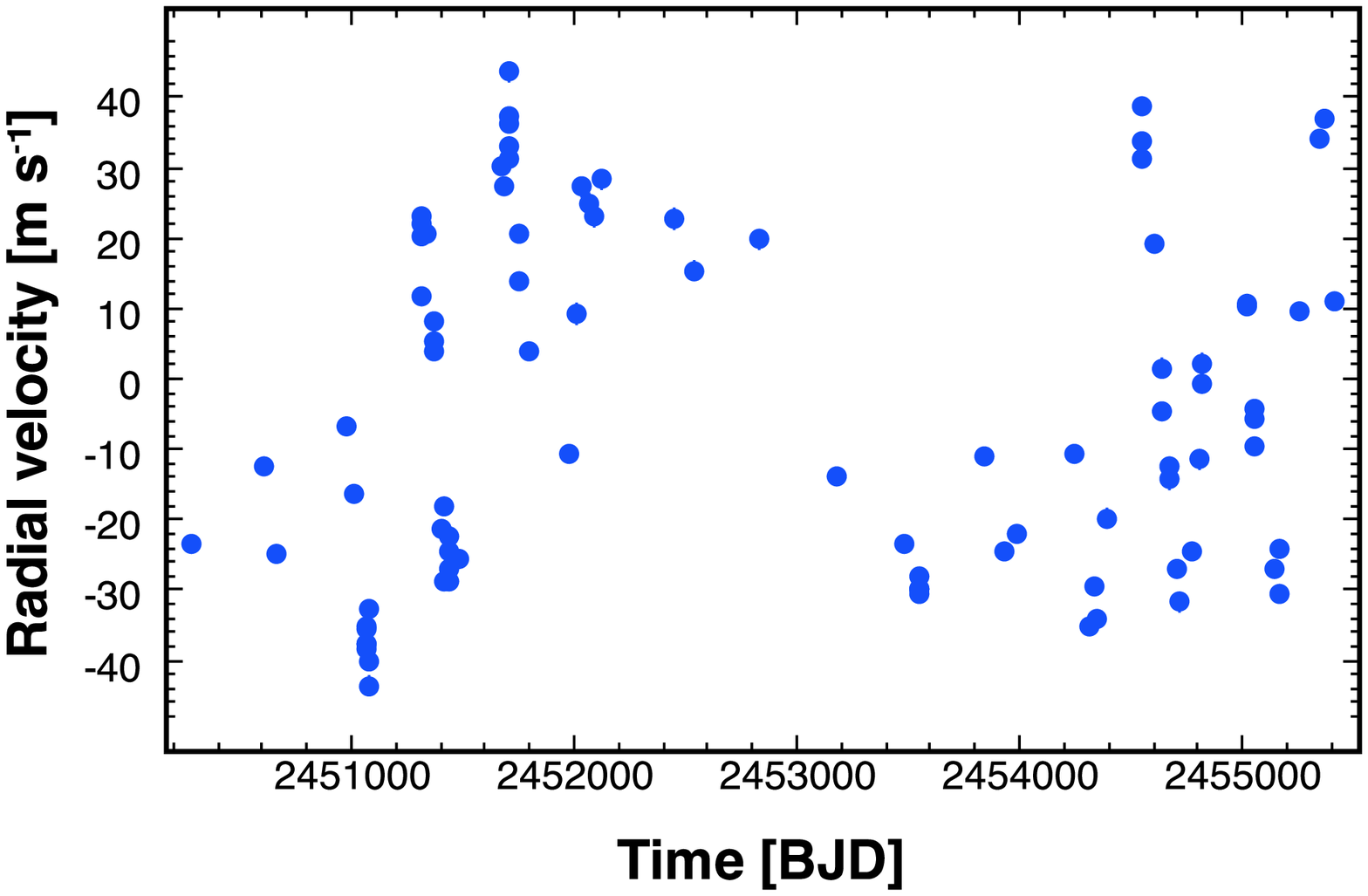}\\
\plotone{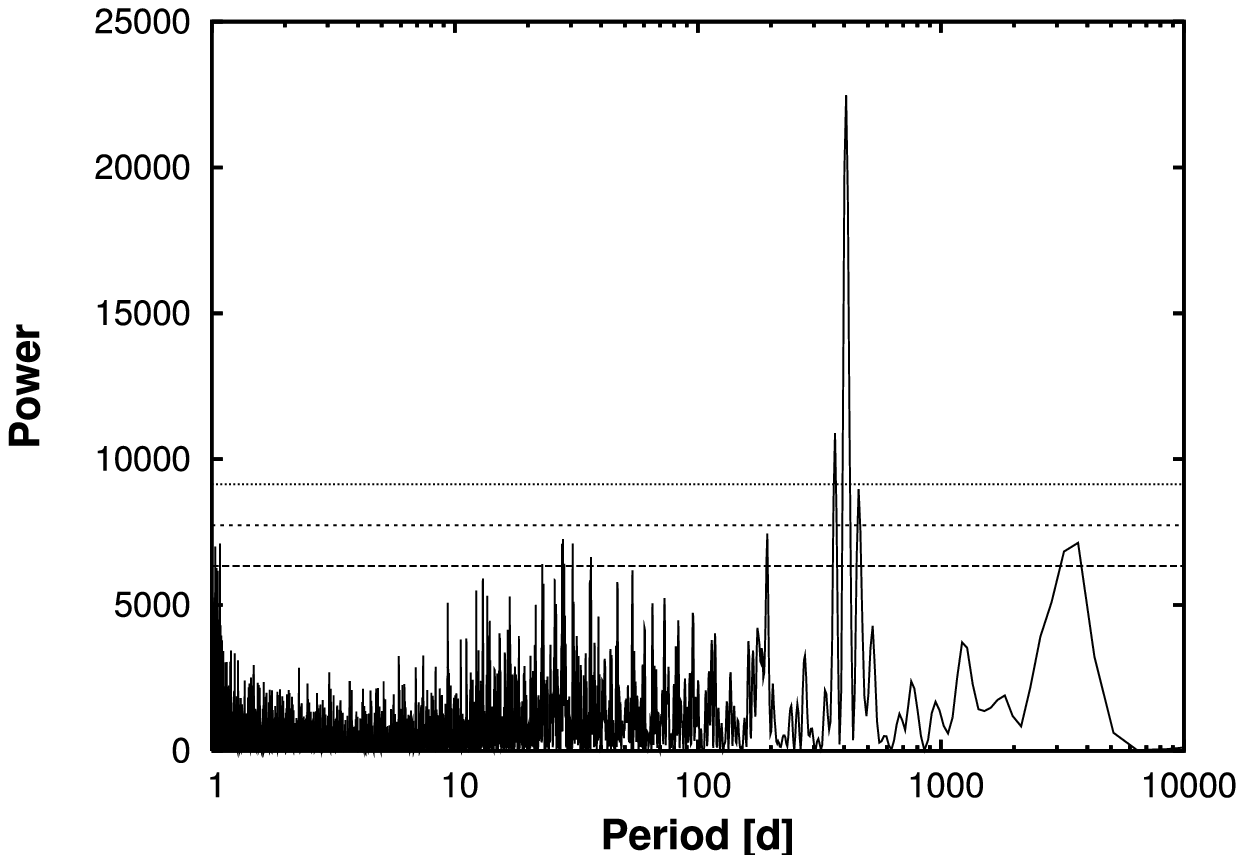}\\
\plotone{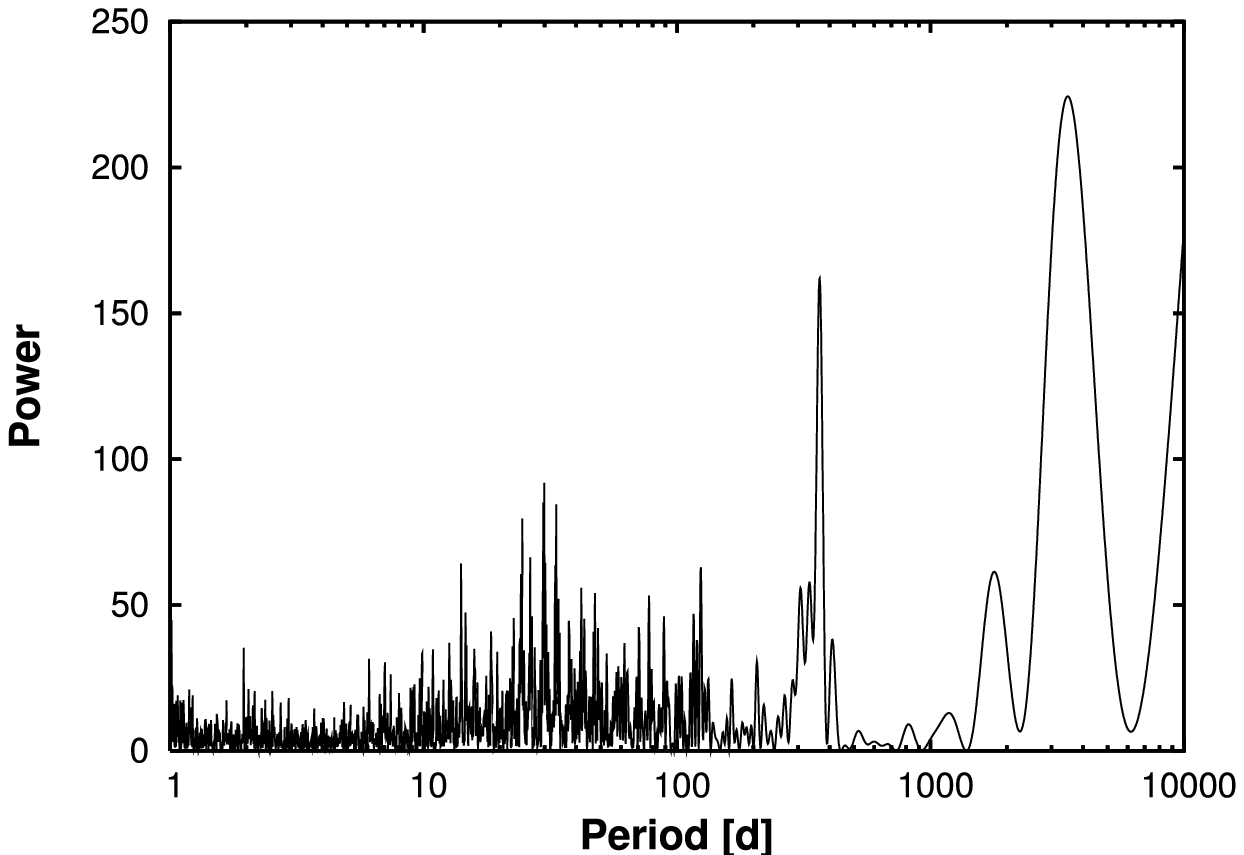}
\caption{Radial velocity data and periodograms for HD 177830. \textit{Top panel:} Relative radial velocity data obtained by KECK. \textit{Middle panel: } Error-weighted Lomb-Scargle periodogram of the radial velocity data. \textit{Bottom panel: } Power spectral window.}\label{fig:data_HD177830}
\end{figure}

\begin{figure}
	\plotone{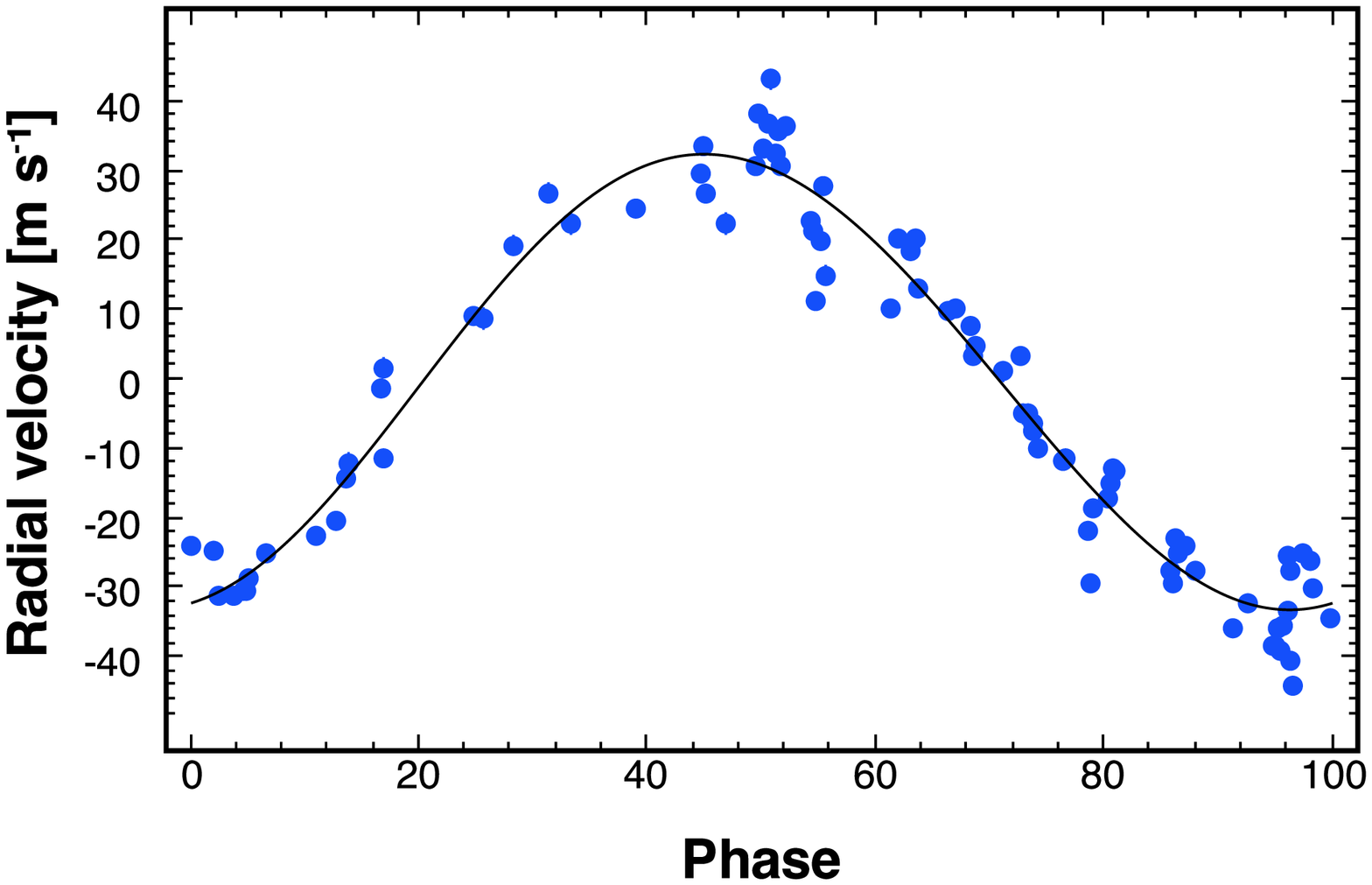}\\
	\plotone{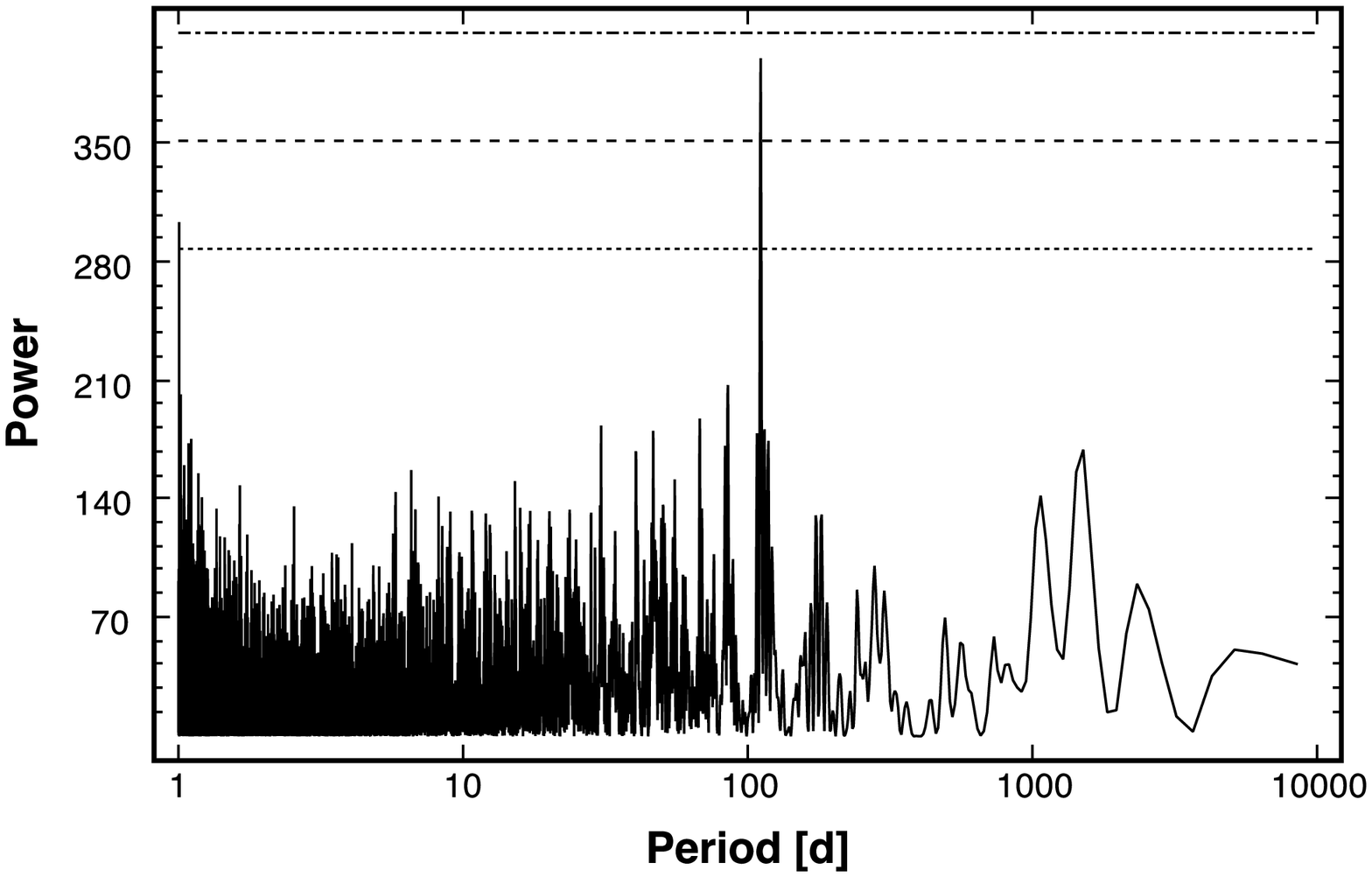}
	\caption{One-planet Keplerian solution and residuals periodogram for HD 177830.
	\textit{Top panel:} Phased Keplerian fit. \textit{Bottom panel: } Periodogram of the residuals to the 1-planet best-fit solution.}\label{fig:1pfit_HD177830}
\end{figure}

The individual RV observations for HD 177830 are shown in the top panel of Figure \ref{fig:data_HD177830}. 
The middle panel shows the error-weighted Lomb-Scargle (LS) periodogram of the full RV dataset.
Finally, the lower panel of Figure \ref{fig:data_HD177830} shows the spectral window.
The strongest peak in the periodogram is well-fit with a Keplerian model with period 407.31
days, semi-amplitude $K = 31.17$ \ms{} and estimated FAP $ < 3\times 10^{-6}$. 
Together with the assumed stellar mass of 1.48 $\msun$,
this amplitude corresponds to a minimum mass  
of $\mass \sin i = 1.48 \mjup$. The best-fit orbit for the planet is essentially circular. 
This 1-planet fit achieves a reduced $\chi^2 = 27.53$, with an RMS of 5.24 \ms. The top panel of Figure \ref{fig:1pfit_HD177830} shows the phased Keplerian fit for the 407-d planet.

The bottom panel of Figure \ref{fig:1pfit_HD177830} shows the periodogram of the residuals to the single-planet fit and the corresponding 
FAPs. The dominant peak at P = 110.98 with a FAP 
$\approx 5 \times 10^{-5}$ indicates the 
rather secure
presence of an additional planet. Our 
best combined 2-planet fit indicates a new planet with $P = 110.91$ days, $K = 5.11 $ \ms{} and a minimum 
mass of $\mass \sin i = 0.15 \mjup$. The orbit of the second planet is moderately eccentric ($e \approx 0.36$). With this revised fit, 
we obtain a reduced $\chi^2 = 15.31$ and an RMS of the residuals of approximately 3.85 \ms. 
The expected jitter of HD 177830 (that is, the amount of jitter required to bring the reduced $\chi^2$ of the best-fit solution to 1.0) is 3.71 \ms. 

\begin{figure}
    \plotone{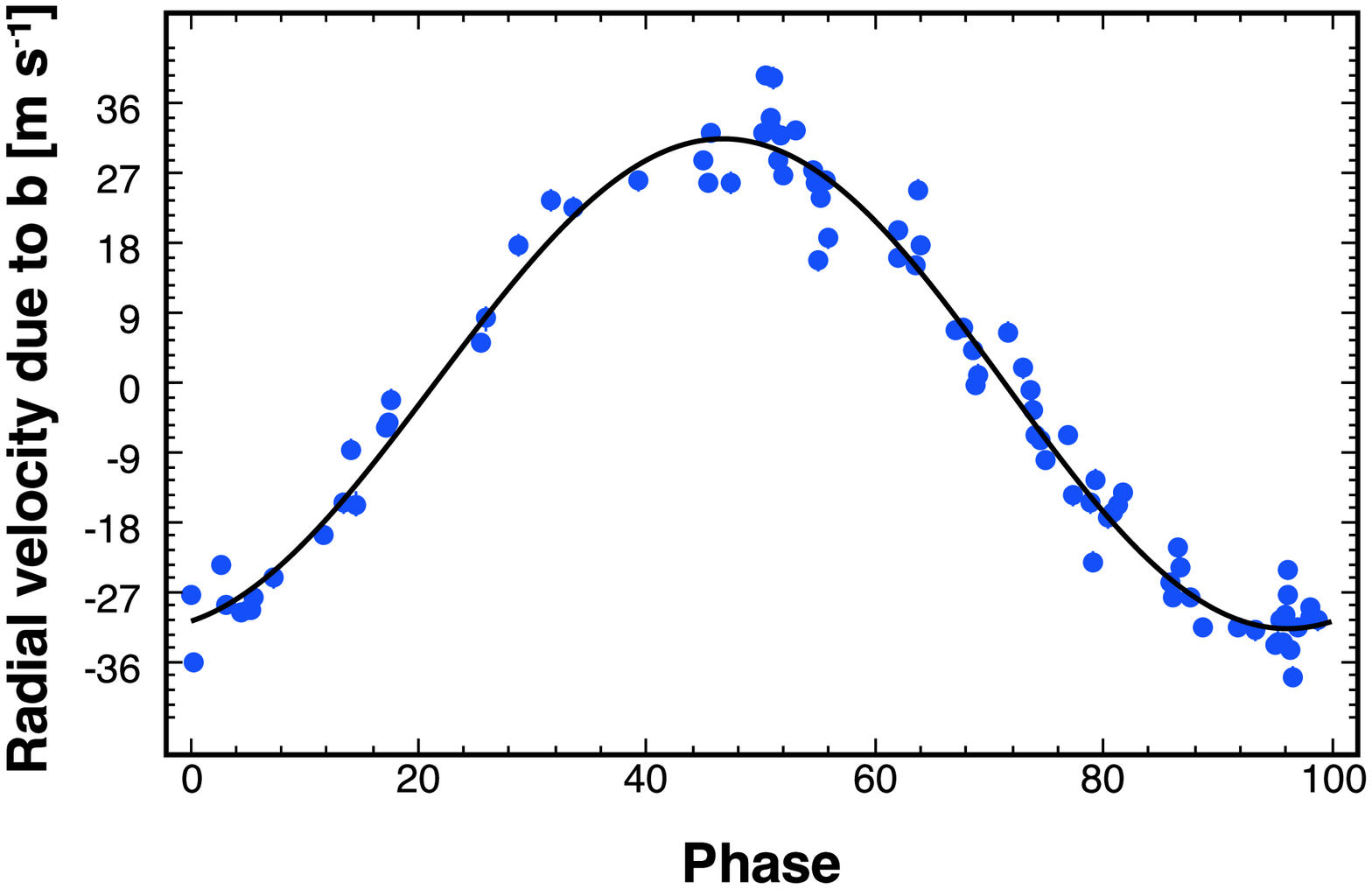}
    \plotone{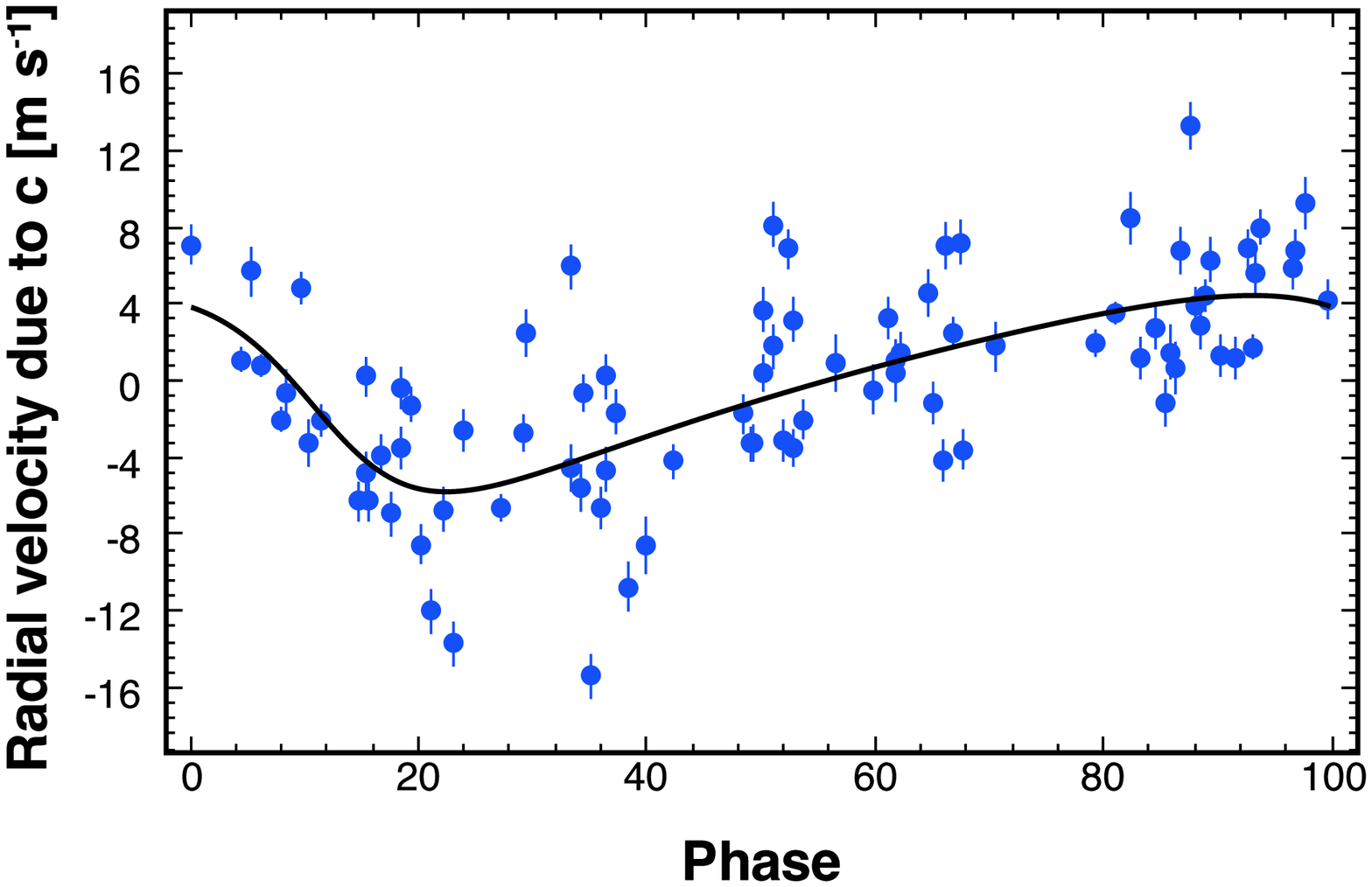}
\plotone{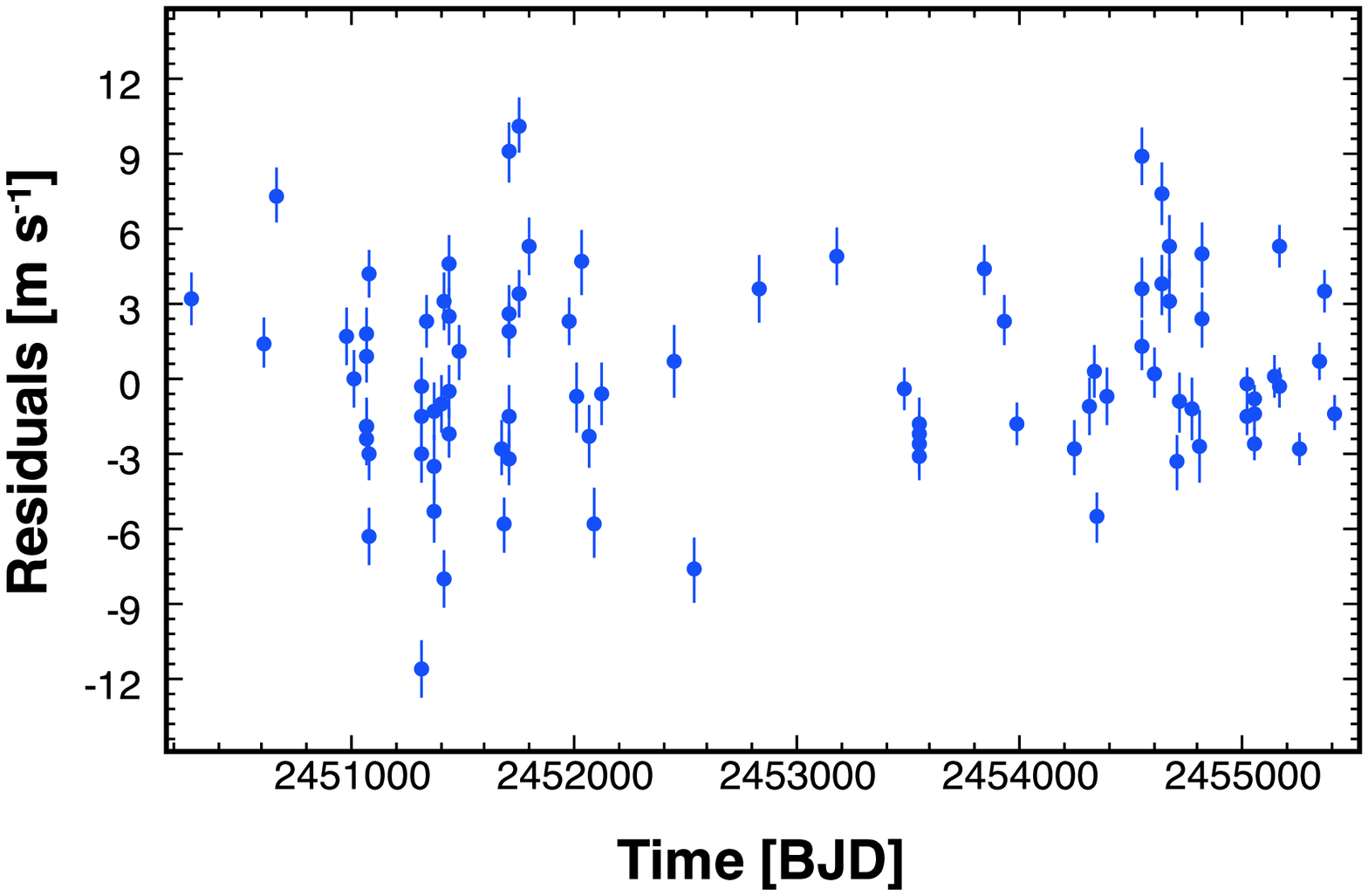}
\plotone{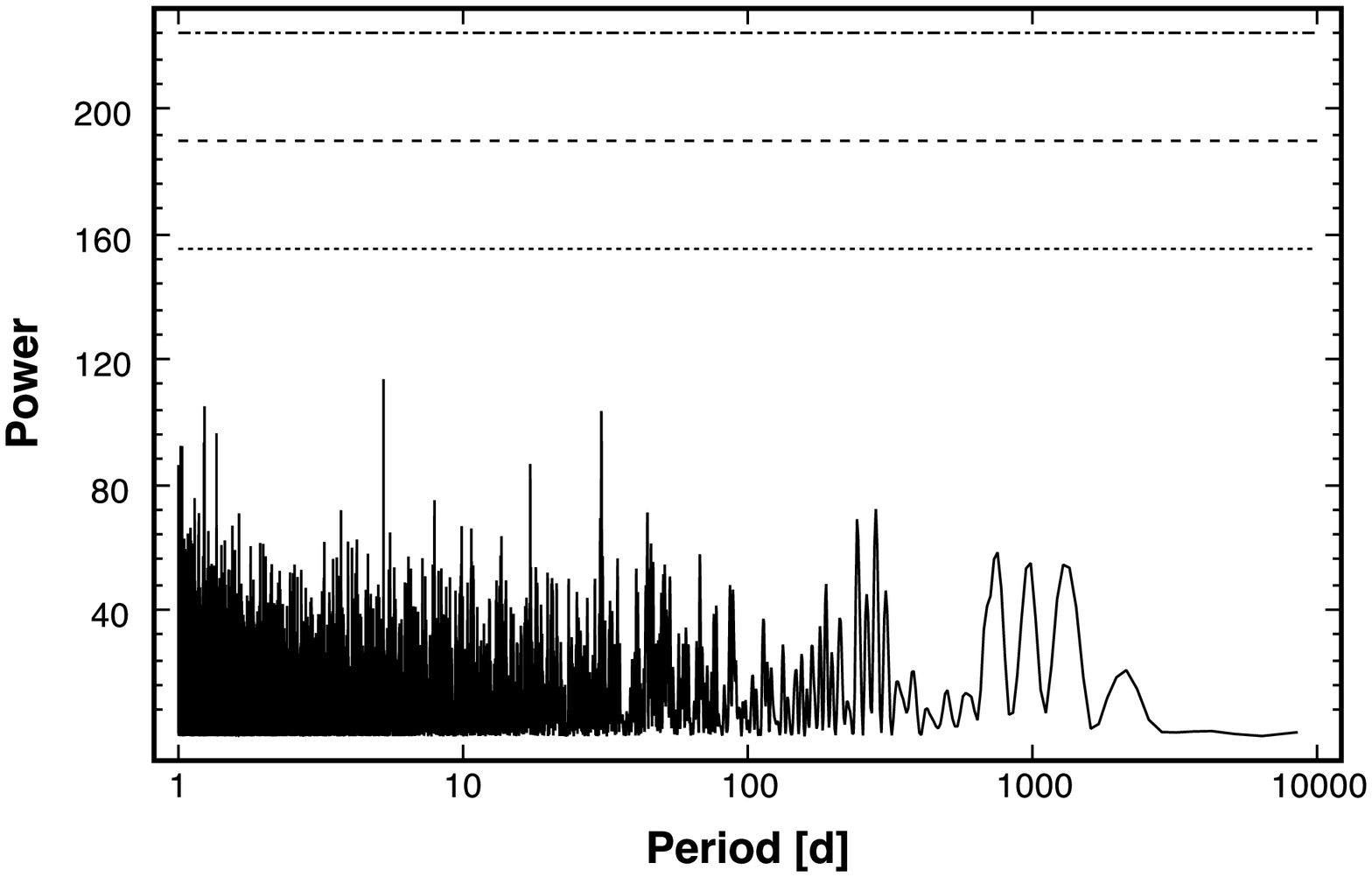}
\caption{Keplerian solution and residuals periodogram for HD 177830.
\textit{1st panel: } Phased Keplerian fit of the 407-d component b.
\textit{2nd panel: } Phased Keplerian fit of the 111-d component c.
\textit{3rd panel: } Residuals to the 2-planet fit. \textit{4th panel: } Periodogram of the residuals to the 2-planet best fit solution.}
\label{fig:bestfit_HD177830}
\end{figure}

The top and 2nd panels of Figure \ref{fig:bestfit_HD177830} show the phased stellar reflex velocity of HD 177830 due to each companion as compared to the 
RV dataset. The 3rd panel shows the residuals to the 2-planet solution, while the bottom panel shows the periodogram of the residuals of the best-fit solution. No compelling peaks are evident in the current Keck dataset, indicating that the present data offers no strong support for additional planets in the system.

		\begin{deluxetable}{ccc}
		\tablewidth{0pt}
		\tablecaption{KECK radial velocities for HD 177830 (\textit{Sample: full table in electronic version})
		\label{tab:rvdata_HD177830}}
		\tablecolumns{3}
		\tablehead{{Barycentric JD}&{RV [\ms]}&{Uncertainty [\ms]}}
		\startdata
		2450276.03 & -16.32 & 0.94\\ 
2450605.04 & -5.29 & 0.95\\ 
2450666.89 & -17.61 & 1.03\\ 
2450982.94 & 0.32 & 1.11\\ 
2451009.93 & -9.34 & 1.04\\ 
2451068.82 & -30.65 & 0.95\\ 
2451069.85 & -30.62 & 1.05\\ 
2451070.90 & -28.30 & 1.00\\ 
2451071.83 & -31.40 & 1.01\\ 
2451072.82 & -27.97 & 1.00\\ 
2451073.82 & -25.70 & 0.89\\ 
2451074.81 & -33.00 & 0.98\\ 
2451075.90 & -36.45 & 1.08\\ 
2451311.11 & 30.25 & 1.07\\ 
2451312.11 & 29.00 & 1.17\\ 
2451313.11 & 18.91 & 1.02\\ 
2451314.13 & 27.42 & 1.09\\ 
2451341.95 & 27.81 & 0.96\\ 
2451367.91 & 15.43 & 1.03\\ 
2451368.91 & 11.00 & 1.17\\ 

		\enddata
		\end{deluxetable}

{\begin{figure}
\epsscale{1}
\plotone{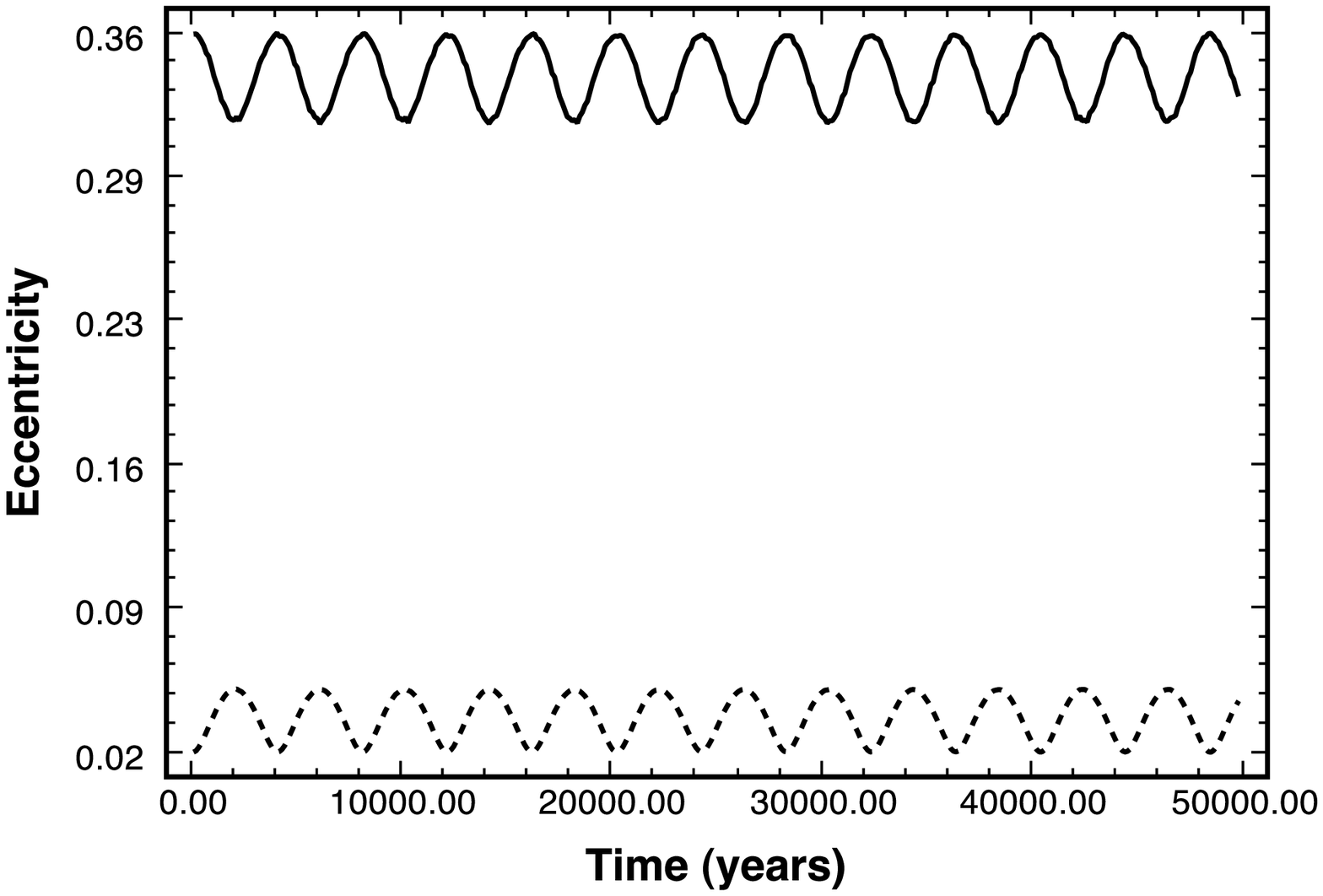}
\caption{Eccentricity evolution of planets HD177830 b (dashed line) and c (solid line) within $5\times 10^4$ years.\label{fig:HD177830_stab}}
\epsscale{\scale}
\end{figure}

The 2-planet fit shows a very slight amount of dynamical interaction between planets b and c, which we accounted for in the modeling using the Bulirsch-Stoer integration scheme in the Systemic console \citep{Meschiari10}; we verified that the best-fit orbital model is stable for at least $10^6$ years. The time evolution of the eccentricity is shown in Figure \ref{fig:HD177830_stab}.

}

\section{HD 99492 (HIP 55848)}\label{sec:4}
\subsection{Stellar properties}\label{sec:HD99492_star}
HD 99492 is a V = 7.383 magnitude star of spectral type 
K2V. 
{A recent determination of many of its fundamental stellar parameters was given by \citet{Marcy05} and is included in Table \ref{tab:allstars}, supplemented 
with additional values from the NStED database. \citet{Marcy05} found this star to be a middle-aged star of average chromospheric activity, with an age of 2-6 Gyr. They report an implied stellar rotation period of about 45 days ($\pm$ 30\%) based on the star's chromospheric activity index. \citet{Marcy05} also reported a 17.1-day 36 \mearth planet orbiting this star. Compared to the Sun, HD 99492 is quite metal-rich ([Fe/H] = 0.36).  
}

\subsection{Keplerian solution}
		Table \ref{tab:rvdata_HD99492} shows the 93 relative radial 
velocity measurements for HD 99492.  The radial velocity coverage spans 
almost 14 years of RV monitoring.
The median internal uncertainty for our observations is 1.36 \ms, and the peak-to-peak velocity 
variation is 28.32 \ms. The velocity scatter around the average RV in our 
observations is 6.39 \ms. 

\begin{figure}
\plotone{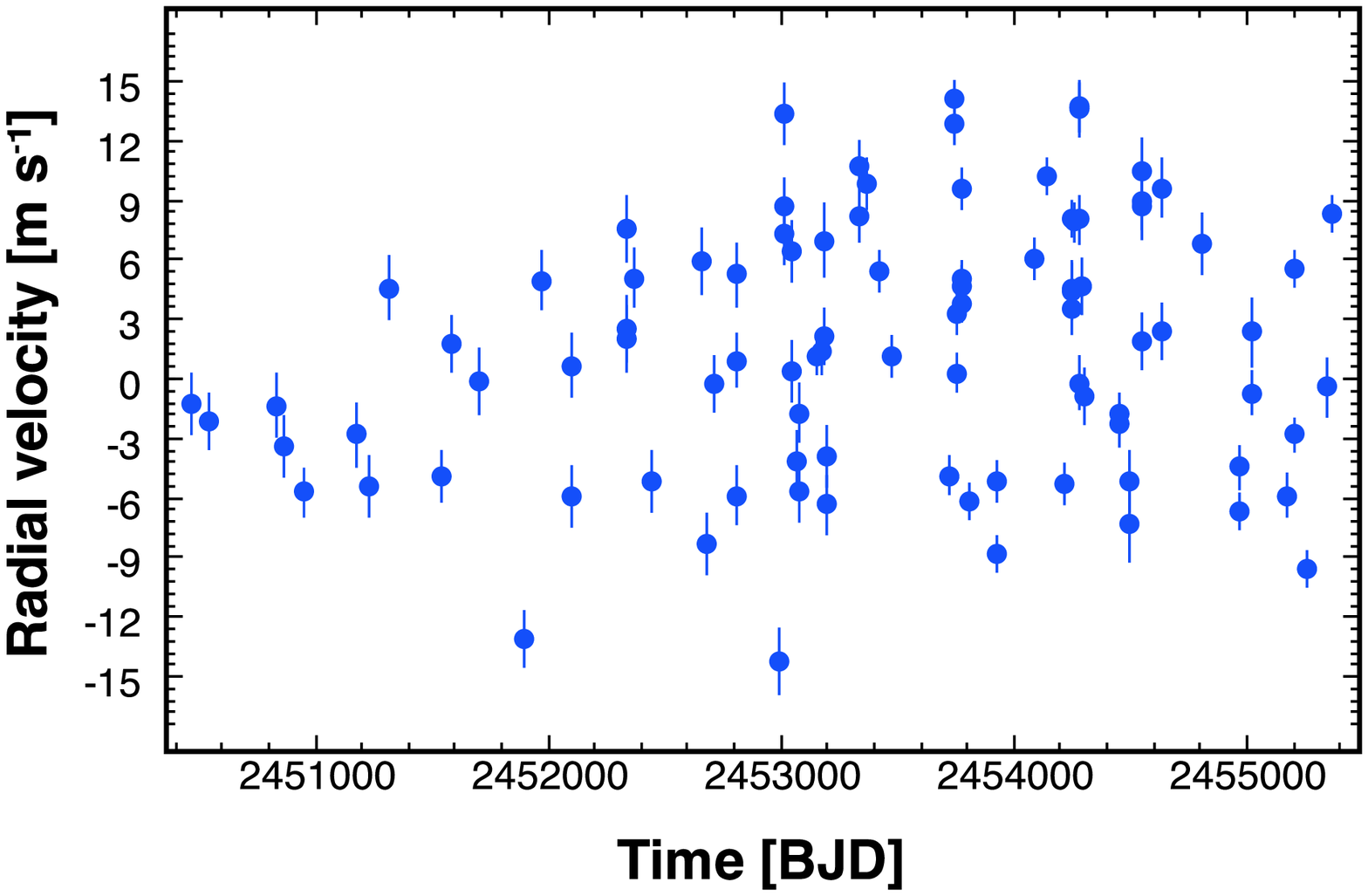}\\
\plotone{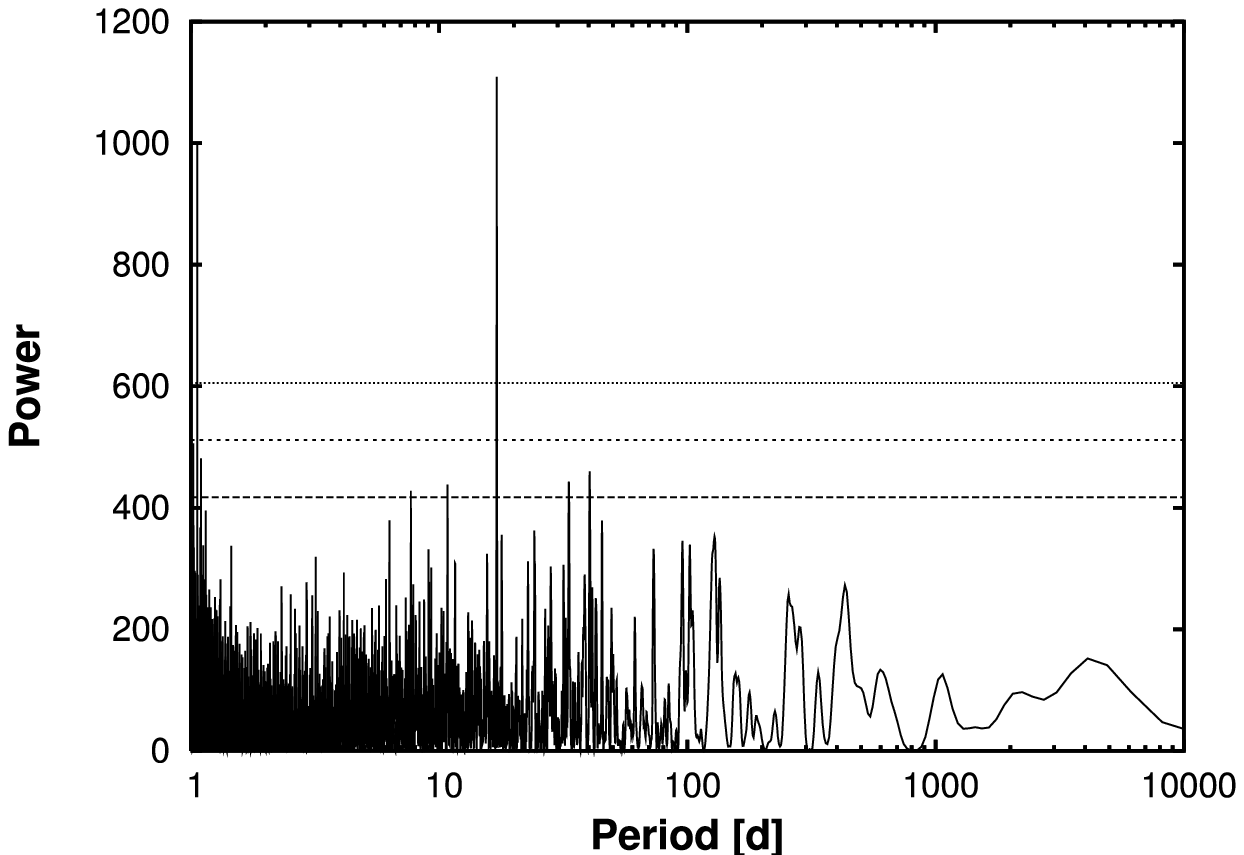}\\
\plotone{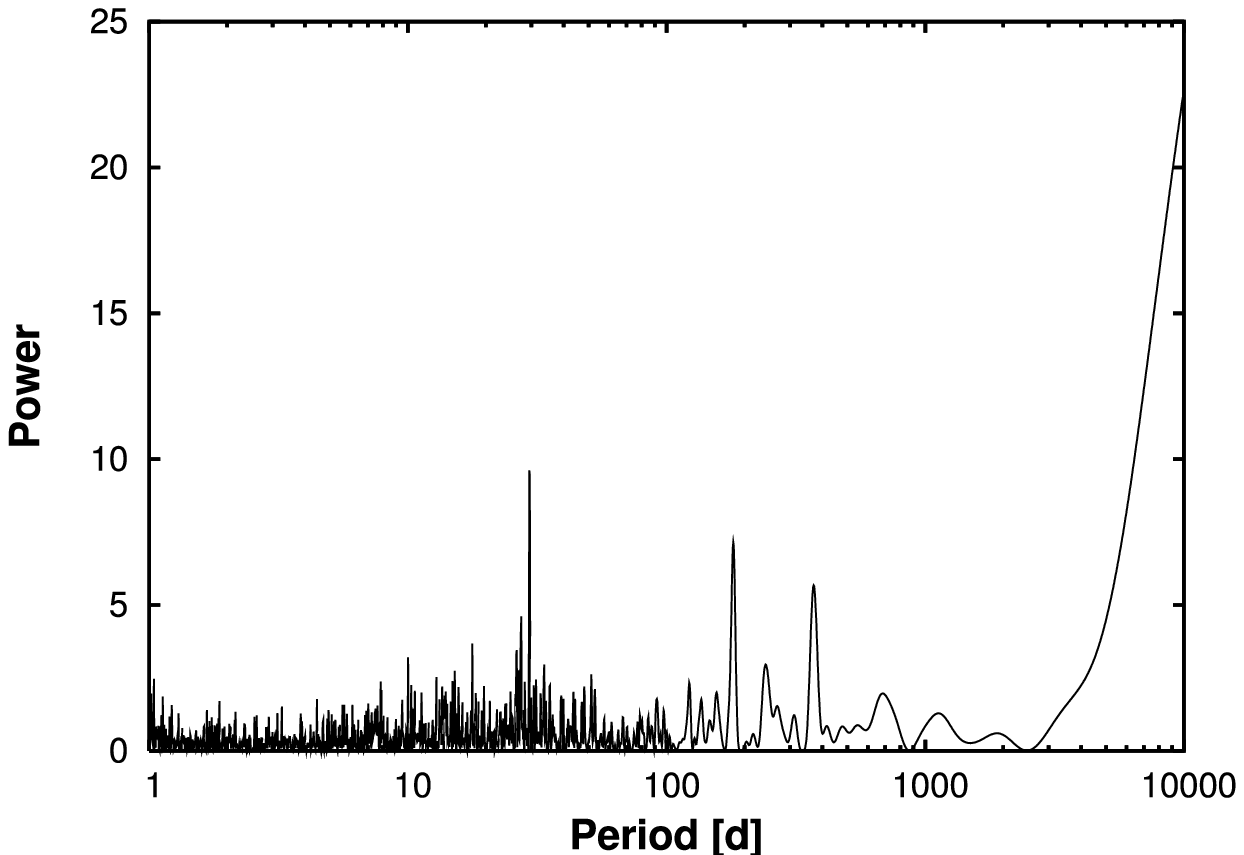}
\caption{Radial velocity data and periodograms for HD 99492. \textit{Top panel:} Relative radial velocity data obtained by KECK. \textit{Middle panel: } Error-weighted Lomb-Scargle periodogram of the radial velocity data. \textit{Bottom panel: } Power spectral window.}\label{fig:data_HD99492}
\end{figure}

\begin{figure}
	\plotone{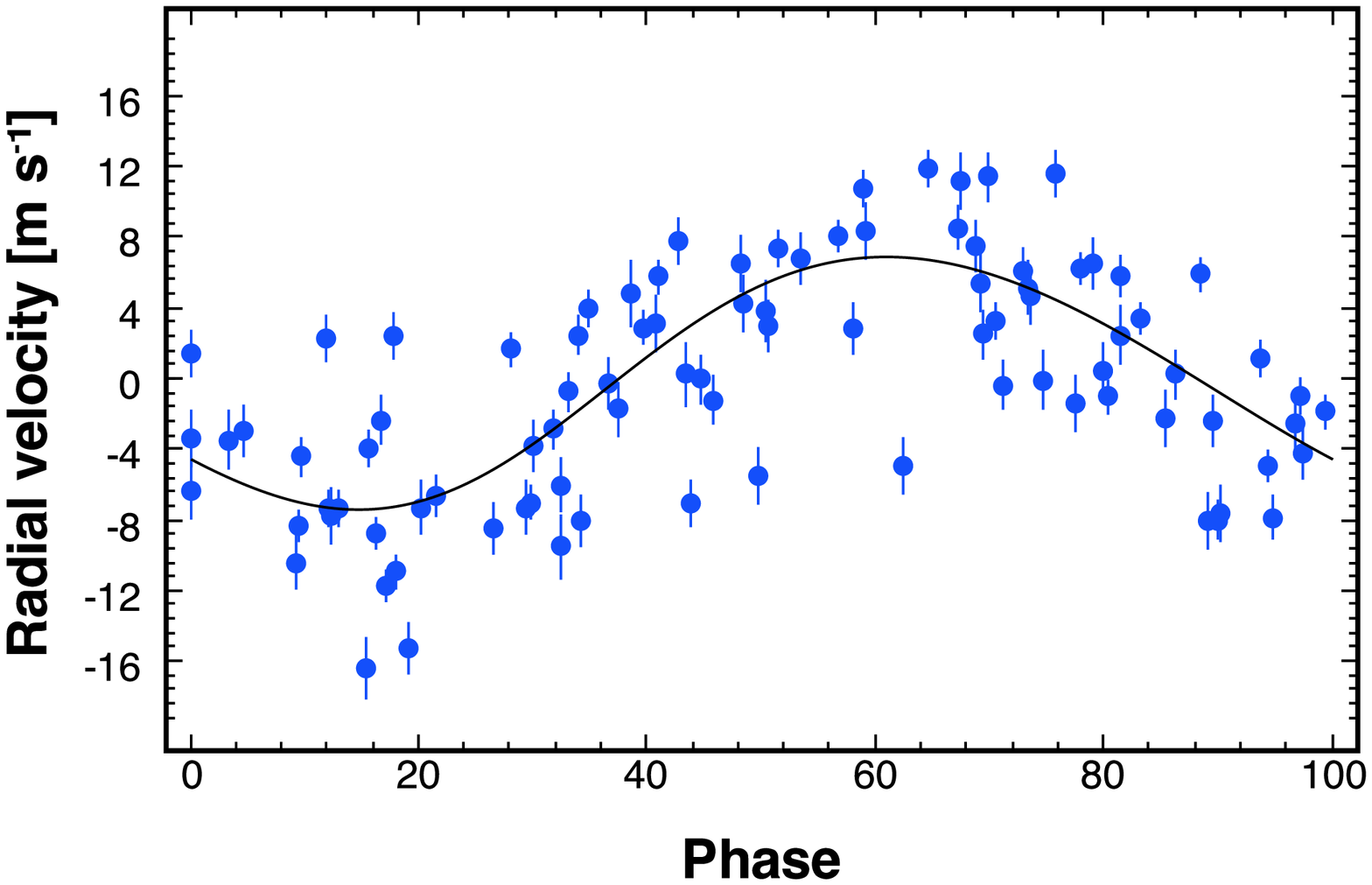}\\
	\plotone{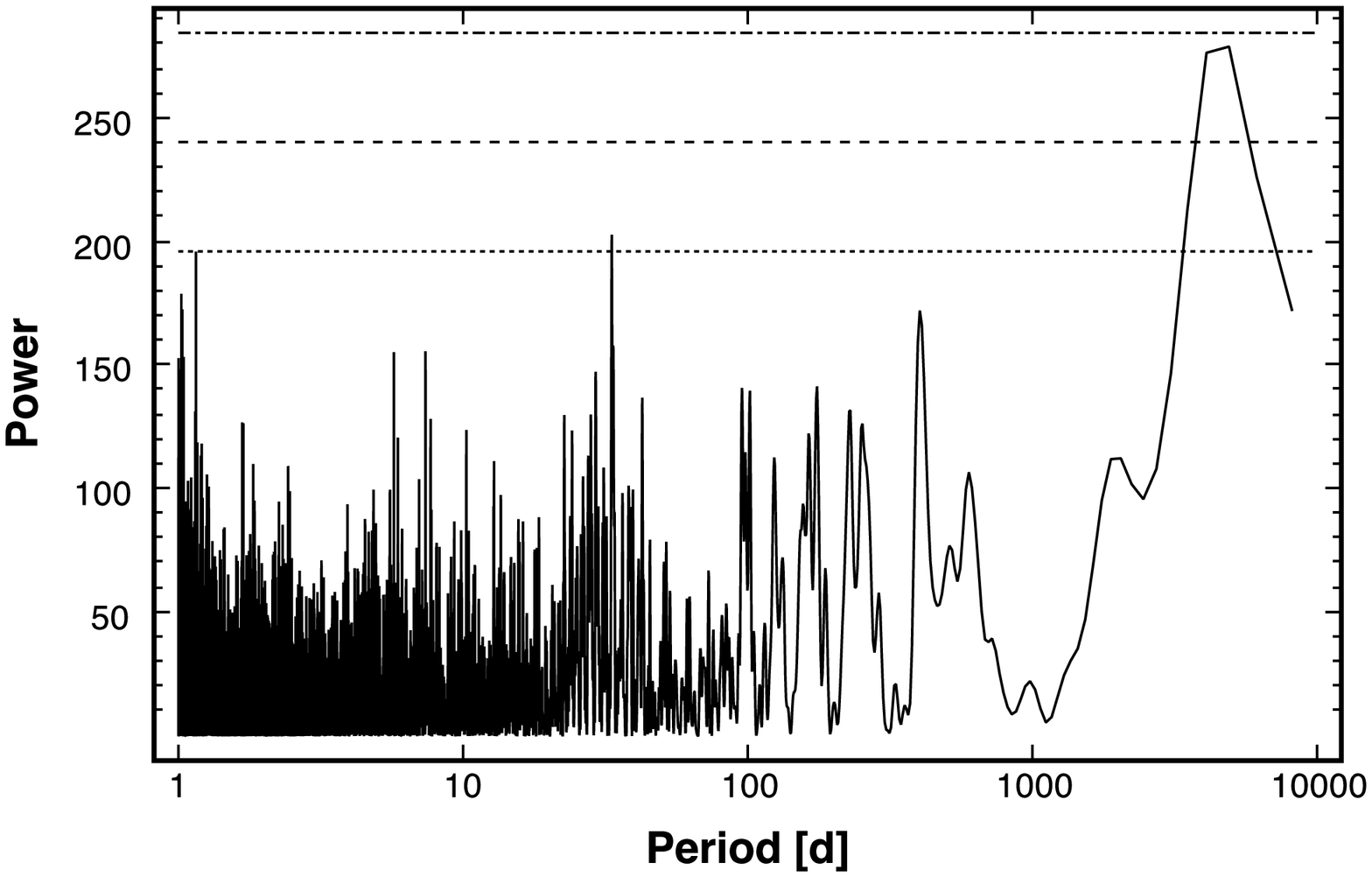}
	\caption{One-planet Keplerian solution and residuals periodogram for HD 99492.
	\textit{Top panel:} Phased Keplerian fit. \textit{Bottom panel: } Periodogram of the residuals to the 1-planet best-fit solution.}\label{fig:1pfit_HD99492}
\end{figure}

The top panel of Figure \ref{fig:data_HD99492} shows the individual RV observations for HD 99492. The middle panel shows the error-weighted Lomb-Scargle (LS) periodogram of the full RV dataset, while the bottom panel shows the spectral window.
The FAP calculation for the strong Keplerian signal at $P  = $ 
17.06 days in the  RV dataset indicates an estimated FAP $< 3\times 10^{-6}$. 
The dominant peak in the periodogram is well-fit by a Keplerian fit of period 17.05
days and semi-amplitude $K = 7.86 $ \ms. 
Together with the assumed stellar mass of 0.83 $\msun$,
this amplitude suggests a minimum mass  
of $\mass \sin i = 27.76 \mearth$. The best-fit orbit for the planet shows a small amount of eccentricity ($e \approx 0.13$). 
This 1-planet fit achieves a reduced $\chi^2 = 12.71 $, with an RMS of 4.39 \ms. 
The top panel of Figure \ref{fig:1pfit_HD99492} shows the phased Keplerian fit for the 17-d planet, while the bottom panel shows the periodogram of the residuals to the single-planet fit and the corresponding 
FAPs. 

The additional peak in the periodogram of residuals with P = 4908.67 reveals the secure detection of an additional planet,  with a FAP 
$\approx 4 \times 10^{-4}$. Our 
best combined 2-planet fit suggests a new planet with $P = 4969.73$ days, $K = 4.88 $ \ms{} and a minimum 
mass of $\mass \sin i = 0.36 \mjup$; the orbit of the second planet is somewhat eccentric ($e \approx 0.11 $). Using this revised fit, 
we obtain a reduced $\chi^2 = 7.17 $ and an RMS of the residuals of approximately 3.22 \ms. 
The expected jitter of HD 99492 (that is, the amount of jitter required to bring the reduced $\chi^2$ of the best-fit solution to 1.0) is 2.94 \ms. 

\begin{figure}
    \plotone{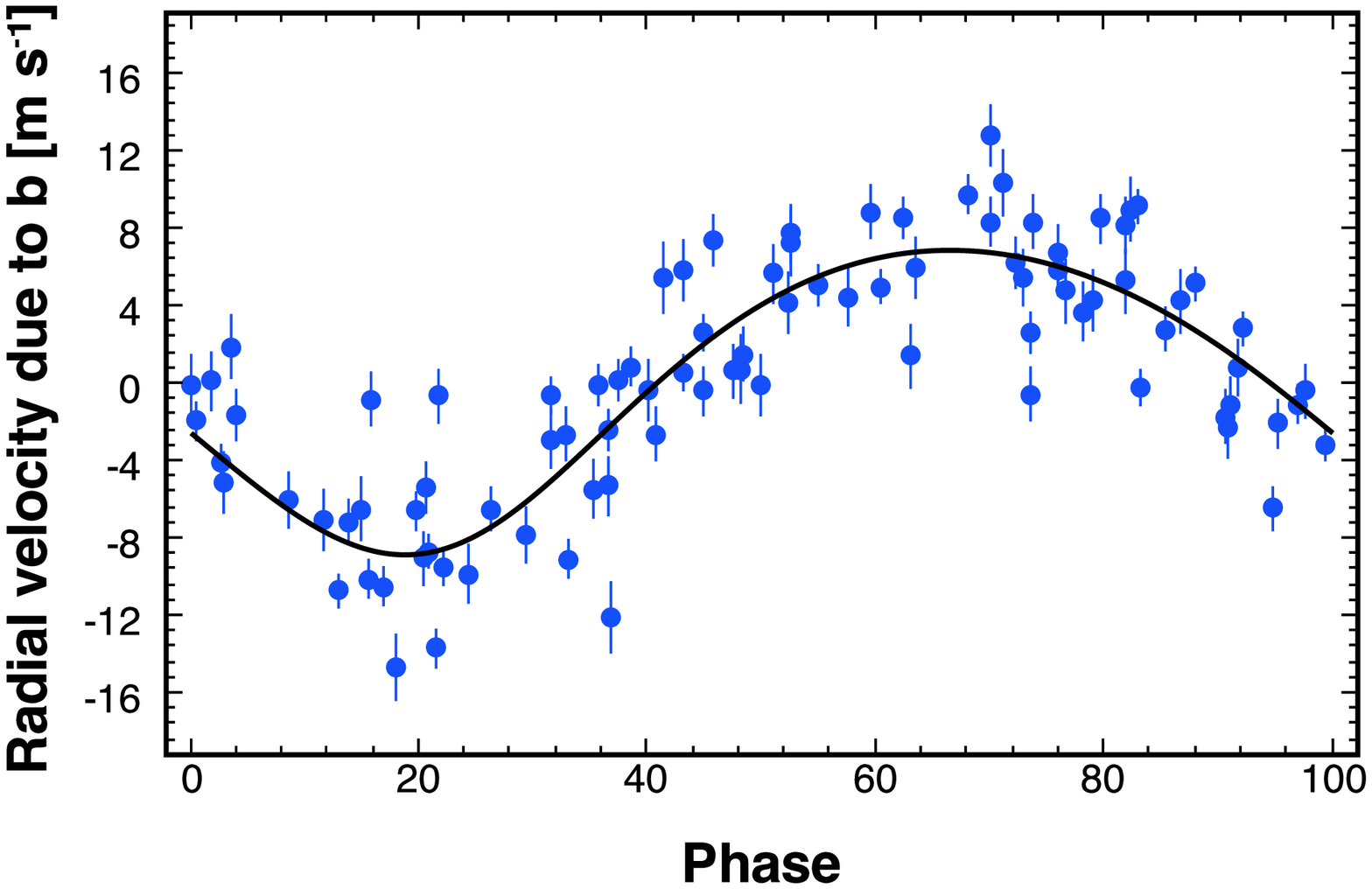}
    \plotone{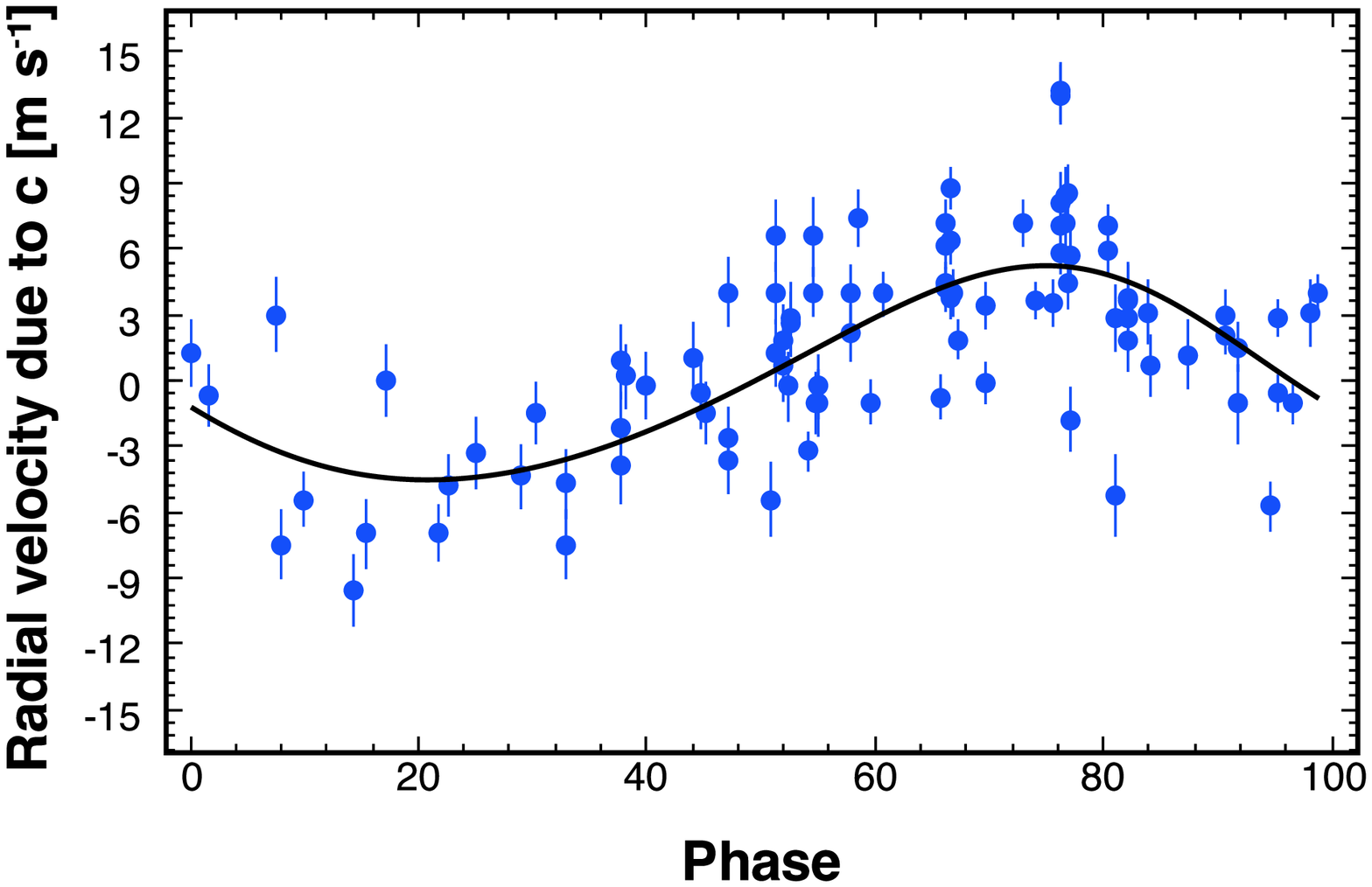}
\plotone{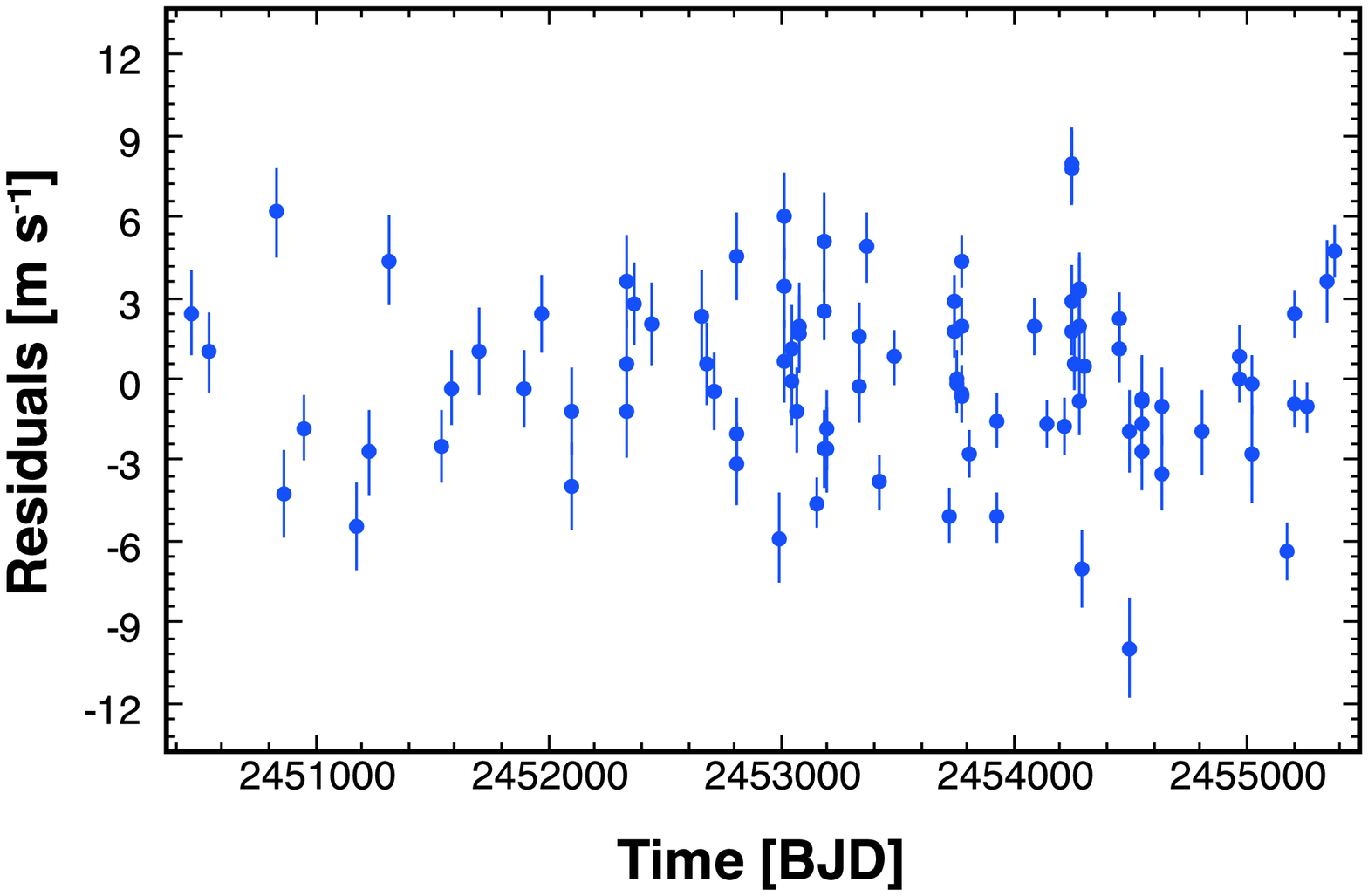}
\plotone{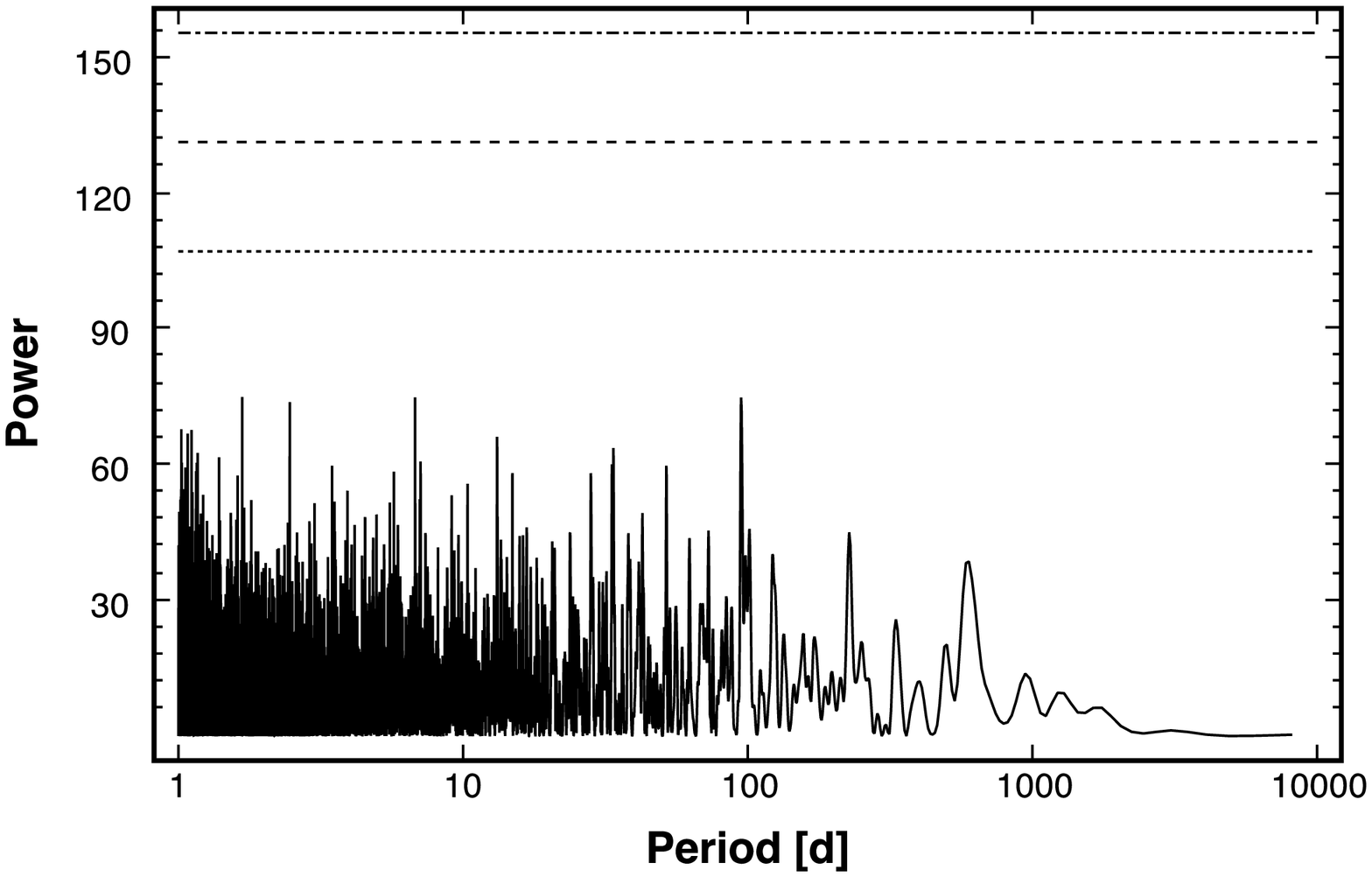}
\caption{Keplerian solution and residuals periodogram for HD 99492.
\textit{1st panel: } Phased Keplerian fit of the 17-d component b.
\textit{2nd panel: } Phased Keplerian fit of the 4697-d component c.
\textit{3rd panel: } Residuals to the 2-planet fit. \textit{4th panel: } Periodogram of the residuals to the 2-planet best fit solution.}
\label{fig:bestfit_HD99492}
\end{figure}

The top and 2nd panels of Figure \ref{fig:bestfit_HD99492} show the phased stellar reflex velocity of HD 99492 from each planet compared to the 
RV dataset. The 3rd panel shows the residuals to the 2-planet solution, while the bottom panel shows the periodogram of the residuals of the best-fit solution. No significant peaks are evident, indicating that the present data set offers no strong support for additional planets in the system.

{}

\begin{deluxetable}{ccc}
		\tablewidth{0pt}
		\tablecaption{KECK radial velocities for HD 99492 (\textit{Sample: full table in electronic version})
		\label{tab:rvdata_HD99492}}
		\tablecolumns{3}
		\tablehead{{Barycentric JD}&{RV [\ms]}&{Uncertainty [\ms]}}
		\startdata
		2450462.11 & -2.62 & 1.51\\ 
2450546.99 & -3.46 & 1.39\\ 
2450837.93 & -2.66 & 1.58\\ 
2450862.90 & -4.70 & 1.51\\ 
2450955.88 & -7.05 & 1.18\\ 
2451172.10 & -4.12 & 1.59\\ 
2451228.04 & -6.80 & 1.53\\ 
2451311.82 & 3.26 & 1.63\\ 
2451544.17 & -6.23 & 1.27\\ 
2451582.97 & 0.43 & 1.34\\ 
2451704.81 & -1.45 & 1.60\\ 
2451898.15 & -14.46 & 1.41\\ 
2451973.05 & 3.64 & 1.37\\ 
2452095.75 & -0.65 & 1.55\\ 
2452097.75 & -7.23 & 1.56\\ 
2452333.14 & 6.24 & 1.65\\ 
2452334.08 & 0.68 & 1.62\\ 
2452334.97 & 1.21 & 1.59\\ 
2452364.07 & 3.76 & 1.44\\ 
2452445.77 & -6.52 & 1.50\\ 

		\enddata
		\end{deluxetable}

\section{HD 74156 (HIP 42723)}\label{sec:5}
\subsection{Stellar properties}\label{sec:HD74156_star}
HD 74156 is a V = 7.614 magnitude star of spectral type 
G1V. In comparison to the Sun, HD 74156 is modestly metal-rich ([Fe/H] = 0.13).

{HD 74156 is a well-studied star, known already to have both a 52-day and a 2500-day planet \citep{Naef04}. The star was claimed by \citet{Bean08} to also harbor a 3rd planet (``d'') at 336 days, in apparent support of the so-called ``Packed Planetary Systems'' hypothesis \citep[PPS; ][]{Barnes04}. Indeed, \citet{Barnes07} cited the discovery of d as a successful prediction of the PPS hypothesis. However, the reality of HD 74516d was called into question by \citet{Baluev09} as a false detection made due to annual systematic errors in the HET RV data. \citet{Wittenmyer09} also found no evidence of HD 74156 d in their follow-up study. 

We have had HD 74156 under precise radial velocity monitoring at Keck for the past 8.9 years and here add 21 new velocities to the mix, combined with previously published data from CORALIE, ELODIE, and HET, bringing the total number of observations to 198. We re-analyzed the compound dataset from scratch, looking for evidence of further planetary companions. As usual, we allowed a floating offset between each data set in the Keplerian fitting process to compensate for the different zero-points of each observatory.
}

\subsection{Keplerian solution}

\begin{figure}
\plotone{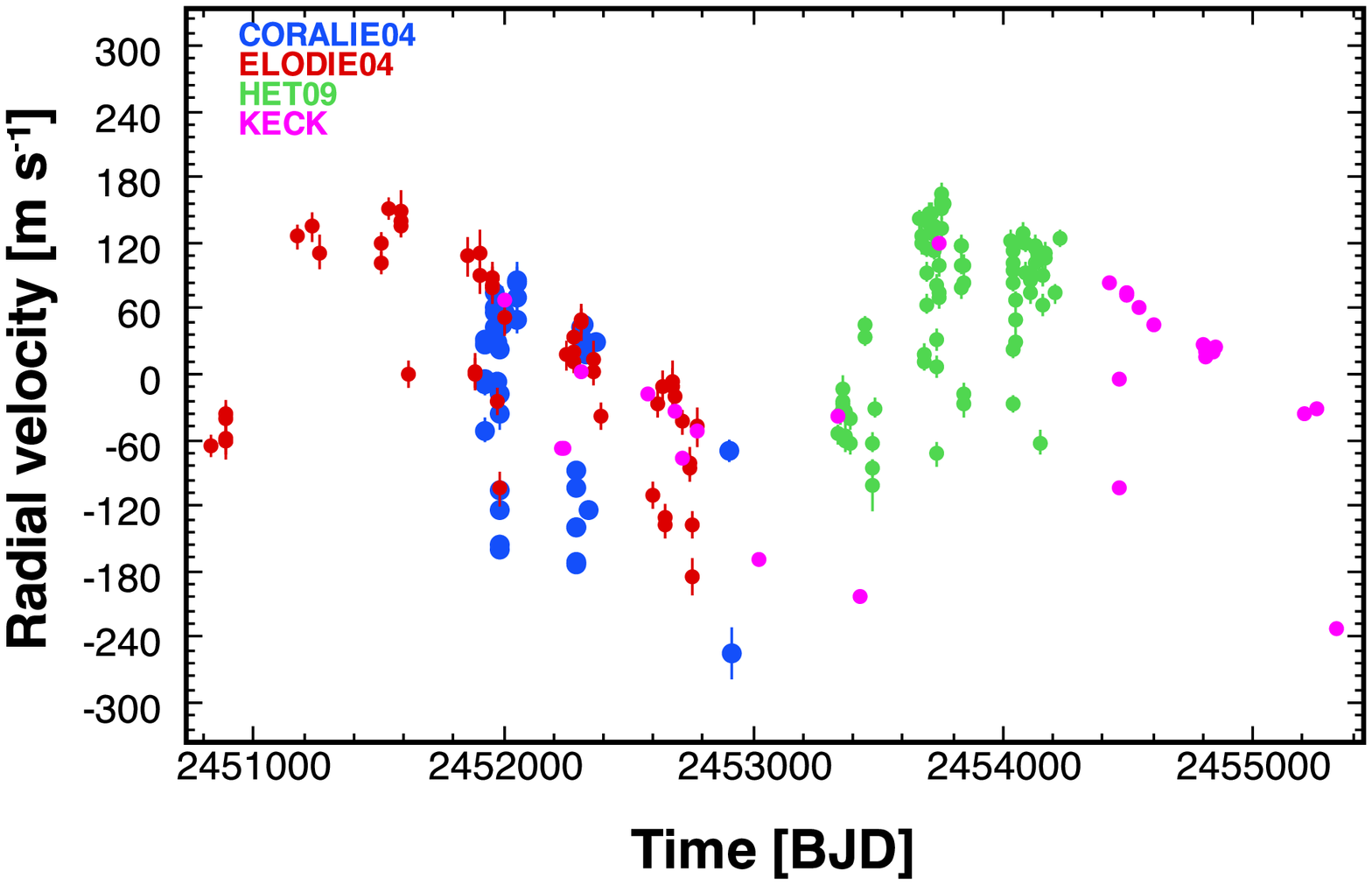}\\
\plotone{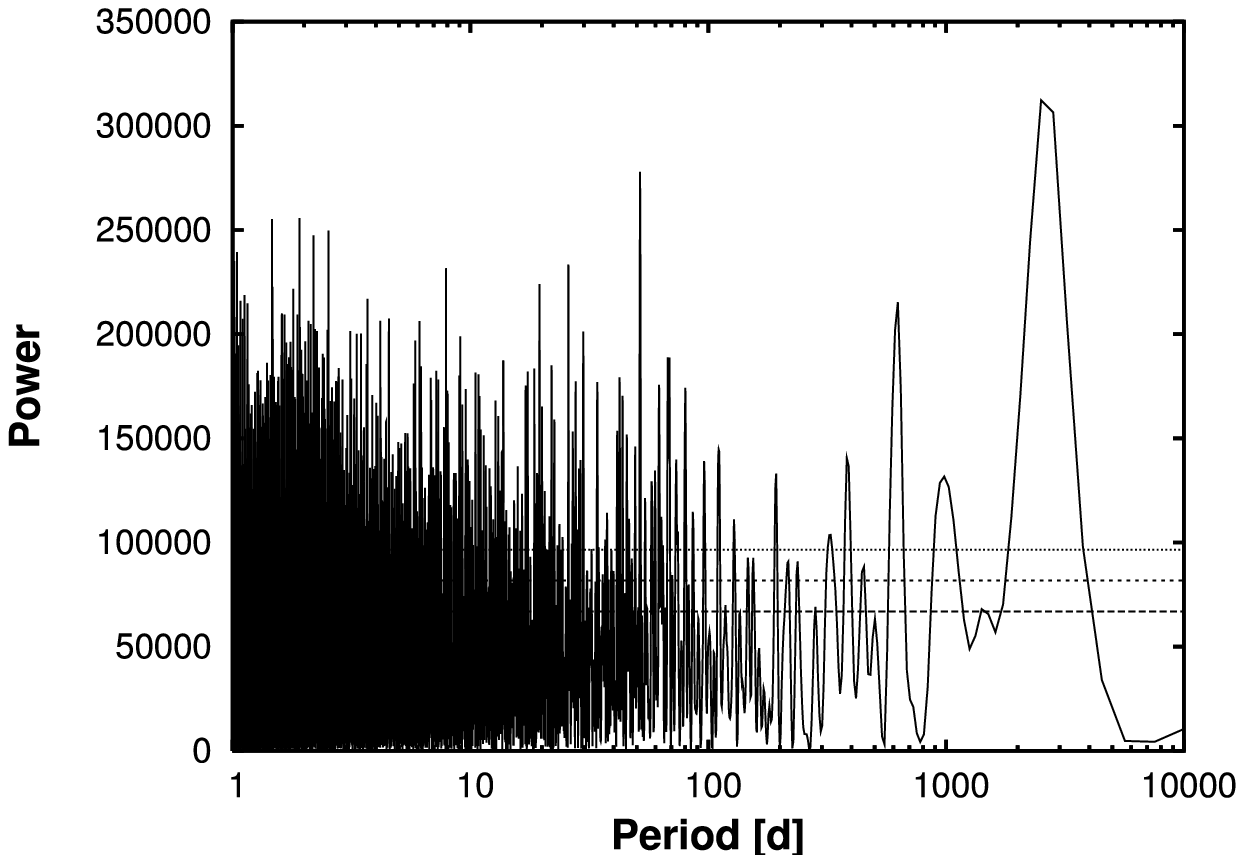}\\
\plotone{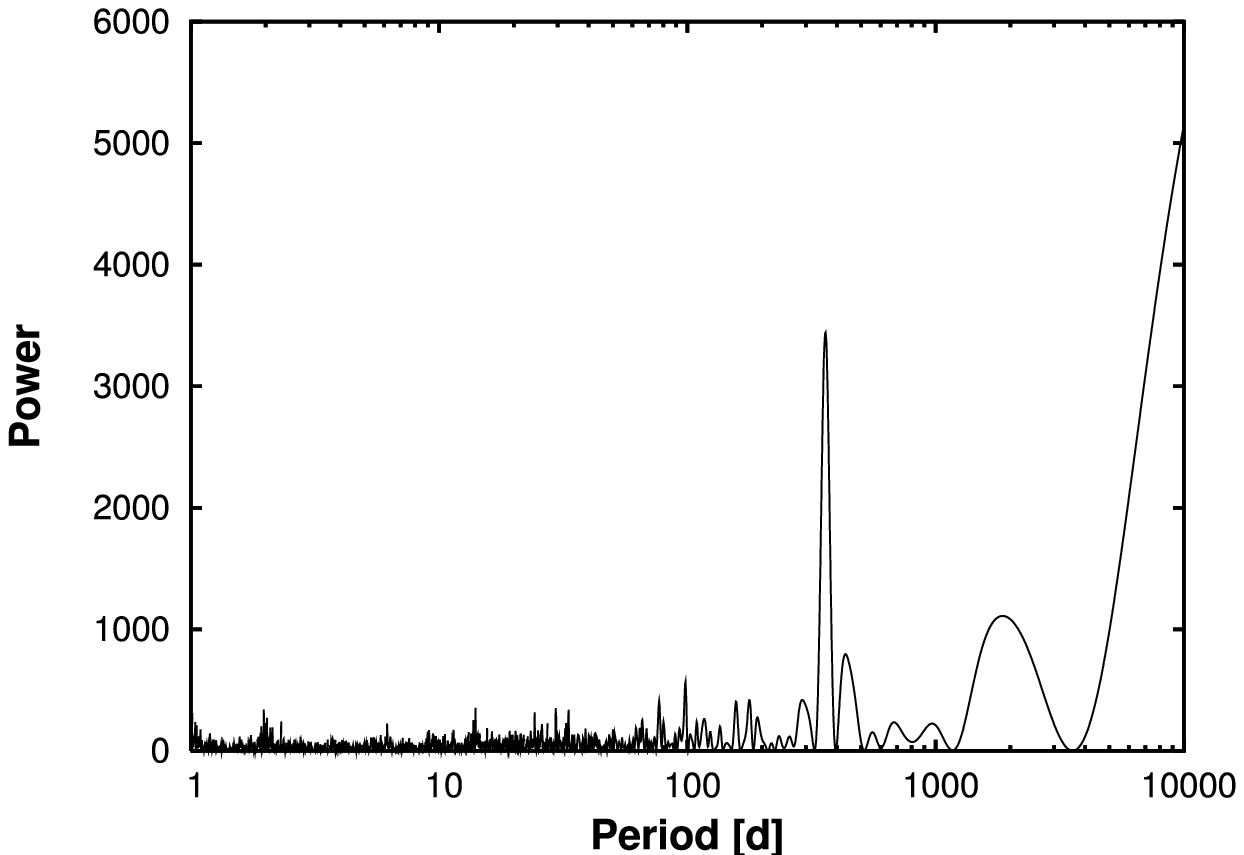}
\caption{Radial velocity data and periodograms for HD 74156. \textit{Top panel:} Relative radial velocity data obtained by CORALIE04, ELODIE04 \citep{Naef04}, HET09 \citep{Wittenmyer09} and KECK. \textit{Middle panel: } Error-weighted Lomb-Scargle periodogram of the radial velocity data. \textit{Bottom panel: } Power spectral window.}\label{fig:data_HD74156}
\end{figure}

\begin{figure}
	\plotone{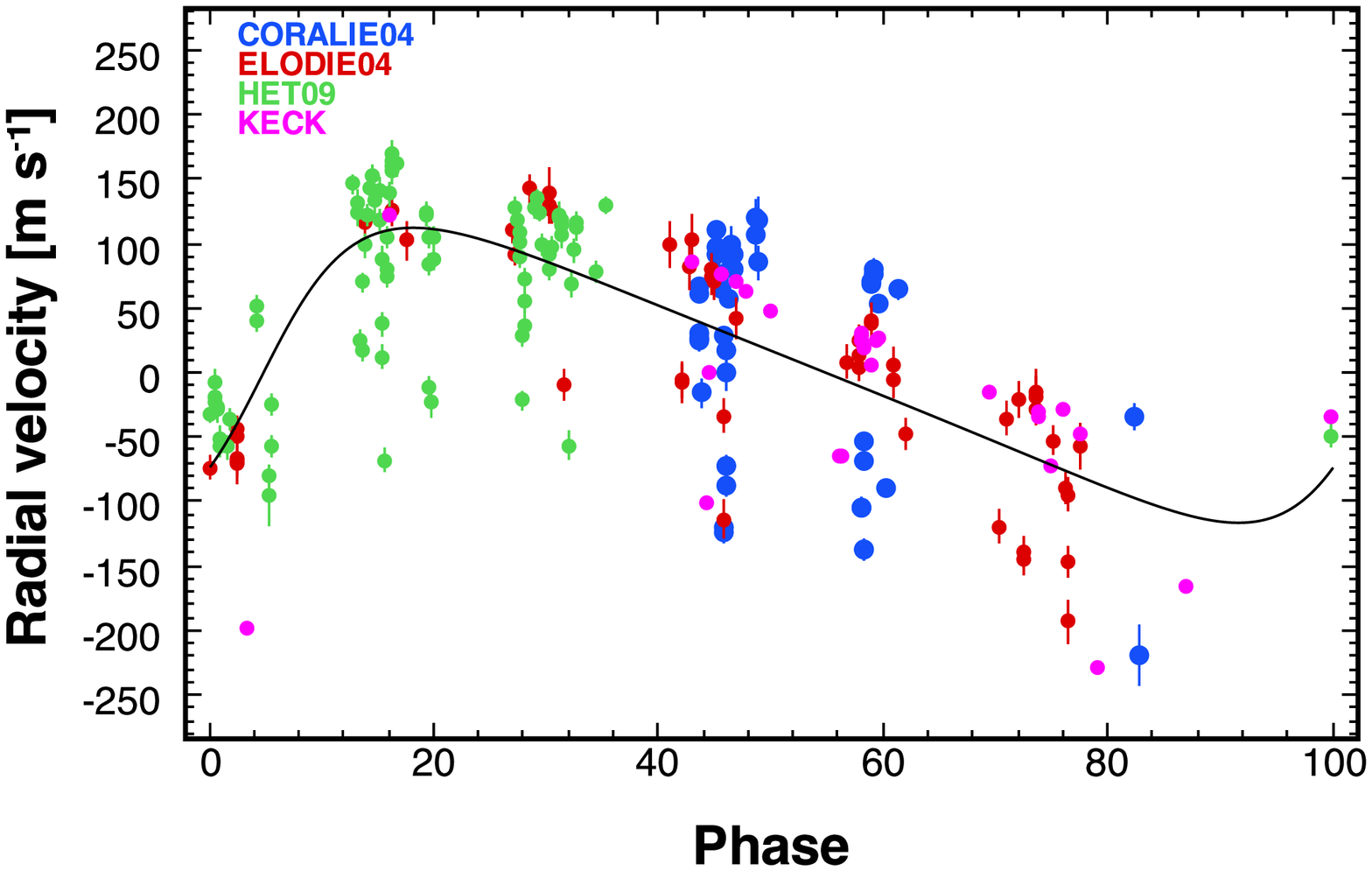}\\
	\plotone{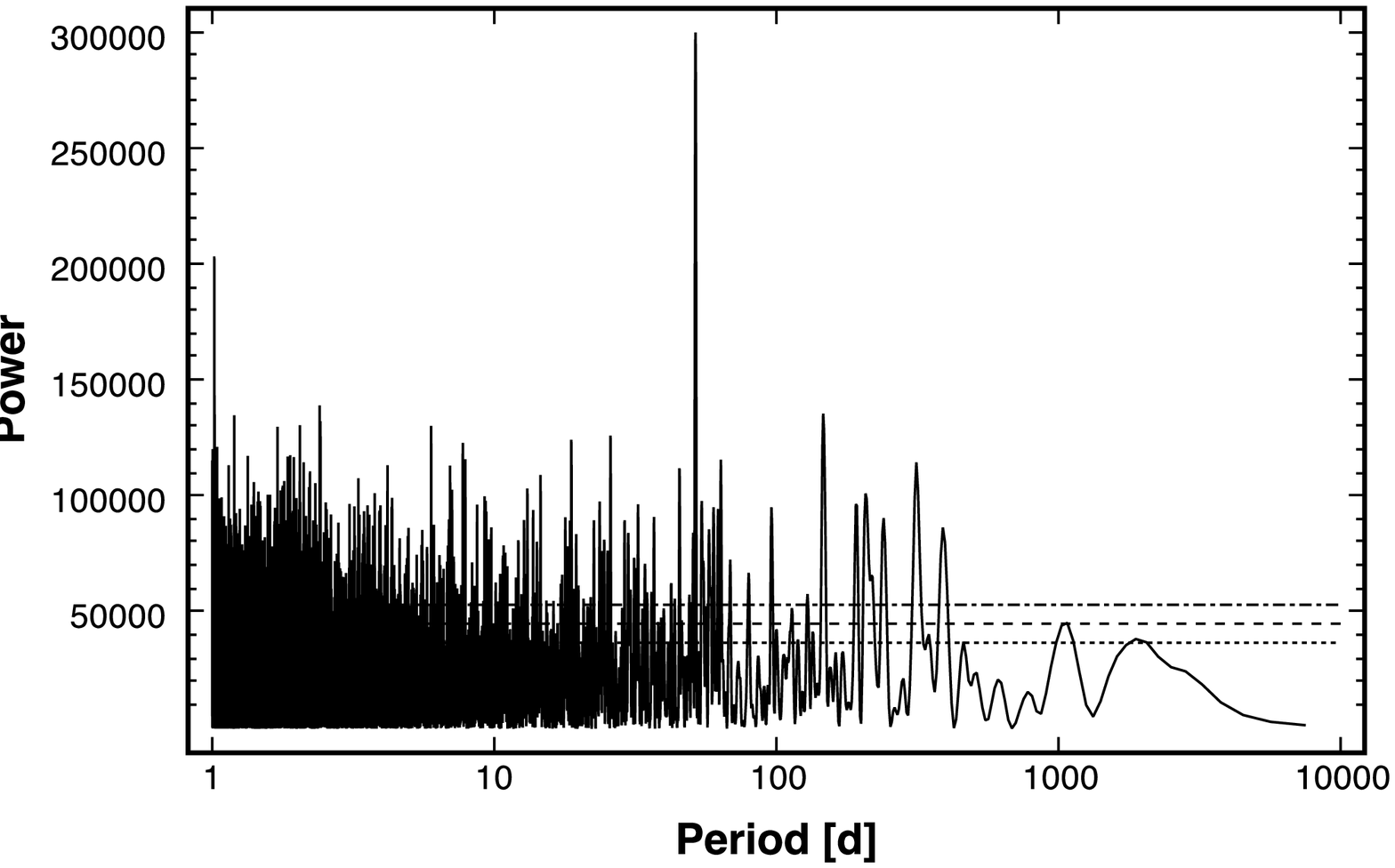}
	\caption{One-planet Keplerian solution and residuals periodogram for HD 74156.
	\textit{Top panel:} Phased Keplerian fit. \textit{Bottom panel: } Periodogram of the residuals to the 1-planet best-fit solution.}\label{fig:1pfit_HD74156}
\end{figure}

\begin{figure}
    \plotone{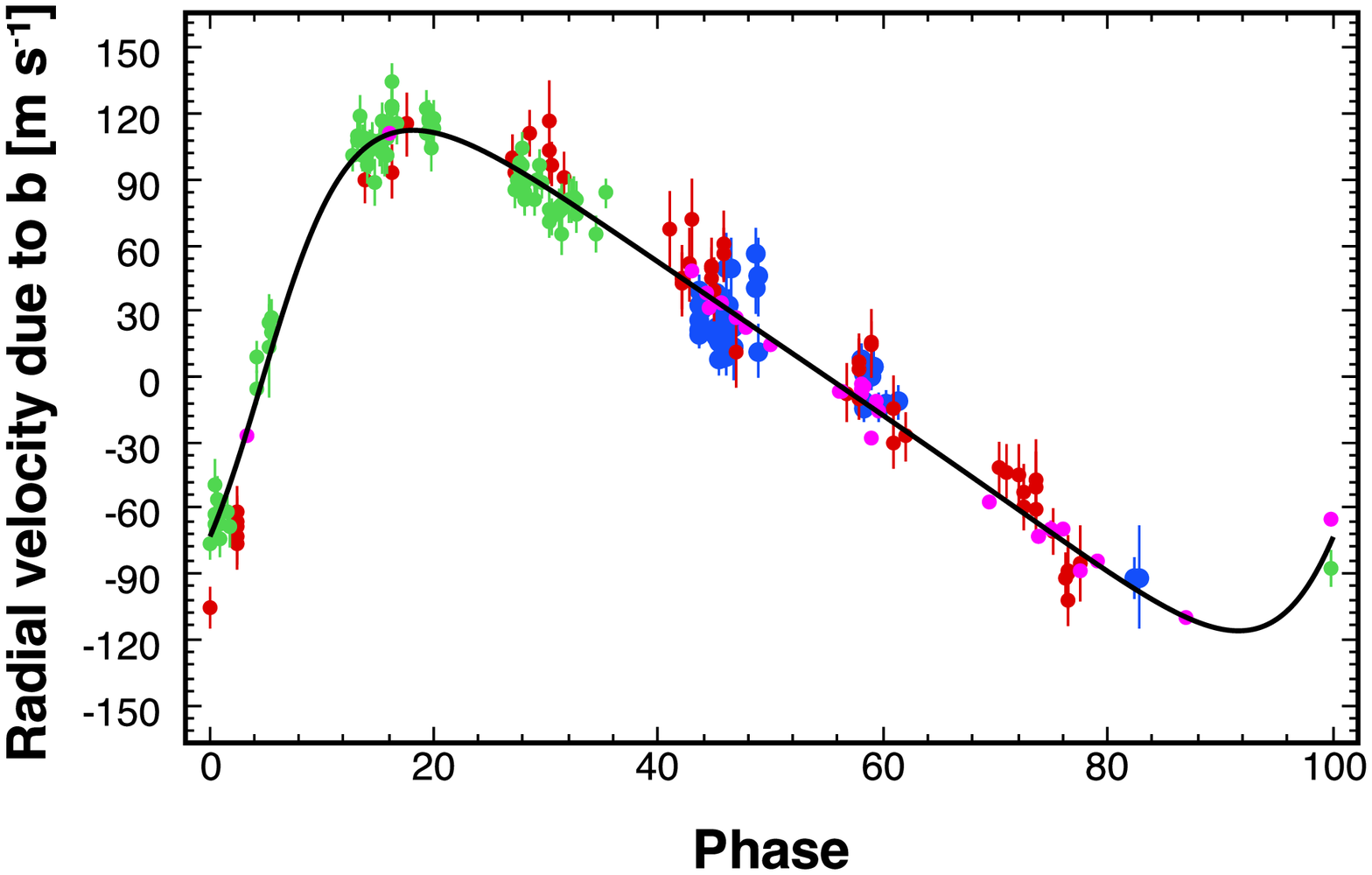}
    \plotone{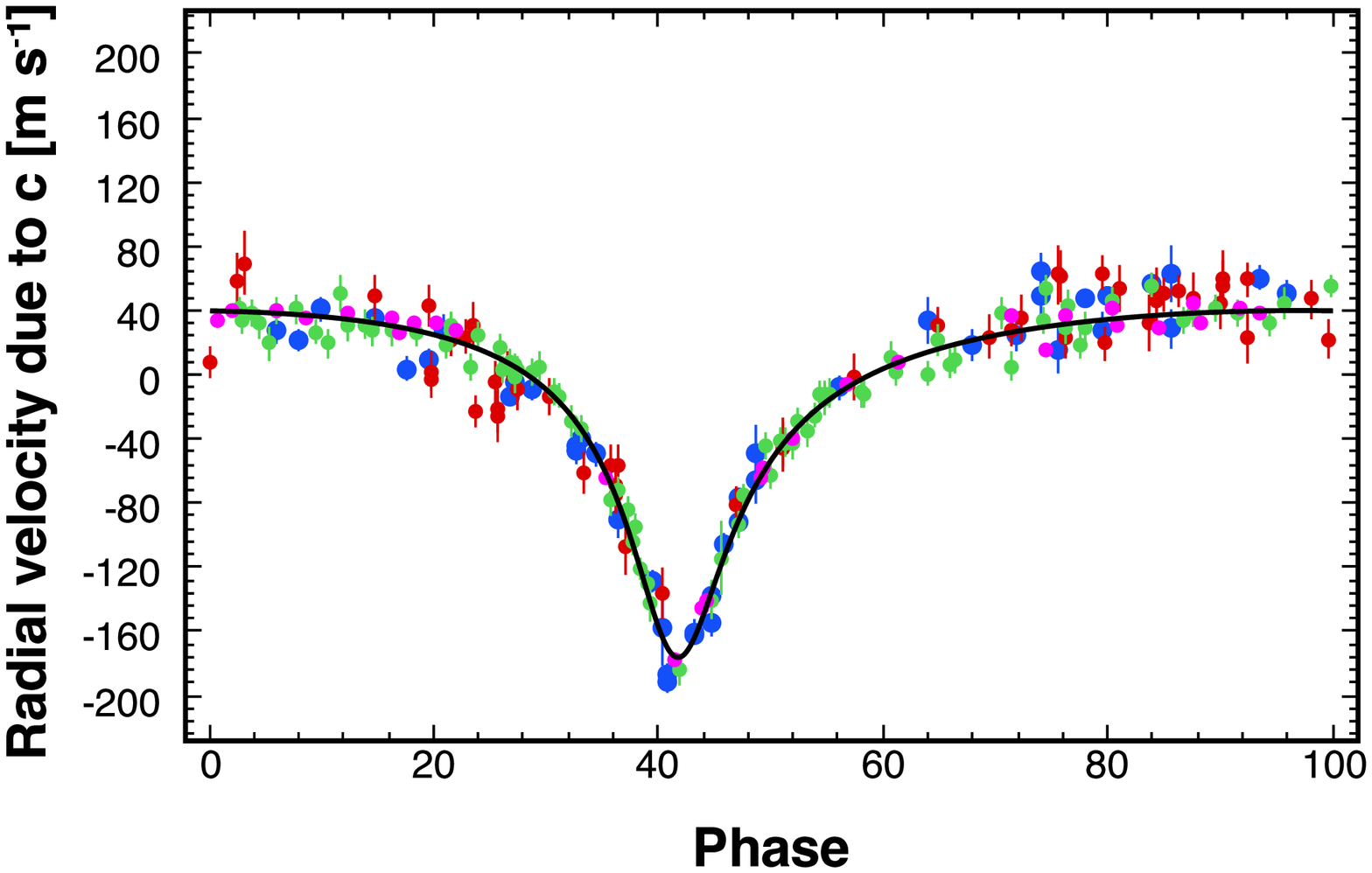}
\plotone{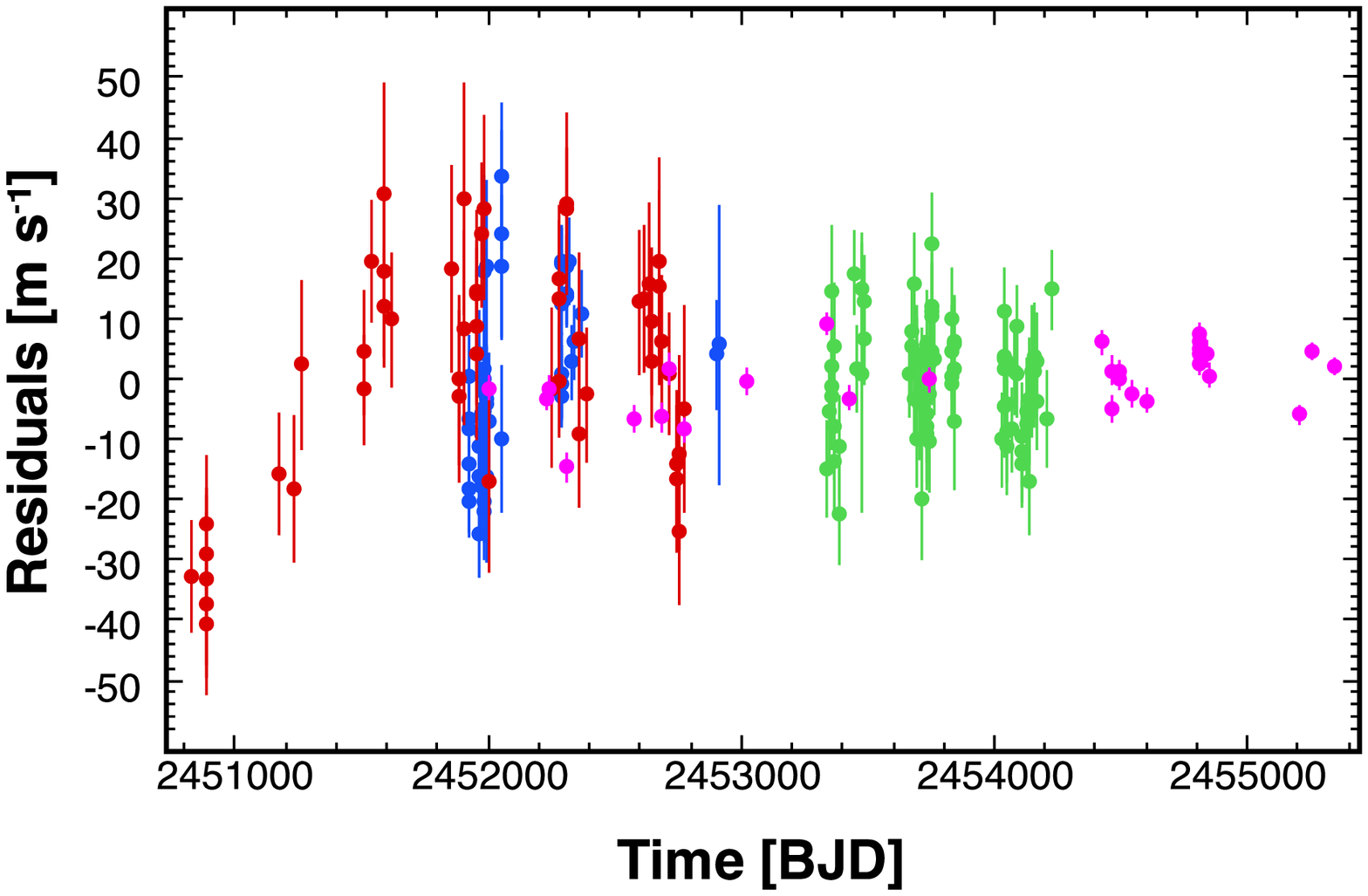}
\plotone{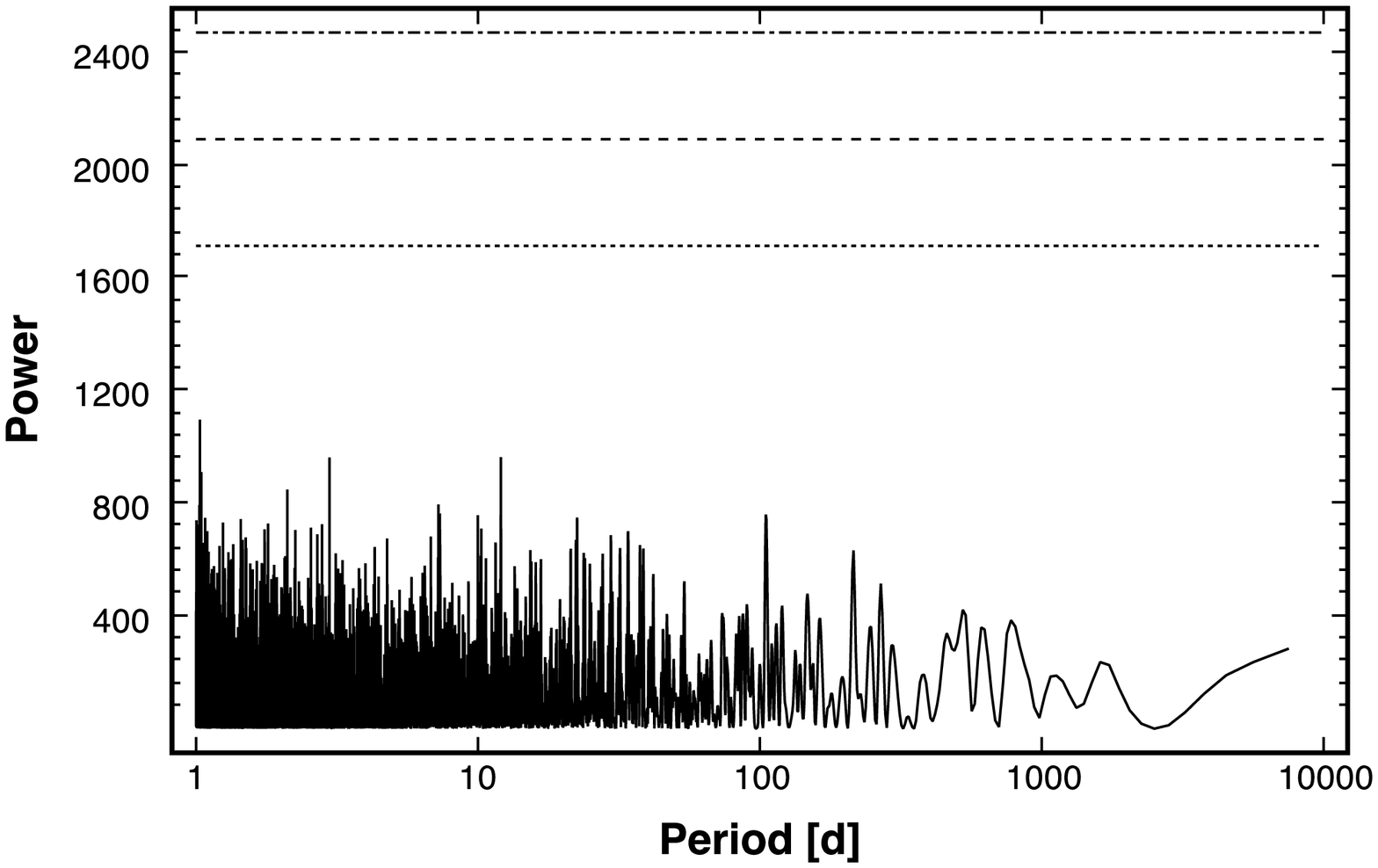}
\caption{Keplerian solution and residuals periodogram for HD 74156.
\textit{1st panel: } Phased Keplerian fit of the 52-d component b.
\textit{2nd panel: } Phased Keplerian fit of the 2514-d component c.
\textit{3rd panel: } Residuals to the 2-planet fit. \textit{4th panel: } Periodogram of the residuals to the 2-planet best fit solution.}
\label{fig:bestfit_HD74156}
\end{figure}

Table \ref{tab:rvdata_HD74156} shows Keck/HIRES relative radial 
velocity observations for HD 74156.  The radial velocity coverage spans 
almost 13 years of RV monitoring. The top panel of Figure \ref{fig:data_HD74156} shows the individual RV observations for HD 74156 
(CORALIE04, ELODIE04 \citet{Naef04}, HET09 \citet{Wittenmyer09} and KECK; each RV dataset has been offset to yield the best-fit solution). The middle panel shows the error-weighted Lomb-Scargle (LS) periodogram of the full RV dataset. Figures \ref{fig:1pfit_HD74156} and \ref{fig:bestfit_HD74156} show the best 1-planet and 2-planet fits, respectively. The best 2-planet fit (derived using the full set of RV observations) obtains a reduced $\chi^2 = 3.09 $, an RMS of the residuals of approximately 12.80 \ms{} and an expected jitter of 
8.59 \ms. 
{The value of the estimated jitter from this best-fit is considerably higher than the 2.2 \ms{} expected from its \rhk\ activity index. However, in this case the RMS  is dominated by the CORALIE and ELODIE data, with a considerable contribution also from the HET data. The RMS of the fit using only the 29 Keck points is 3.5 \ms{} with jitter of 2.9 \ms, in much closer accord with the expected stellar jitter of 2.2 \ms. 

The periodogram shown in the bottom panel of  Figure \ref{fig:bestfit_HD74156}, shows no
compelling peaks in the residuals, indicating that the present data set offers no significant support for additional planets in the system.
Our results confirm the conclusions of \citet{Wittenmyer09}. The expanded dataset presented in this paper does not support the theoretical and observational evidence for a third planetary companion claimed by \citet{Bean08}. 
}

		\begin{deluxetable}{ccc}
		\tablewidth{0pt}
		\tablecaption{KECK radial velocities for HD 74156 (\textit{Sample: full table in electronic version})
		\label{tab:rvdata_HD74156}}
		\tablecolumns{3}
		\tablehead{{Barycentric JD}&{RV [\ms]}&{Uncertainty [\ms]}}
		\startdata
		2452007.90 & 99.69 & 1.97\\ 
2452236.01 & -35.73 & 2.08\\ 
2452243.12 & -35.62 & 1.89\\ 
2452307.89 & 35.08 & 2.32\\ 
2452573.14 & 14.26 & 2.12\\ 
2452682.96 & 0.00 & 2.31\\ 
2452711.82 & -43.62 & 2.32\\ 
2452777.80 & -18.09 & 1.76\\ 
2453017.86 & -135.63 & 2.06\\ 
2453339.08 & -3.45 & 1.83\\ 
2453426.89 & -169.07 & 1.87\\ 
2453746.94 & 153.42 & 1.91\\ 
2454428.07 & 117.86 & 1.83\\ 
2454461.07 & -68.78 & 2.26\\ 
2454464.99 & 32.97 & 2.02\\ 
2454490.94 & 106.20 & 1.98\\ 
2454492.91 & 105.85 & 1.93\\ 
2454545.89 & 94.84 & 2.01\\ 
2454601.83 & 77.13 & 1.95\\ 
2455202.89 & -1.01 & 1.41\\ 

		\enddata
		\end{deluxetable}

\section{Conclusions}\label{sec:conc}

{\begin{figure}
\plotone{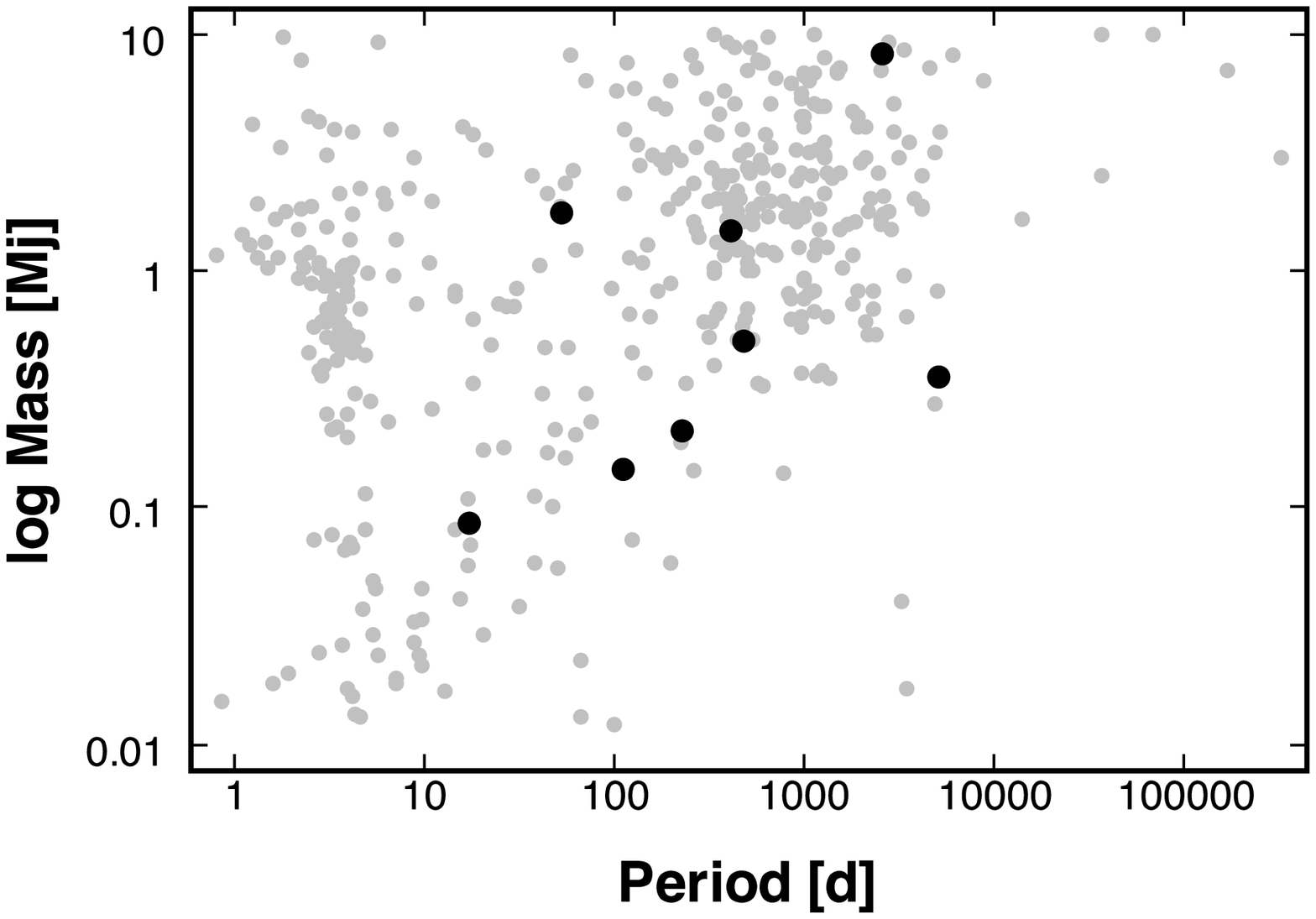}
\plotone{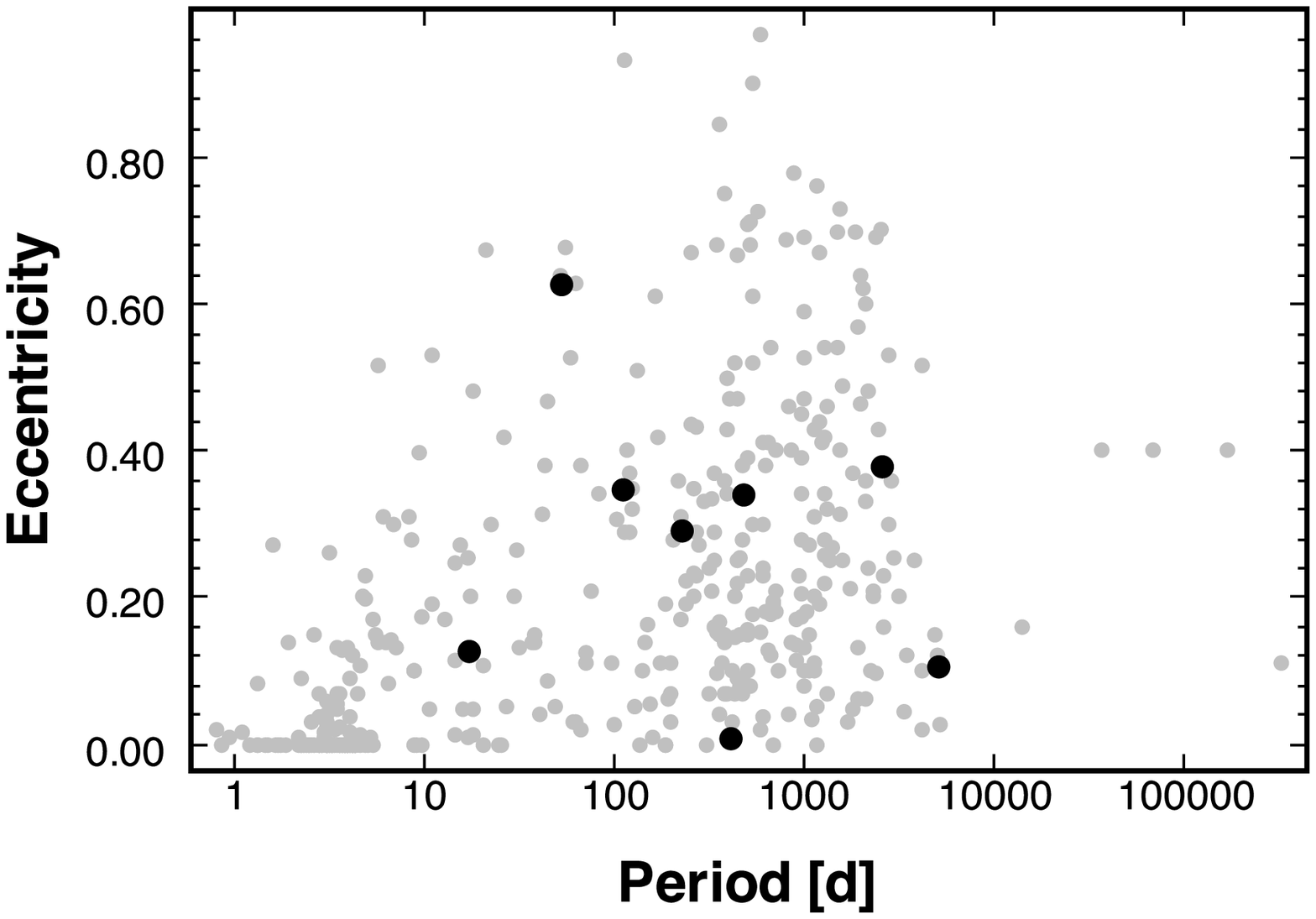}
\caption{Plot of all known extrasolar planets (gray dots) and the orbital elements of all the planets orbiting the host stars presented in this paper (black points).\label{fig:exo}}
\end{figure}

The five systems presented in this paper add to the ever-growing list of single and multiple-components exoplanetary systems. 

Two of the systems, namely, HD 31253 and HD 218566, are well-characterized Saturn-mass planets in $\approx$ 1-year orbits. For both systems, the data presented in this paper do not show prominent peaks in the periodogram of residuals, even at a 10\% FAP level. While several single-planet systems with similar properties have later been characterized with an additional outer, long-period companion \citep[e.g.][]{Jones10}, the absence of any significant linear trend in the current data seem to rule out the presence of additional Jupiter-mass planets with $P <$ 40,000 days.

HD 99492 and HD 177830 each gain a new planetary companion, adding to the previously known planets \citep{Vogt00, Marcy05}. 
The known linear trend in the residuals to HD 99492 b is now fully characterized thanks to the longer phase coverage, indicating the presence of a Saturn-mass planet on a $\sim$5,000-day orbit.  

The RV data we collected for HD 177830 support the existence of an additional inner planet, presenting an interesting case. 
The planets in this system are within a binary with a separation
of approximately 97 AU \citep{Eggenberger07}. Simulations of the formation and stability
of planets in binary star systems imply that the perturbative
effect of the secondary star will be negligible in binaries with
separation larger than 100 AU. The binary system of HD 177830 is
slightly below this limit. This system is also the first binary
with a moderate separation in which multiple planets have been
discovered. Although it is unlikely that the low-mass secondary 
star of this system has had significant effects on the formation
of planets around the primary, it would still be interesting to
study how planets in this system formed and migrated to their 
current stable orbits.

Finally, we analyzed an expanded dataset of Doppler observations of HD 74156, adding 21 Keck RV points to the known data. We repeated the analysis looking for evidence of a third planet, which would lend observational credence to the predictions of the PPS hypothesis \citep[e.g.][]{Barnes04}. Our dataset does not show support for the claimed HD 74156 d planetary companion. Indeed, the residuals periodogram to our best 2-planet fit do not exhibit any promising peaks for future RV follow-ups, strengthening the conclusions of \citet{Wittenmyer09}. 

All the planets presented in this paper lie well within the existing exoplanet parameter envelopes (Fig. \ref{fig:exo}). Several of them lie in the so-called ``desert" in the mass and semi-major axis distribution of extrasolar planets \citep{IdaLin04}. Monte-Carlo population synthesis models for extrasolar giant planet formation tend to suggest that planets migrate relatively rapidly through the period range between 10 and 100 days, and, in addition, often grow quickly through the mass range centered on the Saturnian mass. In the context of the overall planetary census, these four new planets help to further elucidate the various statistical properties of exoplanets. In particular, the discovery of multiple-planet systems helps in further characterizing the number of stars hosting multiple planetary companions and any correlations emerging in the distribution of orbital elements as suggested by observational clues \citep[e.g.][]{Wright09}.

}
\clearpage
\acknowledgments

G.L. acknowledges support from the NASA Astrobiology Institute at NASA Ames Research Center, and from the NSF CAREER Program through NSF Grant AST-0449986. S.S.V gratefully acknowledges support from NSF grants AST-0307493 and AST-0908870.  R.P.B. gratefully acknowledges support from NASA OSS Grant NNX07AR40G, the NASA Keck PI program, and from the Carnegie Institution of Washington. N.H. acknowledges support from the NASA Astrobiology Institute under Cooperative Agreement NNA04CC08A at the Institute for Astronomy, University of Hawaii, and the NASA/EXOB grant NNX09AN05G.
The work herein is based on observations obtained at the W. M. Keck Observatory, which is operated jointly by the University of California and the California Institute of Technology, and we thank the UC-Keck, UH-Keck and NASA-Keck Time Assignment Committees for their support. 
We also wish to extend our special thanks to those of Hawaiian ancestry on whose sacred mountain of Mauna Kea we are privileged to be guests.  Without their generous hospitality, the Keck observations presented herein would not have been possible.  
This research has made use of the NStEd database operated by NASA and the SIMBAD database, operated at CDS, Strasbourg, France. This paper was produced using $^BA^M$.

{\it Facilities:} \facility{Keck (HIRES)}


\begin{thebibliography}{45}
\expandafter\ifx\csname natexlab\endcsname\relax\def\natexlab#1{#1}\fi

\bibitem[{{Baluev}(2009)}]{Baluev09}
{Baluev}, R.~V. 2009, \mnras, 393, 969

\bibitem[{{Barnes} {et~al.}(2008){Barnes}, {Go{\'z}dziewski}, \&
  {Raymond}}]{Barnes08}
{Barnes}, R., {Go{\'z}dziewski}, K., \& {Raymond}, S.~N. 2008, \apjl, 680, L57

\bibitem[{{Barnes} \& {Greenberg}(2007)}]{Barnes07}
{Barnes}, R., \& {Greenberg}, R. 2007, \apjl, 665, L67

\bibitem[{{Barnes} \& {Raymond}(2004)}]{Barnes04}
{Barnes}, R., \& {Raymond}, S.~N. 2004, \apj, 617, 569

\bibitem[{{Barnes}(2001)}]{Barnes01}
{Barnes}, S.~A. 2001, \apj, 561, 1095

\bibitem[{{Bean} {et~al.}(2008){Bean}, {Benedict}, {Charbonneau}, {Homeier},
  {Taylor}, {McArthur}, {Seifahrt}, {Dreizler}, \& {Reiners}}]{Bean08}
{Bean}, J.~L., {Benedict}, G.~F., {Charbonneau}, D., {Homeier}, D., {Taylor},
  D.~C., {McArthur}, B., {Seifahrt}, A., {Dreizler}, S., \& {Reiners}, A. 2008,
  \aap, 486, 1039

\bibitem[{{Bean} \& {Seifahrt}(2009)}]{Bean09}
{Bean}, J.~L., \& {Seifahrt}, A. 2009, \aap, 496, 249

\bibitem[{{Benedict} {et~al.}(2002){Benedict}, {McArthur}, {Forveille},
  {Delfosse}, {Nelan}, {Butler}, {Spiesman}, {Marcy}, {Goldman}, {Perrier},
  {Jefferys}, \& {Mayor}}]{Benedict02}
{Benedict}, G.~F., {McArthur}, B.~E., {Forveille}, T., {Delfosse}, X., {Nelan},
  E., {Butler}, R.~P., {Spiesman}, W., {Marcy}, G., {Goldman}, B., {Perrier},
  C., {Jefferys}, W.~H., \& {Mayor}, M. 2002, \apjl, 581, L115

\bibitem[Bennett et al.(2009)]{Bennett09} Bennett, D.~P., et al.\ 
2009, astro2010: The Astronomy and Astrophysics Decadal Survey, 2010, 18 



\bibitem[{{Bryden} {et~al.}(2009){Bryden}, {Beichman}, {Carpenter}, {Rieke},
  {Stapelfeldt}, {Werner}, {Tanner}, {Lawler}, {Wyatt}, {Trilling}, {Su},
  {Blaylock}, \& {Stansberry}}]{Bryden09}
{Bryden}, G., {Beichman}, C.~A., {Carpenter}, J.~M., {Rieke}, G.~H.,
  {Stapelfeldt}, K.~R., {Werner}, M.~W., {Tanner}, A.~M., {Lawler}, S.~M.,
  {Wyatt}, M.~C., {Trilling}, D.~E., {Su}, K.~Y.~L., {Blaylock}, M., \&
  {Stansberry}, J.~A. 2009, \apj, 705, 1226

\bibitem[{{Butler} {et~al.}(1996){Butler}, {Marcy}, {Williams}, {McCarthy},
  {Dosanjh}, \& {Vogt}}]{Butler96}
{Butler}, R.~P., {Marcy}, G.~W., {Williams}, E., {McCarthy}, C., {Dosanjh}, P.,
  \& {Vogt}, S.~S. 1996, \pasp, 108, 500

\bibitem[{{Butler} {et~al.}(2006){Butler}, {Wright}, {Marcy}, {Fischer},
  {Vogt}, {Tinney}, {Jones}, {Carter}, {Johnson}, {McCarthy}, \&
  {Penny}}]{Butler06}
{Butler}, R.~P., {Wright}, J.~T., {Marcy}, G.~W., {Fischer}, D.~A., {Vogt},
  S.~S., {Tinney}, C.~G., {Jones}, H.~R.~A., {Carter}, B.~D., {Johnson}, J.~A.,
  {McCarthy}, C., \& {Penny}, A.~J. 2006, \apj, 646, 505

\bibitem[{{Charbonneau} {et~al.}(2007){Charbonneau}, {Brown}, {Burrows}, \&
  {Laughlin}}]{Charbonneau07}
{Charbonneau}, D., {Brown}, T.~M., {Burrows}, A., \& {Laughlin}, G. 2007, in
  Protostars and Planets V, ed. B.~{Reipurth}, D.~{Jewitt}, \& K.~{Keil},
  701--716

\bibitem[{{Charbonneau} {et~al.}(2000){Charbonneau}, {Brown}, {Latham}, \&
  {Mayor}}]{Charbonneau00}
{Charbonneau}, D., {Brown}, T.~M., {Latham}, D.~W., \& {Mayor}, M. 2000, \apjl,
  529, L45

\bibitem[{{Chauvin} {et~al.}(2005){Chauvin}, {Lagrange}, {Zuckerman}, {Dumas},
  {Mouillet}, {Song}, {Beuzit}, {Lowrance}, \& {Bessell}}]{Chauvin05}
{Chauvin}, G., {Lagrange}, A.-M., {Zuckerman}, B., {Dumas}, C., {Mouillet}, D.,
  {Song}, I., {Beuzit}, J.-L., {Lowrance}, P., \& {Bessell}, M.~S. 2005, \aap,
  438, L29

\bibitem[Dawson 
\& Fabrycky(2010)]{DawsonFabrycky10} Dawson, R.~I., \& Fabrycky, D.~C.\ 2010, \apj, 722, 937 



\bibitem[{{Eggenberger} {et~al.}(2007){Eggenberger}, {Udry}, {Chauvin},
  {Beuzit}, {Lagrange}, {S{\'e}gransan}, \& {Mayor}}]{Eggenberger07}
{Eggenberger}, A., {Udry}, S., {Chauvin}, G., {Beuzit}, J., {Lagrange}, A.,
  {S{\'e}gransan}, D., \& {Mayor}, M. 2007, \aap, 474, 273

\bibitem[{{Gilliland} \& {Baliunas}(1987)}]{Gilliland87}
{Gilliland}, R.~L., \& {Baliunas}, S.~L. 1987, \apj, 314, 766

\bibitem[{{Henry} {et~al.}(2000){Henry}, {Marcy}, {Butler}, \&
  {Vogt}}]{Henry00}
{Henry}, G.~W., {Marcy}, G.~W., {Butler}, R.~P., \& {Vogt}, S.~S. 2000, \apjl,
  529, L41

\bibitem[{{Ida} \& {Lin}(2004)}]{IdaLin04}
{Ida}, S., \& {Lin}, D.~N.~C. 2004, \apj, 604, 388

\bibitem[{{Jones} {et~al.}(2010){Jones}, {Butler}, {Tinney}, {O'Toole},
  {Wittenmyer}, {Henry}, {Meschiari}, {Vogt}, {Rivera}, {Laughlin}, {Carter},
  {Bailey}, \& {Jenkins}}]{Jones10}
{Jones}, H.~R.~A., {Butler}, R.~P., {Tinney}, C.~G., {O'Toole}, S.,
  {Wittenmyer}, R., {Henry}, G.~W., {Meschiari}, S., {Vogt}, S., {Rivera}, E.,
  {Laughlin}, G., {Carter}, B.~D., {Bailey}, J., \& {Jenkins}, J.~S. 2010,
  \mnras, 403, 1703

\bibitem[{{Kalas} {et~al.}(2008){Kalas}, {Graham}, {Chiang}, {Fitzgerald},
  {Clampin}, {Kite}, {Stapelfeldt}, {Marois}, \& {Krist}}]{Kalas08}
{Kalas}, P., {Graham}, J.~R., {Chiang}, E., {Fitzgerald}, M.~P., {Clampin}, M.,
  {Kite}, E.~S., {Stapelfeldt}, K., {Marois}, C., \& {Krist}, J. 2008, Science,
  322, 1345

\bibitem[{{Marcy} {et~al.}(2005){Marcy}, {Butler}, {Vogt}, {Fischer}, {Henry},
  {Laughlin}, {Wright}, \& {Johnson}}]{Marcy05}
{Marcy}, G.~W., {Butler}, R.~P., {Vogt}, S.~S., {Fischer}, D.~A., {Henry},
  G.~W., {Laughlin}, G., {Wright}, J.~T., \& {Johnson}, J.~A. 2005, \apj, 619,
  570

\bibitem[Marois et al.(2008)]{Marois08} Marois, C., Macintosh, 
B., Barman, T., Zuckerman, B., Song, I., Patience, J., Lafreni{\`e}re, D., 
\& Doyon, R.\ 2008, Science, 322, 1348 



\bibitem[{{Mayor} \& {Queloz}(1995)}]{MayorQueloz95}
{Mayor}, M., \& {Queloz}, D. 1995, \nat, 378, 355

\bibitem[{{Mayor} {et~al.}(2009){Mayor}, {Udry}, {Lovis}, {Pepe}, {Queloz},
  {Benz}, {Bertaux}, {Bouchy}, {Mordasini}, \& {Segransan}}]{Mayor08}
{Mayor}, M., {Udry}, S., {Lovis}, C., {Pepe}, F., {Queloz}, D., {Benz}, W.,
  {Bertaux}, J.-L., {Bouchy}, F., {Mordasini}, C., \& {Segransan}, D. 2009,
  \aap, 493, 639

\bibitem[{{Meschiari} {et~al.}(2009){Meschiari}, {Wolf}, {Rivera}, {Laughlin},
  {Vogt}, \& {Butler}}]{Meschiari09}
{Meschiari}, S., {Wolf}, A.~S., {Rivera}, E., {Laughlin}, G., {Vogt}, S., \&
  {Butler}, P. 2009, \pasp, 121, 1016

\bibitem[Meschiari 
\& Laughlin(2010)]{Meschiari10} Meschiari, S., \& Laughlin, G.~P.\ 2010, \apj, 718, 543 


\bibitem[{{Naef} {et~al.}(2004){Naef}, {Mayor}, {Beuzit}, {Perrier}, {Queloz},
  {Sivan}, \& {Udry}}]{Naef04}
{Naef}, D., {Mayor}, M., {Beuzit}, J.~L., {Perrier}, C., {Queloz}, D., {Sivan},
  J.~P., \& {Udry}, S. 2004, \aap, 414, 351

\bibitem[{{Nordstr{\"o}m} {et~al.}(2004){Nordstr{\"o}m}, {Mayor}, {Andersen},
  {Holmberg}, {Pont}, {J{\o}rgensen}, {Olsen}, {Udry}, \&
  {Mowlavi}}]{Nordstrom04}
{Nordstr{\"o}m}, B., {Mayor}, M., {Andersen}, J., {Holmberg}, J., {Pont}, F.,
  {J{\o}rgensen}, B.~R., {Olsen}, E.~H., {Udry}, S., \& {Mowlavi}, N. 2004,
  \aap, 418, 989

\bibitem[{{Press} {et~al.}(1992){Press}, {Teukolsky}, {Vetterling}, \&
  {Flannery}}]{Press}
{Press}, W.~H., {Teukolsky}, S.~A., {Vetterling}, W.~T., \& {Flannery}, B.~P.
  1992, {Numerical recipes in C. The art of scientific computing} (Cambridge:
  University Press, |c1992, 2nd ed.)

\bibitem[{{Rivera} {et~al.}(2005){Rivera}, {Lissauer}, {Butler}, {Marcy},
  {Vogt}, {Fischer}, {Brown}, {Laughlin}, \& {Henry}}]{Rivera05}
{Rivera}, E.~J., {Lissauer}, J.~J., {Butler}, R.~P., {Marcy}, G.~W., {Vogt},
  S.~S., {Fischer}, D.~A., {Brown}, T.~M., {Laughlin}, G., \& {Henry}, G.~W.
  2005, \apj, 634, 625

\bibitem[{{Silvotti} {et~al.}(2007){Silvotti}, {Schuh}, {Janulis}, {Solheim},
  {Bernabei}, {{\O}stensen}, {Oswalt}, {Bruni}, {Gualandi}, {Bonanno},
  {Vauclair}, {Reed}, {Chen}, {Leibowitz}, {Paparo}, {Baran}, {Charpinet},
  {Dolez}, {Kawaler}, {Kurtz}, {Moskalik}, {Riddle}, \& {Zola}}]{Silvotti07}
{Silvotti}, R., {Schuh}, S., {Janulis}, R., {Solheim}, J.-E., {Bernabei}, S.,
  {{\O}stensen}, R., {Oswalt}, T.~D., {Bruni}, I., {Gualandi}, R., {Bonanno},
  A., {Vauclair}, G., {Reed}, M., {Chen}, C.-W., {Leibowitz}, E., {Paparo}, M.,
  {Baran}, A., {Charpinet}, S., {Dolez}, N., {Kawaler}, S., {Kurtz}, D.,
  {Moskalik}, P., {Riddle}, R., \& {Zola}, S. 2007, \nat, 449, 189

\bibitem[{{Takeda} {et~al.}(2007){Takeda}, {Ford}, {Sills}, {Rasio}, {Fischer},
  \& {Valenti}}]{Takeda07}
{Takeda}, G., {Ford}, E.~B., {Sills}, A., {Rasio}, F.~A., {Fischer}, D.~A., \&
  {Valenti}, J.~A. 2007, \apjs, 168, 297

\bibitem[{{Tanner} {et~al.}(2009){Tanner}, {Beichman}, {Bryden}, {Lisse}, \&
  {Lawler}}]{Tanner09}
{Tanner}, A., {Beichman}, C., {Bryden}, G., {Lisse}, C., \& {Lawler}, S. 2009,
  \apj, 704, 109

\bibitem[{{Trilling} {et~al.}(2008){Trilling}, {Bryden}, {Beichman}, {Rieke},
  {Su}, {Stansberry}, {Blaylock}, {Stapelfeldt}, {Beeman}, \&
  {Haller}}]{Trilling08}
{Trilling}, D.~E., {Bryden}, G., {Beichman}, C.~A., {Rieke}, G.~H., {Su},
  K.~Y.~L., {Stansberry}, J.~A., {Blaylock}, M., {Stapelfeldt}, K.~R.,
  {Beeman}, J.~W., \& {Haller}, E.~E. 2008, \apj, 674, 1086

\bibitem[{{Udry} {et~al.}(2007){Udry}, {Fischer}, \& {Queloz}}]{Udry07}
{Udry}, S., {Fischer}, D., \& {Queloz}, D. 2007, in Protostars and Planets V,
  ed. B.~{Reipurth}, D.~{Jewitt}, \& K.~{Keil}, 685--699

\bibitem[{{Valenti} \& {Fischer}(2005)}]{FischerValenti05}
{Valenti}, J.~A., \& {Fischer}, D.~A. 2005, \apjs, 159, 141

\bibitem[{{Vogt} {et~al.}(2000){Vogt}, {Marcy}, {Butler}, \& {Apps}}]{Vogt00}
{Vogt}, S.~S., {Marcy}, G.~W., {Butler}, R.~P., \& {Apps}, K. 2000, \apj, 536,
  902

\bibitem[{{Vogt} {et~al.}(2010){Vogt}, {Wittenmyer}, {Butler}, {O'Toole},
  {Henry}, {Rivera}, {Meschiari}, {Laughlin}, {Tinney}, {Jones}, {Bailey},
  {Carter}, \& {Batygin}}]{Vogt10}
{Vogt}, S.~S., {Wittenmyer}, R.~A., {Butler}, R.~P., {O'Toole}, S., {Henry},
  G.~W., {Rivera}, E.~J., {Meschiari}, S., {Laughlin}, G., {Tinney}, C.~G.,
  {Jones}, H.~R.~A., {Bailey}, J., {Carter}, B.~D., \& {Batygin}, K. 2010,
  \apj, 708, 1366

\bibitem[{{Vogt} {et~al.}(1994)}]{Vogt94}
{Vogt}, S.~S., {et~al.} 1994, in Society of Photo-Optical Instrumentation
  Engineers (SPIE) Conference Series, Vol. 2198, Society of Photo-Optical
  Instrumentation Engineers (SPIE) Conference Series, ed. D.~L. {Crawford} \&
  E.~R. {Craine}, 362--+

\bibitem[Wittenmyer et al.(2009)]{Wittenmyer09} Wittenmyer, R.~A., 
Endl, M., Cochran, W.~D., Levison, H.~F., 
\& Henry, G.~W.\ 2009, \apjs, 182, 97 


\bibitem[{{Wright}(2005)}]{Wright05}
{Wright}, J.~T. 2005, \pasp, 117, 657

\bibitem[{{Wright} {et~al.}(2004){Wright}, {Marcy}, {Butler}, \&
  {Vogt}}]{Wright04}
{Wright}, J.~T., {Marcy}, G.~W., {Butler}, R.~P., \& {Vogt}, S.~S. 2004, \apjs,
  152, 261

\bibitem[{{Wright} {et~al.}(2007){Wright}, {Marcy}, {Fischer}, {Butler},
  {Vogt}, {Tinney}, {Jones}, {Carter}, {Johnson}, {McCarthy}, \&
  {Apps}}]{Wright07}
{Wright}, J.~T., {Marcy}, G.~W., {Fischer}, D.~A., {Butler}, R.~P., {Vogt},
  S.~S., {Tinney}, C.~G., {Jones}, H.~R.~A., {Carter}, B.~D., {Johnson}, J.~A.,
  {McCarthy}, C., \& {Apps}, K. 2007, \apj, 657, 533

\bibitem[{{Wright} {et~al.}(2009){Wright}, {Upadhyay}, {Marcy}, {Fischer},
  {Ford}, \& {Johnson}}]{Wright09}
{Wright}, J.~T., {Upadhyay}, S., {Marcy}, G.~W., {Fischer}, D.~A., {Ford},
  E.~B., \& {Johnson}, J.~A. 2009, \apj, 693, 1084

\end{thebibliography}
\end{document}